\documentclass[
journal=jctcce,
manuscript=article,
layout=twocolumn
]{achemso}

\usepackage[version=4]{mhchem}

\usepackage{graphicx}
\usepackage{dcolumn}
\usepackage{bm}
\usepackage[T1,T2A]{fontenc}
\usepackage{svg} 
\usepackage[justification=raggedright]{caption}
\usepackage{subcaption}
\usepackage{comment} 
\usepackage{siunitx}
\usepackage{hyperref}
\usepackage{booktabs,siunitx,array} 
\usepackage{dsfont}
\usepackage{relsize}
\usepackage{stfloats} 

\providecommand{\etal}{\emph{et al.}}
\DeclareSIUnit{\cal}{\text{cal}}
\DeclareSIUnit{\angstrom}{\text{\AA}}
\DeclareSIUnit{\amu}{\text{amu}}

\let\oldmaketitle\maketitle
\let\maketitle\relax

\author{Austin Rodriguez}
\affiliation{Department of Chemical Engineering \& Materials Science, Michigan State University, East Lansing, Michigan 48824, United States.}
\author{Justin S. Smith}
\affiliation{NVIDIA Corp., 2788 San Tomas Expy, Santa Clara, California 95051, United States.}
\author{Jose L. Mendoza-Cortes}%
 \email{jmendoza@msu.edu}
\affiliation{Department of Chemical Engineering \& Materials Science, Michigan State University, East Lansing, Michigan 48824, United States.}
\alsoaffiliation{Department of Physics and Astronomy, Michigan State University, East Lansing, MI, 48823, USA.}

\title{Does Hessian Data Improve the Performance of Machine Learning Potentials?}

\begin{document}

\twocolumn[
    \begin{@twocolumnfalse}
    \oldmaketitle
    \vspace{0.5cm}
    \begin{abstract}
    The integration of machine learning into reactive chemistry, materials discovery, and drug design is transforming the development of novel molecules and materials. Machine Learning Interatomic Potentials (MLIPs) predict potential energies and forces with quantum chemistry accuracy, surpassing traditional approaches. Incorporating force fitting in MLIP training enhances potential-energy surface predictions and improves model transferability and reliability.
    
    This article introduces and evaluates the integration of Hessian matrix training in MLIPs, which encodes second-order information about the PES curvature. Our evaluation focuses on models trained only to stable points on the potential surface, demonstrating their ability to extrapolate to non-equilibrium geometries. This integration improves extrapolation capabilities, allowing MLIPs to accurately predict energies, forces, and Hessian predictions for non-equilibrium geometries. Hessian-trained MLIPs enhance reaction pathway modeling, transition state identification, and vibrational spectra predictions, benefiting molecular dynamics (MD) simulations and Nudged Elastic Band (NEB) calculations.
    
    By analyzing models trained with varying combinations of energy, force, and Hessian data on a small molecule reactive dataset, we demonstrate that models including Hessian information not only extrapolate more accurately to unseen molecular systems, improving accuracy in reaction modeling and vibrational analysis, but also reduce the total amount of data required for effective training. However, the primary trade-off is increased computational expense, as Hessian training requires more resources than conventional energy-force training. Our findings provide comprehensive insights into the advantages and limitations of Hessian integration in MLIP training, allowing practitioners in computational chemistry to make informed decisions about employing this method in accordance with their specific research objectives and computational constraints.
    \end{abstract}
\end{@twocolumnfalse}
]
\clearpage

\begin{figure*}[h!]
\centering
\includegraphics[scale=0.95]{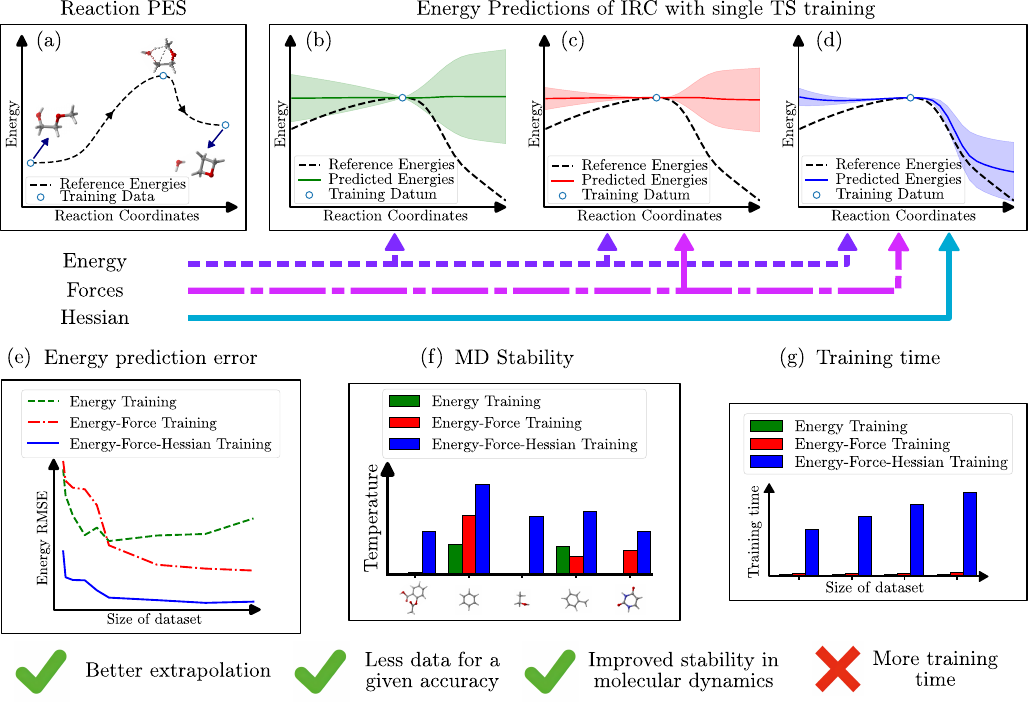}
\caption{Overall view of the effects of using energies, forces, and Hessian information in the MLIP's prediction accuracy of the energetics of molecules outside of the training dataset. The two-dimensional Potential Energy Surface (PES) is shown in plot (a). Plots (b), (c), and (d) show the average predictions of the energies of intermediate structures in a sample chemical reaction's Intrinsic Reaction Coordinate (IRC) path along with their standard deviation (colored areas above and below the average). These predictions are made by ensembles of MLIP models fitted to the energies (b); energies and forces (c); and energies, forces, and Hessian data (d) of only the Transition State (TS) geometry of the same sample reaction calculated using Density Functional Theory (DFT). Plot (e) depicts the energy prediction errors for non-equilibrium structures from models fitted to energies; energy and forces; and energy, forces, and Hessian data trained with different percentages of a reactants, TS, and products dataset. Plot (f) shows the temperature reached until failure of Molecular Dynamics (MD) simulations for a subset of molecules using the different models as force fields. Finally, plot (g) shows the increasing training time for the different fittings with an increasing amount of training data.}
\label{fig:main_fig}
\rule[1ex]{\textwidth}{0.1pt}
\end{figure*}

\section{Introduction}

The application of machine learning to drug and material design promises to revolutionize the development of new molecules and materials for real-world applications \cite{westermayr2021,gomez2018,nandy2021,back2019,smith2019,zuo2020,kulichenko2021,batzner2023,musaelian2023}. Machine learning interatomic potentials (MLIPs), which predict accurate potential energies and forces with quantum chemistry accuracy, are enabling new science by removing computational barriers to many applications, including reactive chemistry \cite{behler2017,segler2018,coley2017,kang2021,schwaller2021}, materials discovery and characterization \cite{pollice2021,rosen2021,ryan2018,graser2018}, and drug design \cite{vamathevan2019,lavecchia2015,gupta2021}. MLIPs are typically trained to density functional theory (DFT) or post Hartree-Fock-calculated potential energies and forces.

The potential energy and atomic forces of a system of atoms are mathematically related; the forces are the negative gradient of the potential energy surface, i.e., the forces are the slope of the potential energy surface at a given point in atom position space. Since most MLIPs are constructed to meet the requirements of a mathematical potential \cite{mueller2020,chan2019}, force fitting through a modified loss function acts as a natural regularization for learning the potential energy surface of a system of atoms \cite{gastegger2017,lee2019,amin2025}. The shift from energy-only training to integrating force training in MLIP development marked a significant advancement around 2018, enhancing the fidelity of atomistic simulations with MLIPs. This transition, highlighted by efforts to simulate infrared spectra using machine learning molecular dynamics \cite{gastegger2017}, underscores a wider trend towards more accurate and transferable MLIPs. Similarly, incorporating stress in addition to force and energy in the loss function has been shown to be crucial for accurately reproducing specific phenomena such as phase transitions, underscoring the nuanced methodological enhancements in MLIP development \cite{shimamura2022}. Furthermore, MLIP packages, such as TorchANI, make force training capabilities widely accessible, encouraging broader application and experimentation within the scientific community \cite{gao2020}.

The Hessian matrix, representing second-order partial derivatives, offers deeper insights into the curvature of the underlying surface for a given geometry. When applied to a potential energy function, such as one describing the potential energy of 3D molecules and materials, the Hessian matrix provides critical information about how a geometry maps to the local surface of the potential: specifically, it can delineate areas of concavity and convexity, identify local minima, maxima, and saddle points. These characteristics are pivotal in the prediction of molecular stability, reaction pathways, and understanding the intrinsic properties of materials at the atomic and molecular levels. Incorporating the Hessian matrix into MLIP training, along with forces, represents a significant step towards refining the accuracy and transferability of these models. The Hessian matrix's ability to pinpoint local minima and saddle points directly enhances MLIPs' capability to accurately model complex energy landscapes. This approach not only captures the forces acting on the atoms (first-order derivatives) but also provides a detailed view of the shape of the energy landscape (through second-order derivatives). Such an enriched representation allows for a more nuanced understanding of chemical reactions, including the identification of transition states and mechanisms of phase transitions.

In this paper, we implement the Hessian loss term for fitting to a reactive chemistry dataset including Hessian-labeled data. Our goal is to evaluate the advantages and disadvantages of including Hessian fittings for developing MLIPs within the context of reactive chemistry. The dataset we deploy in this evaluation includes 35,087 data points from 11,961 reactions \cite{grambow2020}. We test models trained with different combinations of Hessian, force, and energy loss terms and different subsets of our reactive dataset to analyze the impact of using Hessian data in training. As shown in Figure \ref{fig:main_fig}, we demonstrate that, while training with Hessian loss is more expensive, models trained with energy, force, and Hessian data extrapolate better than models trained only with energy or force data. The result of better extrapolation is that fewer overall data are needed in the training process, partially offsetting the higher computational cost. We further show that adding Hessian information allows a model trained only to the stable points (reactant, transition state, and product) of an intrinsic reaction coordinate (IRC) pathway to perform reasonably well on the entire IRC, on perturbed structures along the IRC, and on perturbed structures involved in molecular dynamics (MD) simulations. The most significant downside to adding second derivative information to MLIP training is the increased training time, making the method less ideal for active learning dataset generation techniques. In presenting these findings, we aim to provide computational chemistry practitioners with a thorough understanding of the benefits and limitations of using Hessian data in the training of the MLIP model. This knowledge equips them to make informed decisions about generating and incorporating Hessian data based on their specific research needs and computational resources.

\section{Methodology}

The primary objective of this work is to analyze the impact of incorporating Hessian matrix data from each molecular geometry into machine learning interatomic potential (MLIP) training, such as the ANI model. This incorporation aims to enhance the predictive accuracy and extrapolation capabilities of the models, enabling these models to accurately estimate the potential energies of molecular systems involved in chemical reactions compared to those of DFT calculations, while minimizing the reliance on a vast number of data points.

\begin{table*}[h]
    \centering
        \begin{tabular}{l l l l l}
            \toprule
            & Property & Dimensions & Representation & Units\\
            \midrule
            Inputs & Species & $M \times N$ & $Z_l$ & None\\
            & Coordinates & $M \times N \times 3$ & $x_j$ & \si{\angstrom}\\
            \midrule
            Loss targets & Energy & $M \times 1$ & $E_i$ & \si{\kilo\cal\per\mol}\\
            & Atomic Forces & $M \times N \times 3$ & $\partial E_i / \partial x_j$ & \si{\kilo\cal\per\angstrom\per\mole}\\
            & Hessian Matrix & $M \times 3N \times 3N$ & $\partial^2 E_i / (\partial x_j \partial x_k)$ & \si{\kilo\cal\per\angstrom\squared\per\mole}\\
            \bottomrule
        \end{tabular}
    \caption{Inputs and loss targets used in training the MLIP model}
    \label{tab:features}
\end{table*}

Specifically, each molecular configuration in the dataset that is fed into the model will include the inputs and loss targets provided in Table \ref{tab:features}. The symbol $M$ delineates the aggregate count of molecules present within a given batch or dataset, and $N$ specifies the number of atoms within an individual molecule, or the maximum number of atoms in any one molecule within a batch or dataset (for smaller molecules in the batch or dataset, the excess elements are padded). The term $E_i$ is used to signify the molecular energy associated with molecule $i$, where $i$ ranges from 1 to $M$. Furthermore, the notation $x_j$ and $x_k$ is used to represent coordinates within a three-dimensional Cartesian plane, with $j$ and $k$ ranging from 1 to $3N$.

\subsection{Data Collection and Preparation}

The initial step involves systematically collecting a diverse dataset that encompasses a broad spectrum of molecular geometries relevant to chemical reactions. While this work was in-progress, a database of molecules derived from the QM9 dataset was published containing numerical Hessian matrices. These molecular configurations were exclusively equilibrium geometries \cite{williams2025,ruddigkeit2012,ramakrishnan2014}. In this work we focus on developing a database of Hessian calculations from structures corresponding to minima points as well as saddle points (transition states) of a large set of reactions. We generated a comprehensive initial data set that contains tens of thousands of elementary chemical reactions based on the data set of Grambow \etal \cite{grambow2020}. Each reaction in the dataset comprises DFT-optimized geometries of reactants, products, and transition states, providing detailed information on the molecular structures and energetics involved. The geometry optimizations in that original dataset were performed using the $\omega$B97XD functional and the def2-TZVP basis set.

Starting with optimized structures from this initial dataset, we conducted further geometry optimizations for each reactant, transition state, and product to ensure that the geometries are in the local minima / saddle point consistent with our selected software implementation and basis set. Furthermore, we performed frequency analysis calculations to derive the analytical Hessian matrix via DFT. All calculations were performed using the Gaussian16 software package at the level of theory $\omega$ B97XD / 6-31g (d) \cite{Gaussian16}. The convergence criteria for molecular geometry optimizations in Gaussian16 require the maximum and RMS force components to be below \( 4.5 \times 10^{-4} \) and \( 3.0 \times 10^{-4} \, \text{Hartree/Bohr}\), respectively, and the maximum and RMS atomic displacements to be under \( 1.8 \times 10^{-3} \) and \( 1.2 \times 10^{-3} \, \text{Bohr}\), respectively. 

The choice of the $\omega$B97XD functional and the 6-31g(d) basis set is motivated by several factors. Firstly, the $\omega$B97XD functional is a widely used hybrid density functional that incorporates the long-range correction to improve the description of non-covalent interactions, such as hydrogen bonding and dispersion forces. \cite{grimme2006,grimme2006semiempirical} This functional has shown good performance in reproducing a wide range of molecular properties, making it suitable for studying chemical reactions involving H, C, N, and O. \cite{minenkov2012} Additionally, the 6-31g(d) basis set provides a balanced description of the electronic structure while maintaining computational efficiency. \cite{ditchfield1971,hehre1972,hariharan1973,hariharan1974,gordon1980,francl1982} It includes polarization functions that capture the electron density redistribution around atoms, allowing for an accurate representation of charge distributions and molecular properties. Another important consideration for using the $\omega$B97XD functional and the 6-31g(d) basis set is its compatibility with the ANI-1x model, on which our Hessian-Trained MLIP is based \cite{Smith2017-1,Smith2017-2,Smith2018,Smith2020,gao2020,Devereux2020}.

Our dataset is a rich collection that includes not only 35,087 molecular geometries from 11,961 elementary chemical reactions but also key properties such as electronic energies, atomic forces, and Hessian matrices. In addition, we provide a detailed analysis of the composition of the dataset, including distributions of atom types and frequencies of bond types, in Figures S1 and S2 in the supplementary information. To evaluate the Hessian-trained MLIP ability to extrapolate, we develop a data set of 34,248 structures from 600 Intrinsic Reaction Coordinate (IRC) paths for a randomly selected subset of the reactions. These IRC geometries act as a rigorous benchmark, enabling us to test the accuracy of our MLIP in predicting reaction pathways.  Finally, for testing the ability of the Hessian-trained MLIP to extrapolate its predictions to other non-equilibrium structures outside the IRC path, we generated 62,527 perturbed molecular structures from the intermediate IRC structures of 574 reactions in the randomly selected subset via Normal Mode Sampling (NMS). This comprehensive overview ensures that the diversity and coverage of the data set is suitable to evaluate the machine learning model and to apply the Hessian-trained MLIP.

\subsection{Hessian Matrix Incorporation} 

To achieve a more detailed understanding of the Potential Energy Surface (PES) of molecular systems, extending our dataset to include information beyond merely potential energy and atomic forces is crucial. A pivotal advancement in this work is the integration of the Hessian matrix in the loss function during MLIP model training. The Hessian matrix, detailed in Equation S1, contains the second derivatives of the total molecular energy with respect to the atomic positions for each molecular geometry within our dataset. The Hessian matrix contains information about the dynamic properties of chemical systems such as the vibrational charateristics of molecules and their stability.

\subsection{Machine Learning Model Adaptation}

The architecture of the ANI model makes use of modified Behler and Parrinello symmetry functions to capture the chemical environments that surround individual atoms \cite{Behler2007,Behler2011}. These environments are encoded within Atomic Environment Vectors (AEVs), which serve as a probe into the radial and angular domains surrounding an atom. Upon translating the atomic coordinates of a chemical system into AEVs, these vectors become the inputs for a specialized form of High Dimensional Neural Network Potentials (HD-NNP) \cite{Behler2015}. Different NNPs are deployed for each element, each equipped with its own set of weights and biases tailored to the element's specific characteristics through training. These neural networks undergo optimization (or training) processes, fine-tuning their parameters to align with the high-dimensional details captured in the dataset. Architecturally, these HD-NNPs are structured as feedforward neural networks, featuring multiple hidden layers and a variety of neurons. The outputs of each of these NNPs correspond to a partition per atom of the molecular potential energy. These values are summed up to obtain the potential energy. The general flow of information is visualized in Figure \ref{fig:formaldehyde-HD-NNP}, where the energy of a single formaldehyde molecule is calculated using the MLIP model.

\subsubsection{Force and Hessian Calculation Using Automatic Differentiation}

Calculating atomic forces and Hessian matrices is possible with the use of automatic differentiation, a powerful computational technique that offers a robust and efficient means of determining the gradients of complex functions. This approach leverages the inherent architecture of the neural network, which, by design, facilitates the direct differentiation of the energy output with respect to its inputs. This capability is crucial for accurately modeling the dynamics of molecular systems, as it allows for the prediction of first- and second-order derivatives under a wide range of conditions with minimal computational overhead. 

In practice, we have to feed the MLIP with batches of hundreds of molecules at a time, and the dimensionality of the PyTorch tensors is critical when training these NNP models. As mentioned in Table \ref{tab:features}, the shape of the coordinates, as well as the force tensor, is $[batch\_size, N, 3]$, where $batch\_size$ is the number of molecules in a batch and $N$ is the number of atoms of the molecule with the highest number of atoms in the dataset. If the number of atoms in any particular molecule is less than $N$, the rest of the tensor is padded with zeros. In this manner, the coordinates and force tensors grow along one dimension until they reach the appropriate shape. The atomic forces of a batch are calculated simultaneously for all molecules in a batch via the \textit{autograd.grad} function from PyTorch, resulting in a matrix containing the first derivatives of the energy with respect to the coordinates for each molecule, which we multiply by $-1$ to obtain the forces. Additionally, the Hessian matrix can be calculated from our model in a similar manner. By entering the negative of the forces into the \textit{autograd.grad} function, we can obtain the second derivatives of the energy with respect to each pair of Cartesian coordinates. However, there are some difficulties associated with the Hessian matrix format. For example, the Hessian tensor grows exponentially with the number of atoms (having $3N \times 3N$ elements), where molecules with fewer than $N$ atoms have padded Hessian matrices. This causes the unpadded tensor to grow in two dimensions instead of one, since the Hessian tensor has a shape of $[batch\_size, 3N, 3N]$. Equations \ref{eq:Forces_elements} and \ref{eq:Hessian_elements} are derived using the chain rule and are a simple representation of the calculations made by automatic differentiation.

\vspace{-0.5cm}
\begin{equation}
F_{x_m} = -\frac{\partial E_T}{\partial x_m} = -\sum_{k=1}^3 \left( \frac{\partial E_T}{\partial G_k} \times \frac{\partial G_k}{\partial x_m} \right)
\label{eq:Forces_elements}
\end{equation}

\begin{equation}
\begin{split}
H_{x_m x_n} & = \frac{\partial (\partial E_T / \partial x_m)}{\partial x_n} \\& = \frac{\partial}{\partial x_n} \left[ \sum_{k=1}^3 \left( \frac{\partial E_T}{\partial G_k} \times \frac{\partial G_k}{\partial x_m} \right) \right] \\& = \sum_{k=1}^3 \left( \frac{\partial E_T}{\partial G_k} \times \frac{\partial^2 G_k}{\partial x_n \partial x_m} \right)
\end{split}
\label{eq:Hessian_elements}
\end{equation}

\subsubsection{Loss Function}

Training to Hessian data requires incorporating an error metric, such as the Root Mean Square Error (RMSE), of the Hessian matrix into the loss function of a MLIP as well as error metrics for the energies and forces. The RMSE for energies directly evaluates the model's ability to predict the potential energy of a given geometry, while for forces\textemdash which are derived from the gradient of energy with respect to atomic positions\textemdash it measures the model's accuracy in predicting the high-dimensional slope of the potential energy. The inclusion of the Hessian RMSE extends this further by assessing the model's accuracy in predicting the high-dimensional curvature of the energy landscape, which is crucial for identifying maxima, minima, and stable points. These are crucial for modeling important properties such as transition states, and vibrational frequencies. Equations \ref{eq:energy_loss}, \ref{eq:force_loss}, and \ref{eq:Hessian_loss} are used in the calculation of the energy loss, force loss, and Hessian loss terms, respectively. In this set of equations, $P^{ref}$ represents the reference property used in the training as ground truth (in our case, they are the properties calculated by DFT), while $P^{pred}$ represents the predicted property obtained from our model. The properties can be the molecular potential energy $E_{T,i}$, a force component $F_i$, or a Hessian element $H_i$. Furthermore, $M$ represents the total number of molecules in the training set, $n_F$ represents the total number of force elements in the training set, and $n_H$ represents the total number of Hessian elements in the training set.

\begin{equation}
\mathlarger\varepsilon_E = \sqrt{\frac{\sum_{i=1}^M \left( E_{T,i}^{ref} - E_{T,i}^{pred} \right)^2}{M}}
\label{eq:energy_loss}
\end{equation}

\begin{equation}
\mathlarger\varepsilon_F = \sqrt{\frac{\sum_{i=1}^{n_F} \left( F_{i}^{ref} - F_{i}^{pred} \right)^2}{n_F}}
\label{eq:force_loss}
\end{equation}

\begin{equation}
\mathlarger\varepsilon_H = \sqrt{\frac{\sum_{i=1}^{n_H} \left( H_{i}^{ref} - H_{i}^{pred} \right)^2}{n_H}}
\label{eq:Hessian_loss}
\end{equation}

However, the magnitudes of energies, forces, and elements of the Hessian matrix can vary significantly, both in terms of their physical units and their scales within a given problem. To ensure that each component contributes appropriately to the loss function, normalization factors are essential. These factors adjust the scale of the RMSE values for each term, enabling a balanced optimization that does not disproportionately favor the accuracy of one property over another. By applying normalization factors, we ensure that the model is optimized for an equitable accuracy across these properties, facilitating the development of a more reliable and versatile predictive tool. The final loss function used in our MLIP model is represented in Equation \ref{eq:loss_function}, where $\eta_F$ is the normalization factor for the force loss and $\eta_H$ is the normalization factor for the Hessian loss. The values of $\eta_F = 0.08$ and $\eta_H = 0.02$ were used in our trainings. These values were selected through empirical testing.

\begin{equation}
L(E^{pred},F^{pred},H^{pred}) = \mathlarger{\varepsilon}_{_E} + \eta_{_F} \mathlarger{\varepsilon}_{_F} + \eta_{_H} \mathlarger{\varepsilon}_{_H}
\label{eq:loss_function}
\end{equation}

\subsection{Molecular Dynamics Simulations}

To evaluate the dynamical stability of the Hessian-trained MLIP versus a non-Hessian trained MLIP, MD simulations were conducted. In the following sub-sections we present the methods used in this evaluation.

\subsubsection{Simulation Setup}

MD simulations were performed using the Hessian- and non-Hessian trained MLIPs as the force field, implemented within the Atomic Simulation Environment (ASE) framework. A Langevin thermostat was used to control the temperature during the simulations.

The initial atomic configurations for the simulations were derived from optimized molecular structures. Each system was initialized at a starting temperature of 5 Kelvin. A time step of 0.5 femtoseconds was employed. A simulation is then conducted for 5 picoseconds or 10,000 time steps. After the initial 5-picosecond interval is complete, the temperature is increased by 5 Kelvin. An iterative heating protocol (5 picoseconds simulation followed by 5 kelvin temperature increase) was continued until a predefined failure criterion was met.

\subsubsection{Stability Failure Criteria}

 In this work, we define the failure of stability as shown below. Failure occurs if:
\begin{equation}
\begin{aligned}
    &\exists (i, j) \quad \text{s.t.} \quad\frac{1}{50} \sum_{t=t-49}^{t} d_{ij}(t) > 1.5 d_{ij}^\text{eq}, \\
    &\text{or} \\
    &\exists (i, j) \quad \text{s.t.} \quad\frac{1}{50} \sum_{t=t-49}^{t} d_{ij}(t) < 0.75 d_{ij}^\text{eq}
\end{aligned}
\label{eq:failure_criteria}
\end{equation}

Where $d_{ij}(t)$ is the distance between the atoms $i$ and $j$ at time step $t$ and $d_{ij}^\text{eq}$ is the geometry-optimized equilibrium bond distance between the atoms $i$ and $j$. For the molecules evaluated in this work, which are from the MD17 benchmark, previous literature conducted ab initio MD simulations at 500K for between 50 and 497 picoseconds where such distortions were not observed \cite{chmiela2017,christensen2020}. Hence, we do not expect physically accurate reactions to occur up to a temperature of 500K during our simulations.

\subsection{Computation of Reaction Pathways}

To evaluate the ability of Hessian-trained MLIPs to accurately describe reaction pathways and transition states, Nudged Elastic Band (NEB) calculations were performed. The NEB method determines a minimum energy pathway (MEP) by optimizing a set of interpolated molecular geometries, referred to as images, between the reactant and product states. These images are connected by virtual spring forces, which maintain an even distribution along the reaction coordinate and prevent artificial clustering in low-energy regions.

\subsubsection{Computational Setup}

NEB calculations were carried out using the ASE framework. The reactant and product geometries were first optimized both at the DFT level and at the model level before being used as endpoints for the NEB calculations. The key computational parameters were as follows:
\begin{itemize}
    \item 58 images were used to achieve a smooth representation of the reaction path.
    \item A force constant of 50 eV/\AA$^2$ was applied.
    \item The LBFGS optimizer was used for relaxation.
    \item Climbing Image NEB (CI-NEB) was applied to the highest-energy image to accurately locate the transition state (TS).
\end{itemize}

\subsection{Calculation of Reaction Vibrational Spectra}

To calculate the vibrational spectra of a reaction, the Hessian matrices of each step in the MEP of a chemical reaction is calculated by either a frequency calculation job from Gaussian16 to obtain the DFT-calculated Hessian or by automatic differentiation of the energy predictions of the Hessian-trained MLIP to obtain the model's Hessian prediction. Vibrational frequencies were obtained from the diagonalization of the Hessian matrix at each point along the reaction coordinate. The computed normal modes describe molecular distortions, while their corresponding frequencies provide a spectroscopic fingerprint of the system at each step.

\section{Results and Discussion}

This section evaluates how the ANI machine learning interatomic potential (MLIP) model performs when trained with different complexities of information\textemdash single-point energies (E), atomic force vectors (E-F), and the Hessian matrix (E-F-H)\textemdash through increasingly stringent validation scenarios. The analysis begins with training an ensemble of 100 ANI models to a single reaction transition state (one data point), with the goal of understanding how higher-order data impact the models' ability to extrapolate to the complete IRC pathway. Evaluation expands by training a set of models to a diverse data set of stationary points, that is, reactants, transition states, and products (hitherto referred to as the RTP training set and models), then assessing the extrapolation capabilities of the models across various molecular configurations and reaction pathways, including stable points, IRC, molecular dynamics trajectories, vibrational frequencies, and perturbed structures along the IRC. We then explore the data efficiency and the computational demands of each approach.

\subsection{Impact on Extrapolation}

In this section, we test the hypothesis (see Figure \ref{fig:main_fig}) that Hessian matrix data improve extrapolation performance of the neural network-based ANI MLIP through experimentation. In this experiment, we train three types of models with different loss functions: energy (E), energy-force (E-F), and energy-force-hessian (E-F-H). Based on equation \ref{eq:loss_function}, the E models have $\eta_F = \eta_H = 0$, the E-F models have a non-zero $\eta_F$ and $\eta_H = 0$, while the E-F-H models have non-zero $\eta_F$ and $\eta_H$. An ensemble of 100 ANI models is trained for each loss to a single data point, the transition state of the reaction \ce{H2O + C3H6O -> CH3OCH2CH2OH}. We then compare the three ensembles of models\textemdash E, E-F and E-F-H\textemdash to predict discrete points along the entire IRC from reactant to product.

\noindent\hrulefill
\begin{figure}[h!]
\captionsetup[subfigure]{justification=centering,font={stretch=0.8}}
\centering
\begin{subfigure}{0.46\textwidth}
\centering
\includegraphics[scale=0.42]{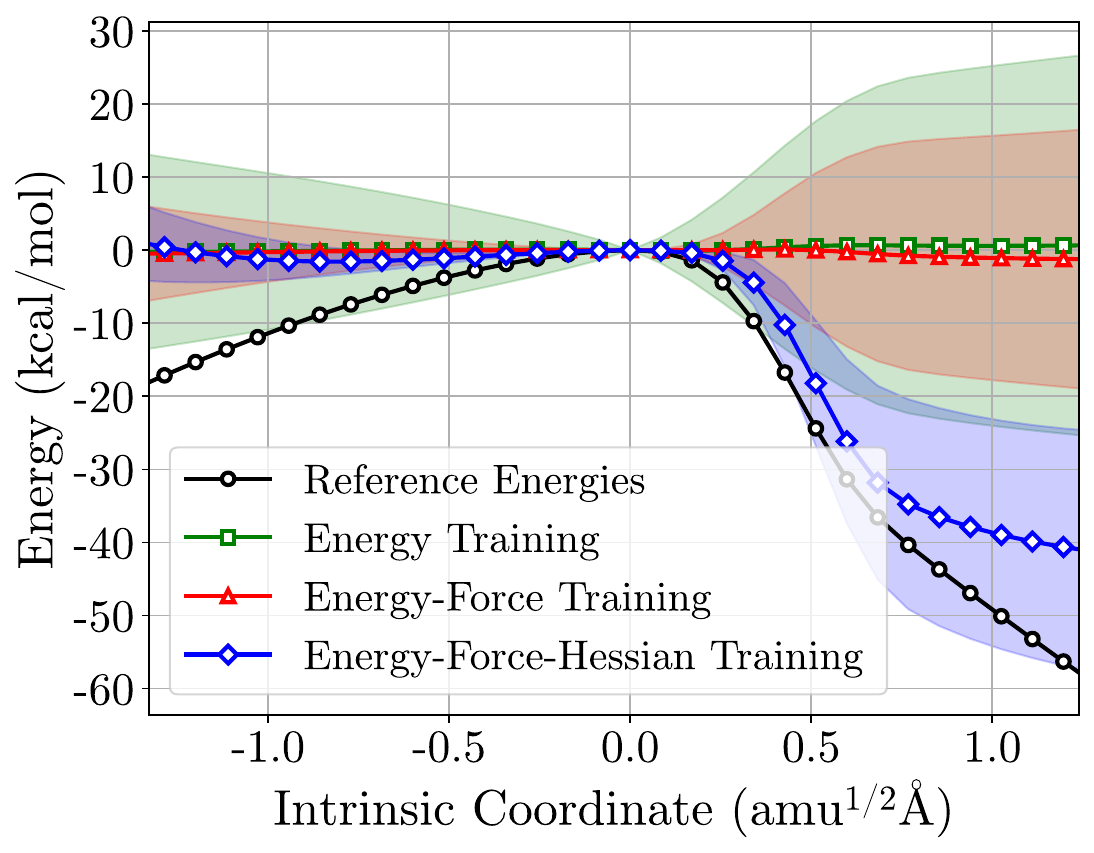}
\caption{Energy predictions for the 30 molecular structures closest to the TS.}
\label{subfig:TS-training-zoomed-in}
\end{subfigure}
\vspace{4pt}
\\
\begin{subfigure}{0.46\textwidth}
\centering
\includegraphics[scale=0.42]{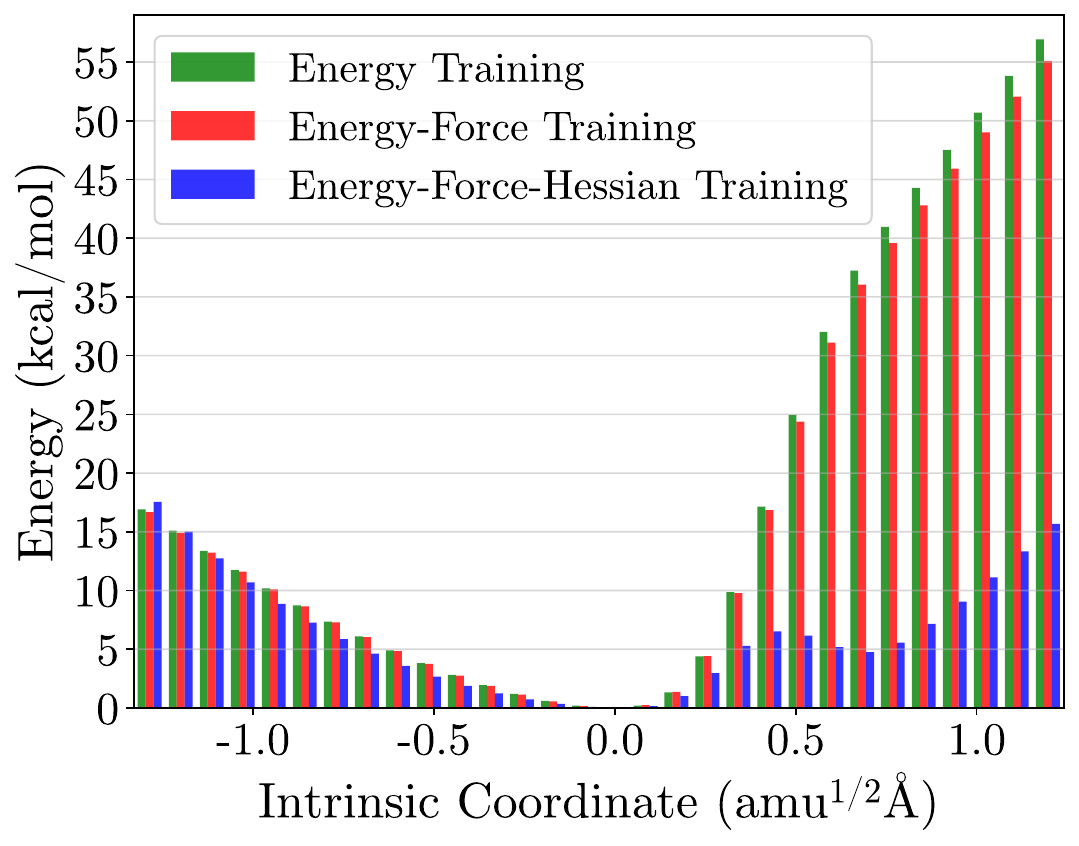}
\caption{Mean error bars for the 30 molecular structures closest to the TS.}
\label{subfig:mean-errors}
\end{subfigure}

\caption{Energy predictions for the 30 molecular structures in the IRC closest to the TS structure (a) with their mean errors along with a stacked bar plot of the mean errors of the models (b). The energy fitting, energy-force fitting, and energy-force-Hessian fitting models are represented with the colored markers, colored error areas, and colored error bars. DFT-calculated energies are shown as black circles in (a).}
\label{fig:TS-training-zoomed-in}
\end{figure}

In Figure \ref{subfig:TS-training-zoomed-in}, the models predict the energy of the TS point well. However, as expected, the farther the test structure is from the TS, the worse the models perform. Figure \ref{subfig:mean-errors} shows that all models perform poorly for predictions far from the TS. Both the E and E-F models show increasing uncertainty and error at $\pm 0.25$ IRC units (\si{\amu\tothe{1/2}\angstrom}) from the TS.

The E-F-H models follow the reference potential energies better, especially on the right side of the TS, with lower uncertainties and errors until $+0.6$ IRC units. However, they infer an erroneous energy increase below $-1.0$ IRC units, as seen in Figure \ref{fig:TS-training}. Since the models were trained on a single data point in the IRC, we do not expect reasonable predictions far from the training point.

Figure \ref{fig:TS-training-zoomed-in} provides a detailed description of energy predictions for the 30 points closest to the TS, including a stacked mean error plot. The E models diverge from the TS energy sooner than the E-F models, indicating that models trained with a force term learn the zero gradient of the PES at the TS and maintain a constant energy around it.

The E-F-H models are the only ensemble that learned the shape of the PES around the TS, with most predictions lower than the TS energy, indicating they learned the TS represents a local maximum. Despite some random behavior in energy predictions due to weight initialization, as shown in Figure \ref{fig:ensemble-predictions}, the majority of E-F-H models retain information about the PES shape near the TS. The evidence provided by this test case shows a strong tendency for Hessian training to enable local extrapolation on the potential energy surface compared to training without it.

\subsection{Testing the models on held-out reactants, transition states, and products}

In this section, the MLIP models are trained with the RTP dataset, consisting of 35,087 molecular geometries from 11,961 elementary chemical reactions, and tested using a reserved set of reactants, transition states, and products from the same dataset. This phase assesses the models' capacity to generalize to unseen chemical systems by predicting the potential energies of stationary points (reactants, transition states, and products). Although specific stationary point geometries in the test set do not appear in the training set, other geometries from the same reactions might. However, we do not expect this to affect the significance of our model comparison.

Table \ref{tab:test-set-RMSE} presents the RMSE for each loss function used in our models. The RMSE for the E-F model's energy predictions is $4.29$ \si{\kilo\cal\per\mole}, higher than the E model's $3.83$ \si{\kilo\cal\per\mole}. This increase is due to the added complexity of training the model to attain the correct slope of the potential surface, minimizing random behavior. However, the E-F-H model shows a lower RMSE of $3.67$ \si{\kilo\cal\per\mole} for energy predictions.

Interestingly, the E-F model outperforms the E-F-H model in force prediction ($4.87$ \si{\kilo\cal\per\mole} vs $5.61$ \si{\kilo\cal\per\mole}) on this set of stationary test points. This unexpected result could be due to overfitting, where the model predicts a flatter potential across phase space. This hypothesis will be tested as benchmark cases move further from equilibrium structures with nonzero reference forces. The E model has significantly higher errors for forces, indicating substantial overfitting, with the slope of the potential surface at stationary points approaching random.

For the Hessian prediction task, both the E and E-F models show an order of magnitude or worse RMSE increase compared to the E-F-H model, providing initial evidence of the E and E-F models high degree of overfitting.

\subsection{Testing the model on IRC structures}

In this benchmark, the E, E-F, and E-F-H models trained on the RTP dataset are evaluated on structures derived from IRC paths, representing the minimum energy path from reactants, through transition states, to products.

Table \ref{tab:IRC-RMSE} quantifies the prediction accuracy for IRC structures using the RMSE metric for energy, force, and Hessian predictions compared to density functional theory reference calculations. The results show that adding higher-order information (forces and Hessian) to the model increases accuracy for energy predictions along the minimum energy pathway. Adding a force term to the loss function reduces the energy prediction error by 31.3\% to 7.30 \si{\kilo\cal\per\mole}, while adding both forces and Hessian terms reduces it by 38.3\% to 6.55 \si{\kilo\cal\per\mole} compared to the E model (10.62 \si{\kilo\cal\per\mole}).

Although the training molecules consist only of stationary points (reactants, transition states, and products), predicting intermediate structures in a reaction is impressive but still an interpolation task. As expected, the E models perform poorly in force prediction with an RMSE of 58.06 \si{\kilo\cal\per\mole}. However, the E-F-H model (RMSE 7.30 \si{\kilo\cal\per\mole}) outperforms the E-F model (8.48 \si{\kilo\cal\per\mole}) by 13.9\% in force prediction, indicating that the E-F model's better performance in the stationary point benchmark was due to overfitting. This trend continues in other benchmark cases.

For Hessian prediction, both the E and E-F models perform poorly, with more than an order of magnitude increase in RMSE compared to the E-F-H model, further indicating overfitting in the E-F model.

\begin{table*}[h!]
    \centering
    \begin{tabular}{l r r r}
        \toprule
        Training & {Energy RMSE} & {Force RMSE} & {Hessian RMSE}\\
        & {(\si{\kilo\cal\per\mole})} & {(\si{\kilo\cal\per\angstrom\per\mole})} & {(\si{\kilo\cal\per\angstrom\squared\per\mole})}\\
        \midrule
        E fitting & $3.83 \pm 0.23$ & $53.14 \pm 2.90$ & $208.58 \pm 15.17$\\
        E-F fitting & $4.29 \pm 0.18$ & $4.87 \pm 0.10$ & $146.32 \pm 1.90$\\
        E-F-H fitting & $3.67 \pm 0.11$ & $5.61 \pm 0.16$ & $12.76 \pm 0.24$\\
        \bottomrule
    \end{tabular}
    \caption{Energy, force, and Hessian RSMEs of each model fitting tested on 35,087 reactants, transition states, and products in the database.}
    \label{tab:test-set-RMSE}
\end{table*}
~~
\begin{table*}[h!]
    \centering
    \begin{tabular}{l r r r}
        \toprule
        Training & {Energy RMSE} & {Force RMSE} & {Hessian RMSE}\\
        & {(\si{\kilo\cal\per\mole})} & {(\si{\kilo\cal\per\angstrom\per\mole})} & {(\si{\kilo\cal\per\angstrom\squared\per\mole})}\\
        \midrule
        E fitting & $10.62 \pm 0.37$ & $58.06 \pm 2.79$ & $218.79 \pm 18.09$\\
        E-F fitting & $7.30 \pm 0.15$ & $8.48 \pm 0.11$ & $143.59 \pm 1.77$\\
        E-F-H fitting & $6.55 \pm 0.14$ & $7.30 \pm 0.10$ & $14.66 \pm 0.12$\\
        \bottomrule
    \end{tabular}
    \caption{Energy, force, and Hessian RSMEs of each model fitting tested on 34,248 structures in the IRCs of around 574 reactions (See SI, Figure \ref{fig:data-efficiency_IRC}).}
    \label{tab:IRC-RMSE}
\end{table*}
~~
\begin{table*}[h!]
    \centering
    \begin{tabular}{l r r r}
        \toprule
        Training & {Energy RMSE} & {Force RMSE} & {Hessian RMSE}\\
        & {(\si{\kilo\cal\per\mole})} & {(\si{\kilo\cal\per\angstrom\per\mole})} & {(\si{\kilo\cal\per\angstrom\squared\per\mole})}\\
        \midrule
        E fitting & $46.38 \pm 2.35$ & $81.97 \pm 3.64$ & $231.07 \pm 23.91$\\
        E-F fitting & $21.67 \pm 0.75$ & $26.53 \pm 0.35$ & $128.09 \pm 1.27$\\
        E-F-H fitting & $13.52 \pm 0.21$ & $13.47 \pm 0.29$ & $37.82 \pm 1.77$\\
        \bottomrule
    \end{tabular}
    \caption{Energy, force, and Hessian RSMEs of ensembles of each model fitting tested on 62,527 perturbed structures along the IRCs of randomly selected reactions via NMS.}
    \label{tab:NMS-RMSE}
\end{table*}

\subsection{Testing the model on NMS structures outside of the IRC path}

This section aims to evaluate the extrapolation capabilities of the MLIP models trained on the RTP dataset. The models are tested on perturbed structures derived from IRC pathways and normal mode sampling (NMS), which introduces thermodynamic variability in structural phase space. NMS generates structures which are randomly displaced along every normal mode of a given structure along the IRC, except the mode corresponding to the IRC path direction, providing a range of configurations from subtle to significant perturbations relative to the original IRC structures.

Table \ref{tab:NMS-RMSE} provides the RMSEs for energy, force, and Hessian predictions for all models. The model incorporating energies, forces, and Hessian terms in the loss function outperforms the others, reducing the energy RMSE to 13.52 \si{\kilo\cal\per\mole}, a 37.6\% reduction compared to the model trained with energy and forces (21.67 \si{\kilo\cal\per\mole}) and a 70.8\% reduction compared to the model trained only with energies (46.38 \si{\kilo\cal\per\mole}). This trend holds for both force and Hessian prediction tasks, where the Hessian-trained model, despite higher RMSEs than in previous benchmarks, greatly outperforms models without Hessian fitting.

The nearly twofold improvement in force RMSE for the E-F-H model compared to the E-F model provides further evidence that the E-F model was overfit when it outperformed the E-F-H model in the held-out reactants, transition states, and products benchmark. This superior performance underscores the value of the Hessian matrix in providing critical information on the energy landscape's curvature, improving extrapolation capability away from the minimum-energy pathway of the IRC. Finally, the E-F-H model greatly outperforms the others on the Hessian RMSE as is expected.

\subsection{An evaluation of extrapolation via stability in molecular dynamics simulations}

\begin{figure*}[h]
    \centering
    \includegraphics[scale=0.50]{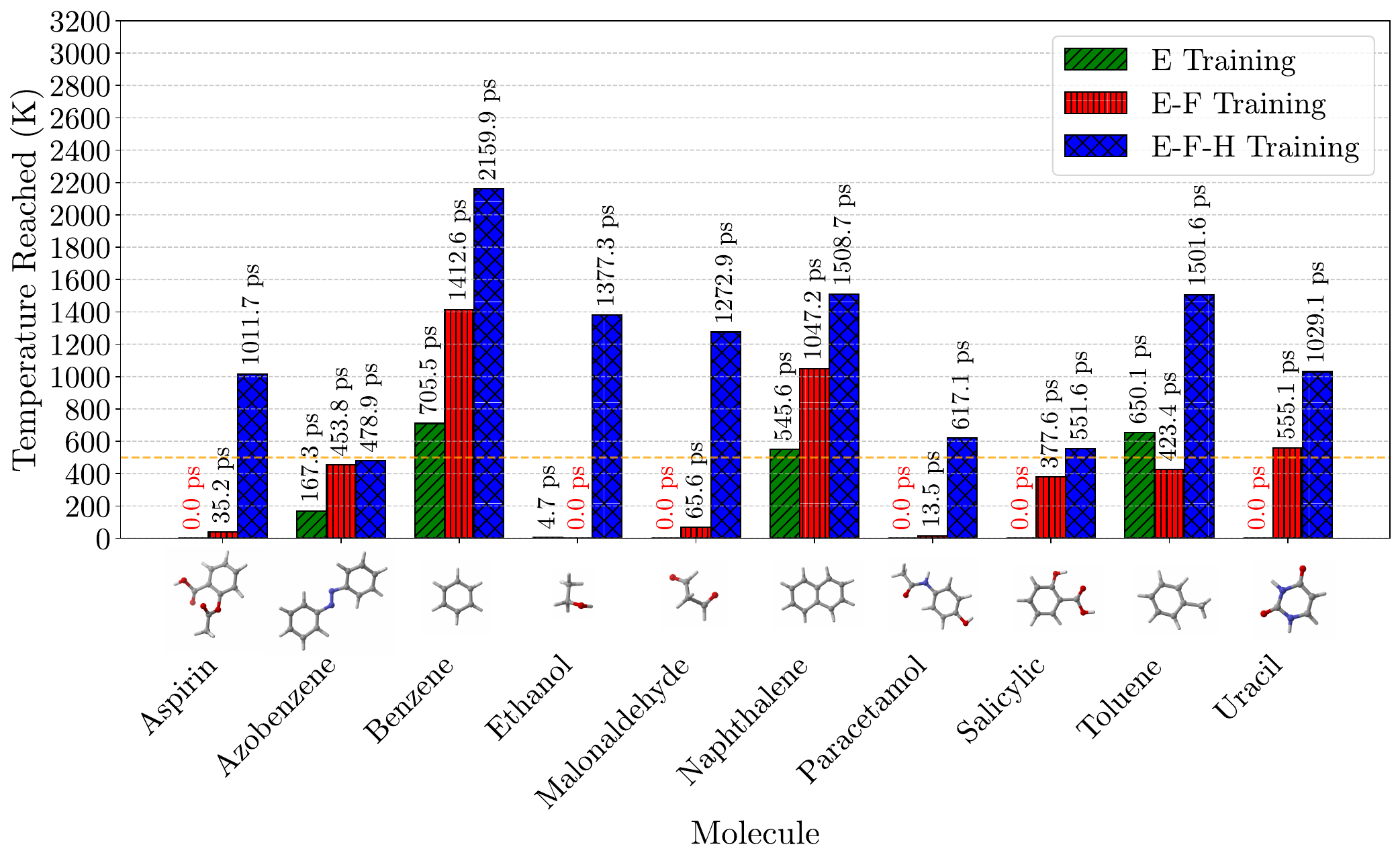}
    \caption{Temperatures reached before failure for each molecule in the MD17 dataset using ensembles of MLIP models trained with energy-only (E), energy-force (E-F), and energy-force-Hessian (E-F-H) loss functions. The height of each bar represents the temperature reached before failure for a particular molecule. Simulation times reached are shown above each bar. Simulations that failed on optimization are shown as having a time of $0.0$ ps in a red color. The dashed orange horizontal line at 500K represents the temperature at which the MD simulations of the MD17 dataset were run at.}
    \label{fig:md_temperatures}
    \rule[1ex]{\textwidth}{0.1pt}
\end{figure*}

In this section, we compare the RTP-trained MLIP models in molecular dynamics simulations to evaluate the dynamical stability of E, E-F, and E-F-H model ensembles on 10 molecules from the MD17 dataset. This approach is inspired by Fu \etal \cite{fu2022}. The MD17 dataset, consisting of small organic molecules with DFT-optimized geometries and properties simulated at 500 K, is commonly used for MLIP performance benchmarks. We selected an optimized configuration of each molecule as the starting point for stability evaluation simulations, with setup and failure criteria described in the Methodology section.

Reaching higher simulation temperatures without observing nonphysical behaviors measures a model's stability and extrapolation capability. In this NVT dynamics test, starting from the optimized initial geometry, the temperature is initialized at 5 K then slowly ramped by 5 K every 5 ps. At higher temperatures, molecular systems sample a wider range of microstates, including those far from equilibrium. Here we propose that a more stable MLIP model is one that sustains simulations at higher temperatures without non-physically breaking bonds or exhibiting close contacts between atoms (i.e. "failing"). This stability measure also reflects the model's ability to generalize beyond its training data, capturing the nuanced dynamics of molecular systems. This is especially true since all models tested are only RTP-trained MLIPs, which are trained only to stable points on the potential energy surface (minima and transition states with atomic forces close to zero).

Figure \ref{fig:md_temperatures} shows the average simulation temperatures reached before failure for each model and molecule in the MD17 data set. The average simulation times before failure are also shown above their corresponding bars, with data tabulated in Table S1. Models trained with energies and forces (E-F) exhibited similar stability to energy-only (E) models. However, significant stability improvements were observed in energy-force-Hessian (E-F-H) models. These models maintained MD simulations of most MD17 molecules at temperatures above 500 K (dashed orange line in Figure \ref{fig:md_temperatures}) and for significantly longer times than other models. The only molecule that failed below 500 K, the temperature at which the MD17 trajectories were generated, was azobenzene, probably due to the lack of examples of hydrogen-benzene interactions in our data set.

The stability improvements highlight the value of Hessian data in enhancing the MLIP’s representation of the PES. By incorporating higher-order information, E-F-H models gain a deeper understanding of the PES around training geometries, enabling better extrapolation to higher-energy configurations encountered during MD simulations. These findings emphasize the critical role of Hessian matrices in developing robust and reliable MLIPs for molecular simulations, particularly under challenging conditions of elevated temperatures and extended timescales with 
a relatively limited training data set size.

\subsection{NEB Analysis: Reaction Pathways and Barrier Predictions}

To assess the accuracy and extrapolation capabilities of the Hessian-trained MLIP in modeling chemical reaction mechanisms, we performed NEB calculations of an intramolecular single hydrogen transfer reaction and compared the predicted reaction barriers to those obtained from DFT calculations. The analysis focused on evaluating the MEP, the accuracy of transition state geometries, and the smoothness of the PES.

\begin{figure}[h]
\centering
\includegraphics[scale=0.42]{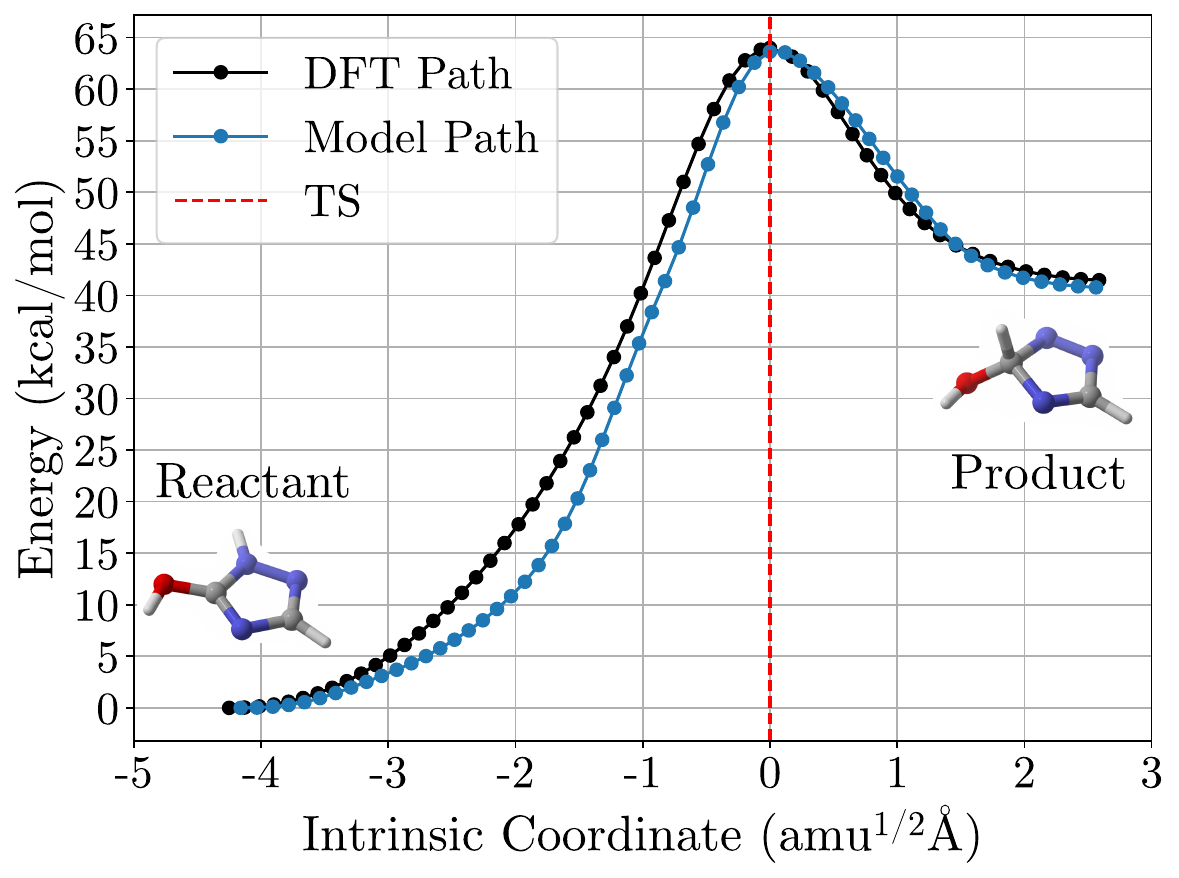}
\caption{Energy profiles obtained from NEB calculations using models trained with an E-F-H loss functions (blue dots) alongside DFT reference values (black dots). The x axis represents the geometric distances between intermediates as Intrinsic Coordinates, where a reference value of zero was assigned to the TS geometry (dashed vertical red line). The y axis represents the energies in \si{\kilo\cal\per\mole}. The atom coloring follows the CPK convention (red for oxygen, blue for nitrogen, grey for carbon, and white for hydrogen).}
\label{fig:NEB_plot}
\hrulefill
\end{figure}

The NEB calculations using the E-only and E-F trained models failed to converge to a stable reaction pathway, producing highly irregular PESs with energy fluctuations and multiple peaks, preventing the identification of a well-defined transition state. The E-only model often collapsed the reaction pathway, skipping intermediate geometries and converging to unphysical structures. The E-F model showed minor improvements but still failed to yield a continuous MEP, indicating that force training alone was insufficient to stabilize the transition state search.

In contrast, the E-F-H trained model, incorporating energy, forces, and Hessian matrix information, successfully converged to a well-defined reaction pathway. The single-step reaction profile and height were accurately reproduced (Figure \ref{fig:NEB_plot}), with the TS structure located at the saddle point. The predicted reaction barrier of 63.63 \si{\kilo\cal\per\mole} closely matched the DFT barrier of 63.99 \si{\kilo\cal\per\mole}. However, for reactions involving simultaneous processes (e.g., 2-proton transfers and a C-O bond breaking), the model predicted sequential multistep processes, reflected in the NEB barrier shape (Figure \ref{fig:NEB_plot_SI}). Despite this, the model accurately reproduced the barrier height. We hypothesize that increasing the dataset to include such reactions could improve the model's accuracy for multi-process reactions.

The convergence failure of the E-only and E-F models indicates immense value for including higher-order derivative information (i.e., the Hessian) in training MLIPs for reaction modeling. The E-F-H model's ability to stably reproduce MEPs and accurately predict reaction barriers demonstrates that Hessian training is essential for transition state identification and reliable extrapolation beyond equilibrium geometries. These results suggest that future MLIP models for reaction simulations should incorporate Hessian information to ensure robustness in applications involving reaction kinetics and catalysis.

\subsection{Vibration Spectra to aid `in-operando' experimental setups}

\begin{figure}[h!]
\captionsetup[subfigure]{justification=centering,font={stretch=0.8}}
\centering
\begin{subfigure}{0.46\textwidth}
\centering
\includegraphics[scale=0.4]{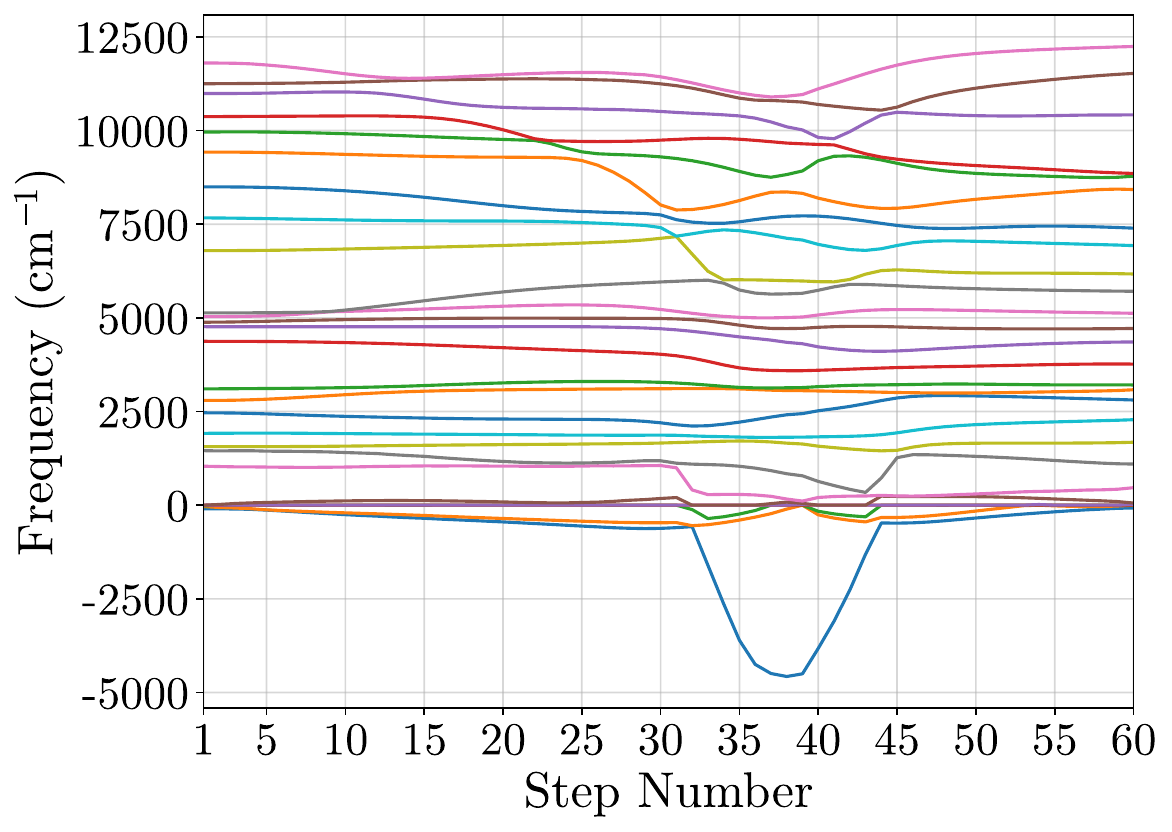}
\caption{Vibrational frequencies obtained from DFT calculations.}
\label{subfig:frequencies-dft}
\end{subfigure}
\vspace{4pt}
\\
\begin{subfigure}{0.46\textwidth}
\centering
\includegraphics[scale=0.4]{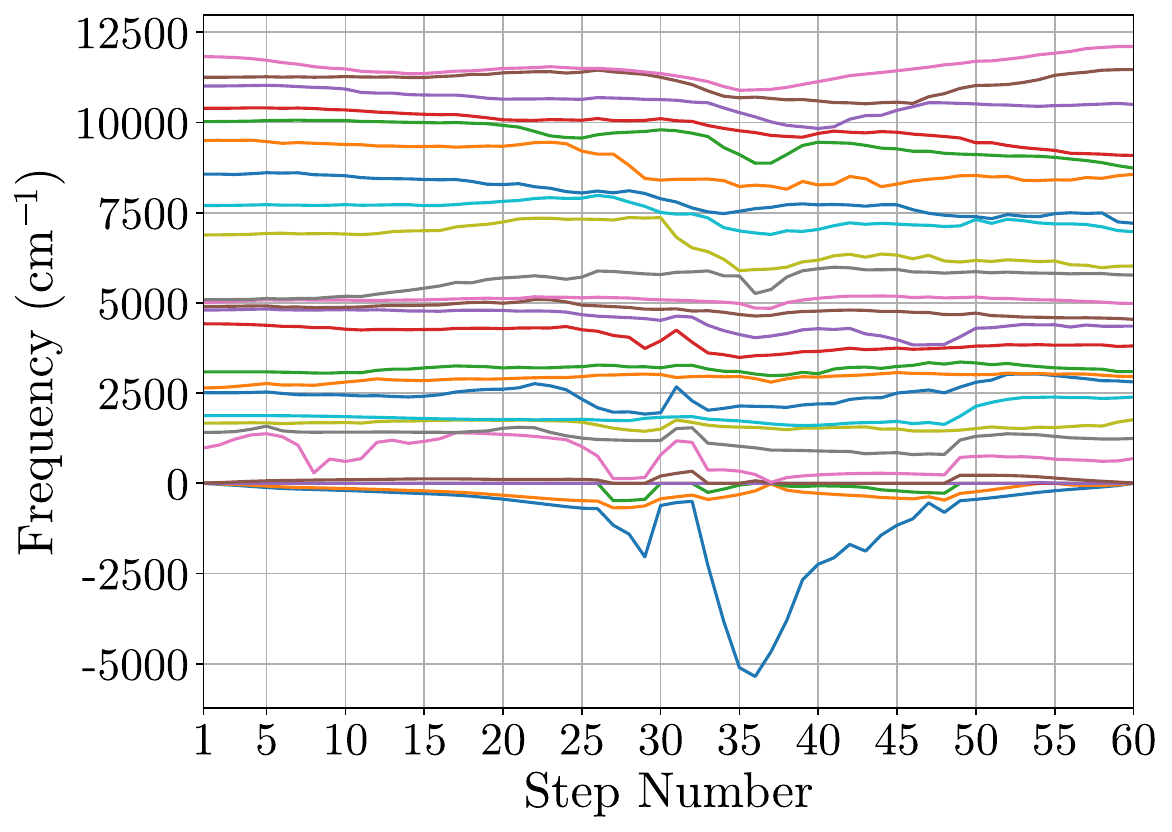}
\caption{Vibrational frequencies obtained from the EFH model.}
\label{subfig:frequencies-model}
\end{subfigure}

\caption{Vibrational frequencies of the molecular system along the reaction steps from the NEB images. The top subplot (a) displays the frequencies of the vibrational modes computed from DFT-calculated Hessians, while the bottom subplot (b) shows vibrational modes obtained from the EFH model predictions. In both cases, one of the vibrational modes exhibits a minimum at the TS structure, reflecting the characteristic softening of the reaction coordinate mode at the saddle point.}
\label{fig:vib_spectra}
\hrulefill
\end{figure}

One of the biggest challenges of current spectroscopic techniques is following reactions as they happen, i.e., `in-operando', because some reactions are too fast to measure or too complex to generate easily interpretable spectra. Spectral techniques usually measure everything at once, making it difficult to deconvolute time-dependent steps. If we can deconvolute these steps, we can interpret the full spectra more straightforwardly. Accordingly, we tested our model by calculating the vibrational spectra (e.g., IR) of the MEP of a chemical reaction using images from our previous NEB analysis. The vibrational spectra obtained from DFT-calculated Hessians and the EFH model's Hessian predictions are shown in Figures \ref{subfig:frequencies-dft} and \ref{subfig:frequencies-model}, respectively.

\begin{figure*}[h]
\captionsetup[subfigure]{justification=centering,font={stretch=0.8}}
\centering
\begin{subfigure}[t]{0.46\textwidth}
\centering
\includegraphics[scale=0.4]{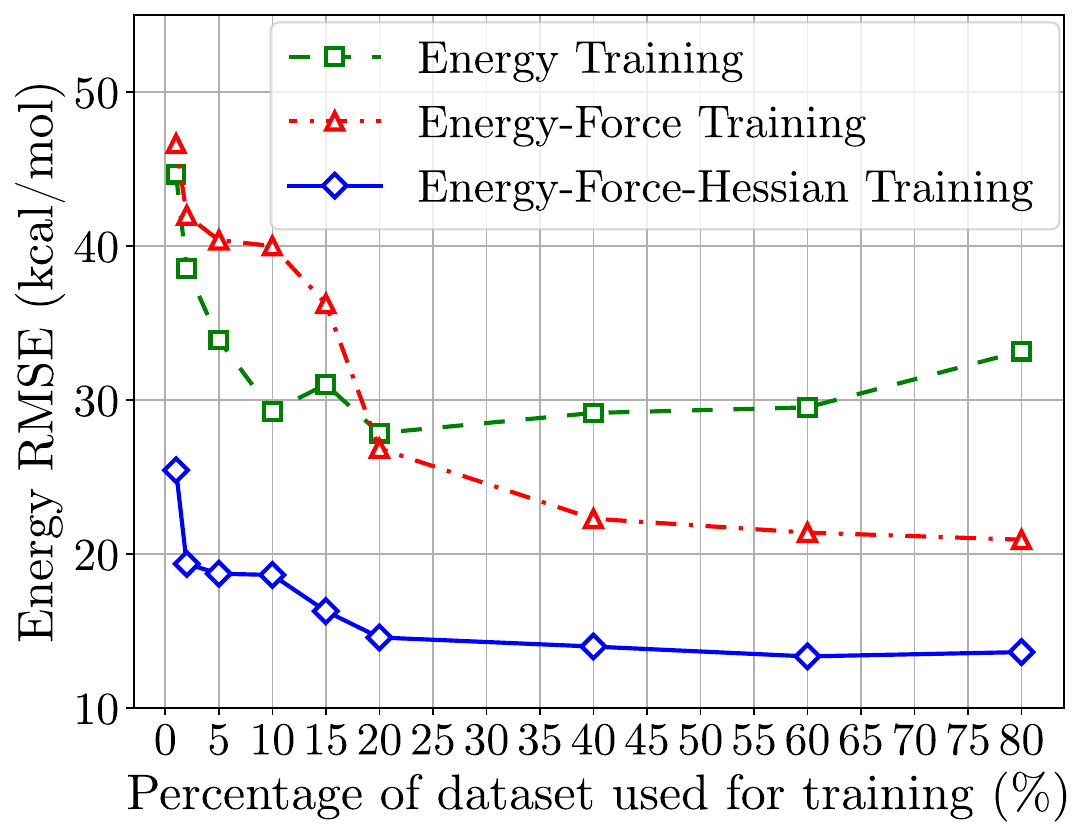}
\caption{Energy RMSEs}
\label{subfig:RMSE-E}
\end{subfigure}
~~
\begin{subfigure}[t]{0.46\textwidth}
\includegraphics[scale=0.4]{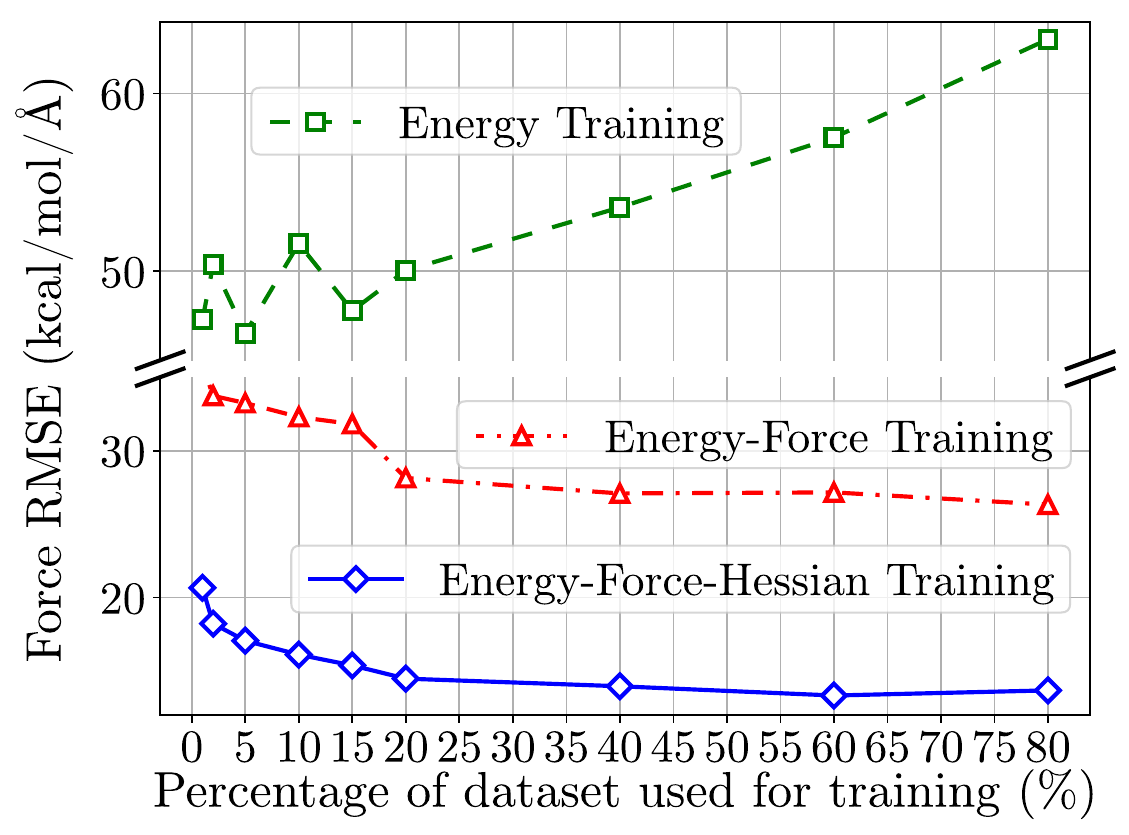}
\caption{Force RMSEs}
\label{subfig:RMSE-F}
\end{subfigure}
\\
\medskip
\begin{subfigure}[t]{0.46\textwidth}
\centering
\includegraphics[scale=0.4]{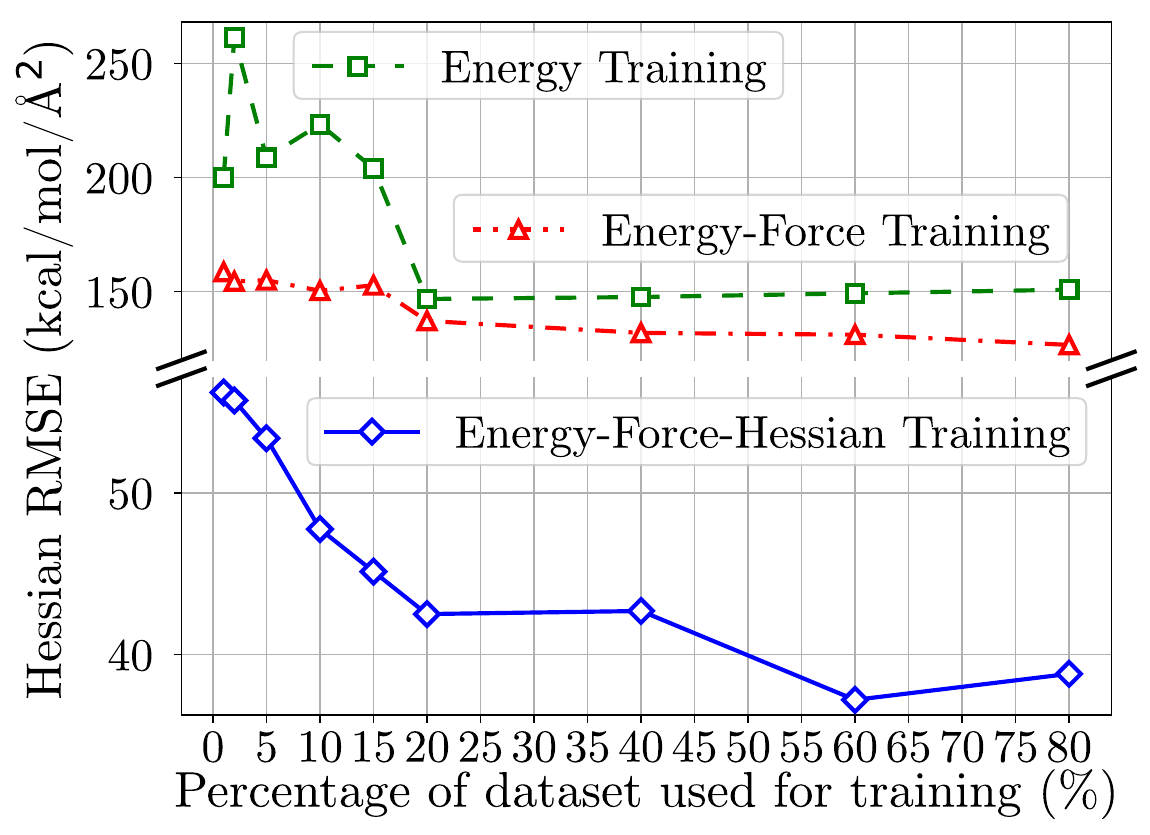}
\caption{Hessian RMSEs}
\label{subfig:RMSE-H}
\end{subfigure}
~~
\begin{subfigure}[t]{0.46\textwidth}
\includegraphics[scale=0.4]{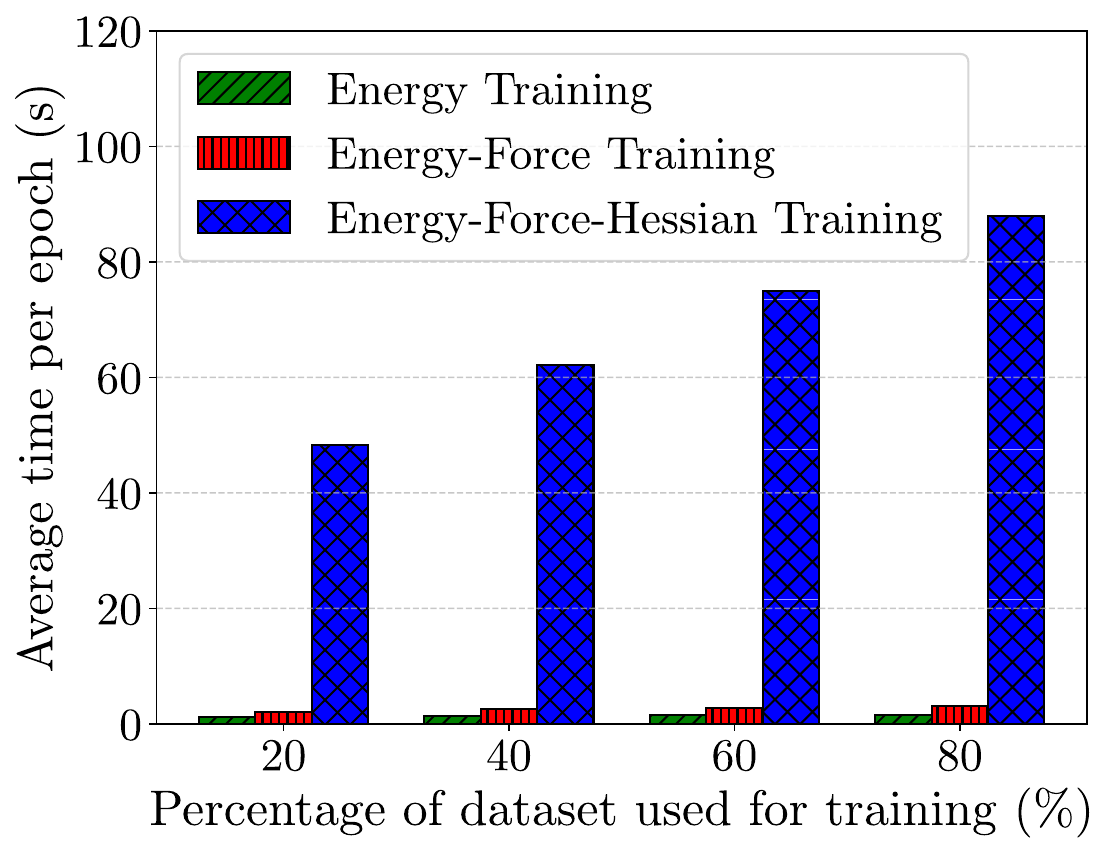}
\caption{Average training time per epoch}
\label{subfig:time-per-epoch}
\end{subfigure}

\caption{Energy, force, and Hessian root mean squared errors and average training time per epoch versus training data volume. The models shown in these figures were trained on reactants, transition states, and product of 11,961 elementary chemical reactions and tested on perturbed structures outside of the minimum energy pathway generated through Normal Mode Sampling (NMS) on the intermediate structures of randomly selected reactions in the dataset (exactly 62,527 NMS structures).}
\label{fig:data-efficiency}
\rule[1ex]{\textwidth}{0.1pt}
\end{figure*}

Our results show that we can reproduce most quantum-level frequencies, which usually require significant resources. For DFT frequency analysis calculations, it took an average of 301 \si{\second} of CPU time per image, totaling 18,071 \si{\second} for the 60 images. However, the EFH model calculated the atomic forces and Hessian matrices for the 60 images simultaneously in 0.478 \si{\second} of CPU time, representing a speedup of about 5 orders of magnitude in generating vibrational spectra. Although our model reproduces the vibrational frequencies from the structural and identity information of the molecule at a fraction of the computational cost compared to DFT calculations, there is extra noise in the frequency predictions, possibly because of uncertainty.

\subsection{Improved data efficiency}

As shown in previous benchmarks, integrating the Hessian matrix into MLIP training data enhances predictive accuracy and data efficiency. By examining learning curves for energy, force, and Hessian RMSE versus training data volume, alongside average time per epoch (Figure \ref{fig:data-efficiency}), we gain insights into the practical implications of incorporating second-order derivatives.

Including the Hessian matrix increases the model's learning complexity but provides richer training information. RMSE plots in Figures \ref{subfig:RMSE-E}, \ref{subfig:RMSE-F}, and \ref{subfig:RMSE-H} show that models trained with Hessian information achieve lower errors with fewer training data compared to models without it. This suggests that the Hessian matrix offers critical insights into the energy landscape, increasing the information yield per example.

Figure \ref{subfig:RMSE-E} shows that including Hessian information significantly improves data efficiency. The E-F-H model, trained on energies, forces, and Hessian matrices, achieves a significantly lower RMSE in energy predictions using only 2\% of the dataset volume compared to models trained with energy and energy-force loss functions using 80\% of the dataset.

However, integrating Hessian information increases computational demands, resulting in longer training times (Figure \ref{subfig:time-per-epoch}). Computing Hessian matrices using PyTorch's \textit{autograd.grad} function involves calculating second derivatives of the energy with respect to atomic coordinates, a complex and time-consuming process. Training times are approximately 25 times longer per epoch, and routine evaluation of Hessian RMSE further exacerbates the computational load.

Despite these challenges, future innovations in computational techniques are promising. New algorithms and optimization methods could streamline Hessian matrix calculations and their integration into MLIP training. Leveraging more efficient methodologies could mitigate current computational challenges \cite{smith2020simple}. Pursuing these advancements is crucial to ensure the benefits in predictive performance are not overshadowed by increased computational demands.

\section{Conclusion}

In conclusion, our investigation into integrating Hessian matrix data within ANI MLIP model training shows significantly enhanced predictive accuracy and extrapolation capabilities. Including Hessian data enables MLIP models to more accurately predict energies, forces, and second-order derivatives with fewer training examples, improving data efficiency.

MD simulations with Hessian-trained MLIPs show enhanced stability and robustness under dynamic conditions, sustaining longer simulation times and higher temperatures before failure. This makes them promising for realistic molecular dynamics studies, especially in reactive environments. NEB analysis highlights the critical role of Hessian training in accurately describing reaction pathways and transition states. While energy-only (E) and energy-force (E-F) models failed to converge to stable MEPs, the E-F-H model successfully reproduced smooth PESs and transition state structures matching DFT calculations. This confirms Hessian-trained MLIPs as efficient alternatives for reaction barrier predictions and mechanistic studies. Additionally, generating vibrational spectra along reaction coordinates underscores the advantages of incorporating Hessian information. The model reproduces vibrational frequency trends, including the softening of the reaction coordinate mode at the transition state, linking computational modeling with in-operando vibrational spectroscopy.

However, incorporating Hessian information increases computational demands, with training times per epoch up to 25 times longer. This trade-off between predictive accuracy and computational resources suggests the need for advancements in computational techniques and algorithms to estimate Hessian errors, ensuring efficient and accurate MLIP models.

Overall, Hessian-trained MLIPs provide a powerful approach for molecular simulations, reaction modeling, and vibrational analysis. Their ability to accurately predict energies, forces, and PES curvatures makes them valuable for computational chemistry, catalysis, and materials science. Future work will focus on enhancing computational efficiency, scalability, and generalizability, expanding their applicability across diverse molecular systems.

\vspace{-0.17cm}
\section*{Acknowledgements}
We would like to acknowledge Leon Alday-Toledo for initial discussions on this project. This work was supported in part through computational resources and services provided by the Institute for Cyber-Enabled Research at Michigan State University. A.R. would like to thank the Chemical Engineering and Material Science (ChEMS) department of Michigan State University and the College of Engineering of the same university for their partial support during this research in the form of a first-year fellowship and a summer fellowship, respectively.

\vspace{-0.17cm}
\section*{Supporting Information Available}

    The supporting information includes detailed descriptions and visual representations of the datasets used in this study, with figures illustrating the composition and distribution of molecules in the dataset containing reactants, TSs and products (Figure \ref{fig:atom-types} and Figure \ref{fig:bond-counts}). MLIP architecture diagrams are also included to illustrate the inner functioning of the models used in this work (Figure \ref{fig:formaldehyde-HD-NNP}). In addition, the ensemble prediction plots for the energy values along the IRC of a single reaction are shown in Figure \ref{fig:TS-training} and Figure \ref{fig:ensemble-predictions}, showcasing the predictions from the ensembles of 100 models for each type of fit. The supporting information also contains correlation plots that compare the predicted values of our models with the reference values obtained from the DFT calculations, provided for three different types of models: energy fitting, energy force fitting, and energy force--Hessian fitting across the three datasets (Figures \ref{fig:energy_fitting_E} to \ref{fig:energy_force_Hessian_fitting_H}). The simulation times reached before failure in the MD simulations for each molecule in the MD17 data set are illustrated in Figure \ref{fig:md_times}, in addition to a table of temperatures and simulation times reached before failure (Table \ref{tab:md-results}) to show the stability differences between the three types of models under MD simulations. The NEB plot of the multi-process reaction is shown in Figure \ref{fig:NEB_plot_SI}. Finally, the data efficiency plots for the IRC dataset are presented in Figure \ref{fig:data-efficiency_IRC} as a comparison to the data efficiency plots for the NMS structures data set in Figure \ref{fig:data-efficiency}.

\bibliography{Arxiv_references}
\clearpage  

\renewcommand{\thesection}{\Alph{section}}  
\renewcommand{\thefigure}{S\arabic{figure}}  
\renewcommand{\thetable}{S\arabic{table}}  
\renewcommand{\theequation}{S\arabic{equation}}  

\setcounter{section}{0}   
\setcounter{figure}{0}    
\setcounter{table}{0}     
\setcounter{equation}{0}  

\onecolumn  

\section*{Supporting Information}

\newcommand*{\Hess}{\mathit{H}}
\newcommand*{\diag}[2]{\dfrac{\partial^2 #1}{\partial #2^2}}
\newcommand*{\ndiag}[3]{\dfrac{\partial^2 #1}{\partial #2 \partial #3}}
\newcommand*{\diagE}[1]{\dfrac{\partial^2 E}{\partial #1^2}}
\newcommand*{\ndiagE}[2]{\dfrac{\partial^2 E}{\partial #1 \partial #2}}

\setstretch{1.5}

\section{Training Dataset Description}

\begin{figure}[h!]
\centering
\rule[1ex]{\textwidth}{0.1pt}

\begin{subfigure}[t]{0.3\textwidth}
\centering
\includegraphics[width=\textwidth]{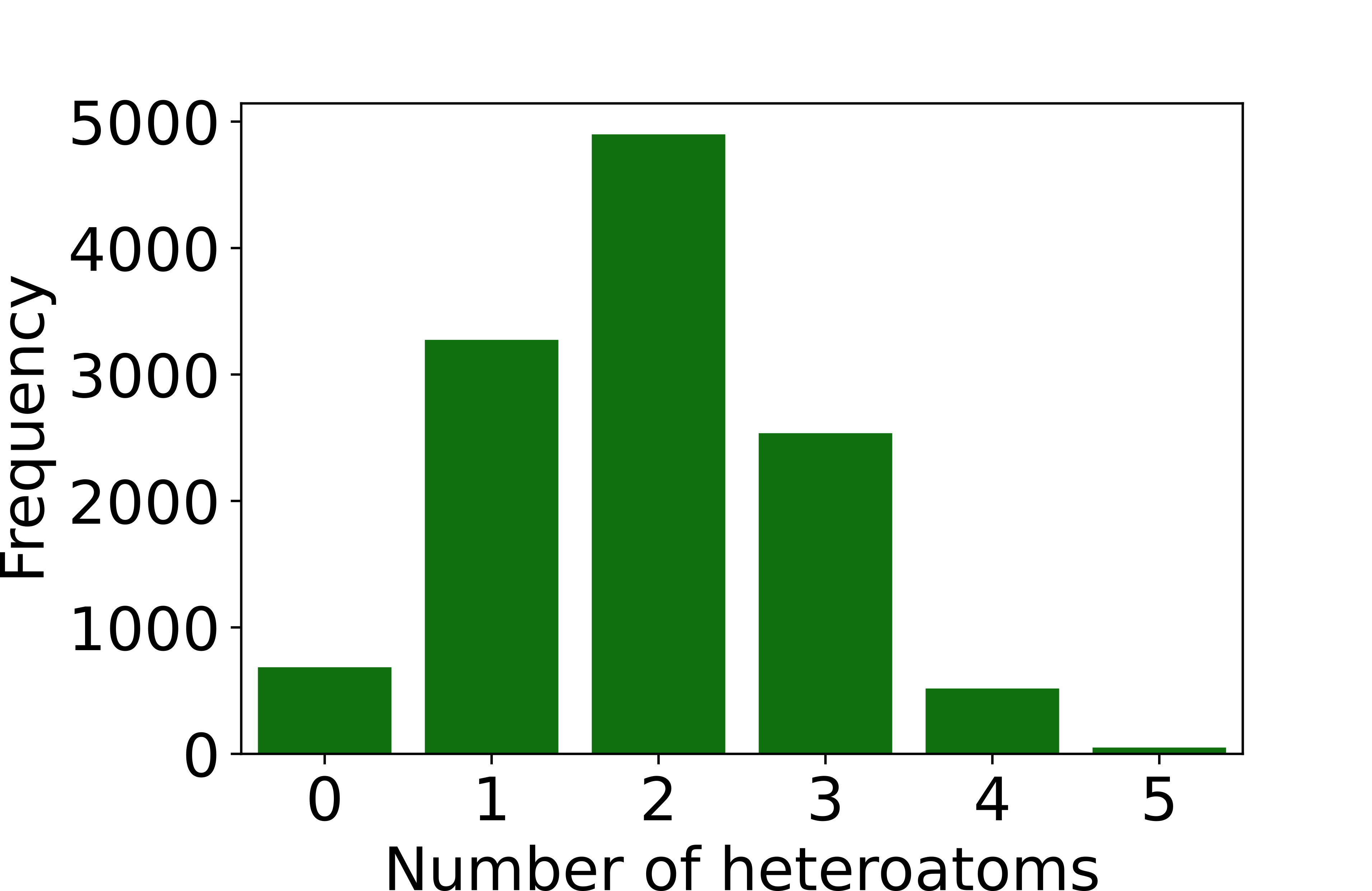}
\caption{Heteroatom population}
\label{subfig:nhet}
\end{subfigure}%
~~
\begin{subfigure}[t]{0.3\textwidth}
\centering
\includegraphics[width=\textwidth]{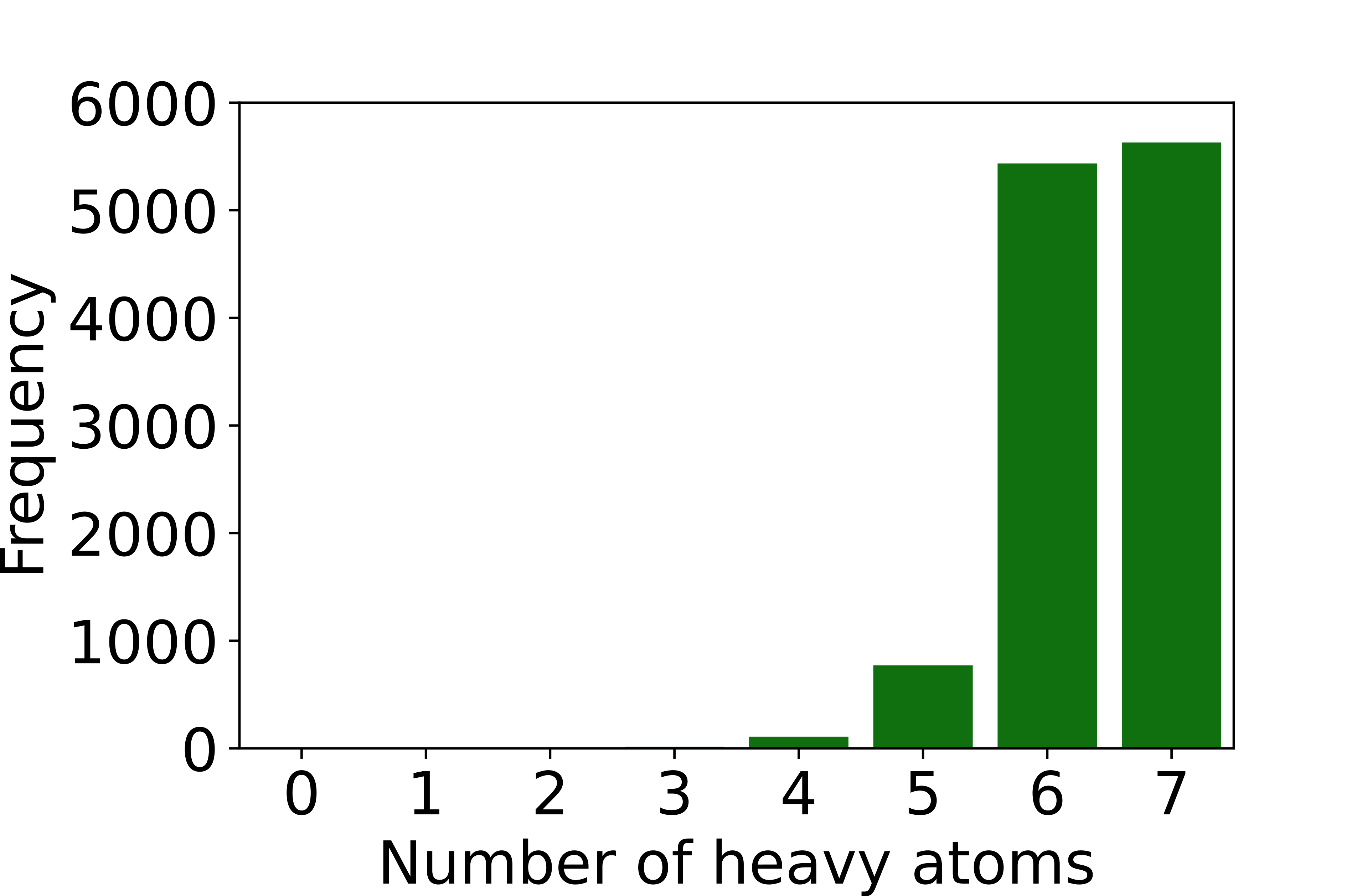}
\caption{Heavy atom population}
\label{subfig:nha}
\end{subfigure}
~~
\begin{subfigure}[t]{0.3\textwidth}
\centering
\includegraphics[width=\textwidth]{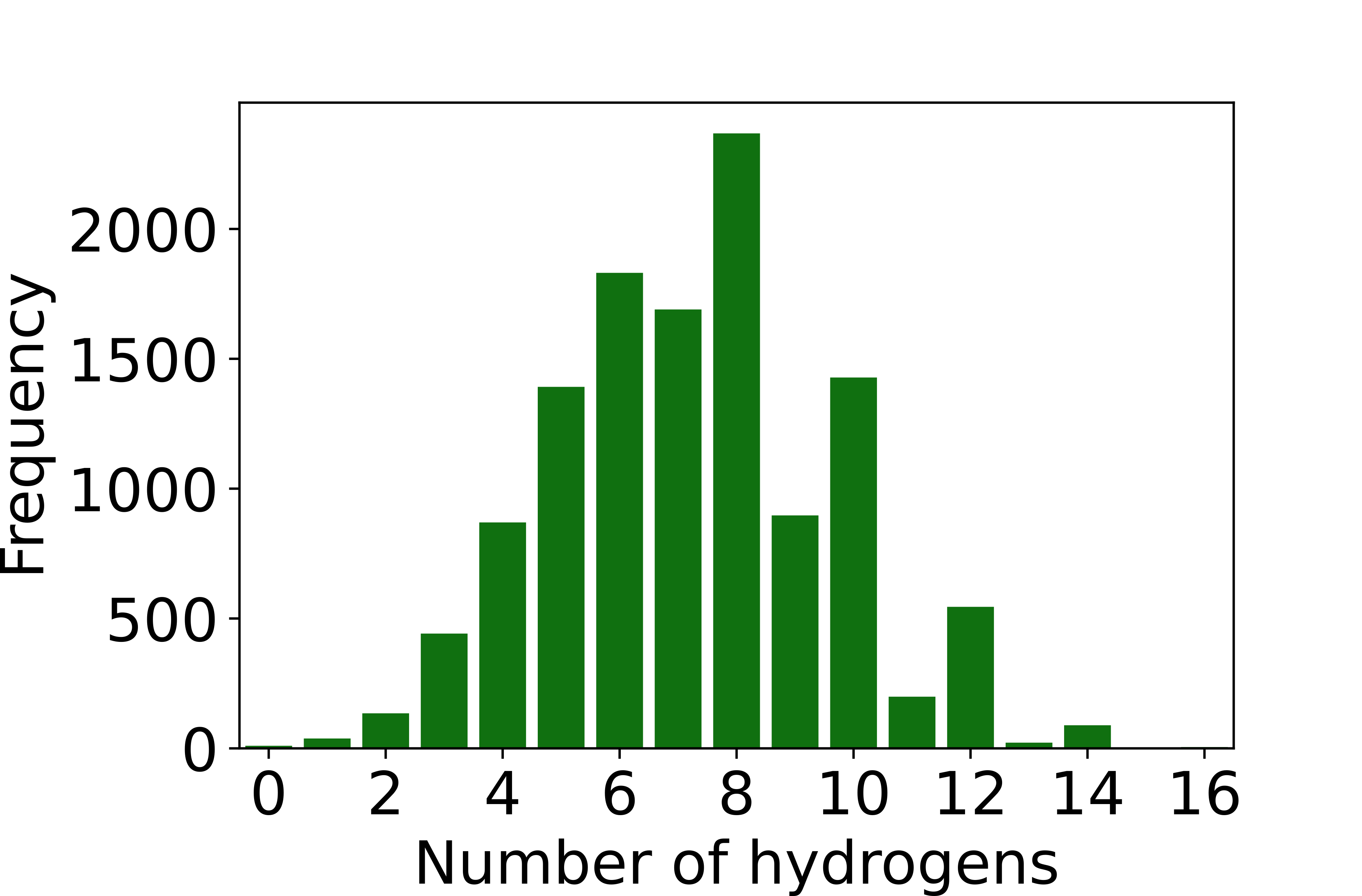}
\caption{Hydrogen population}
\label{subfig:nH}
\end{subfigure}
~~
\begin{subfigure}[t]{0.3\textwidth}
\centering
\includegraphics[width=\textwidth]{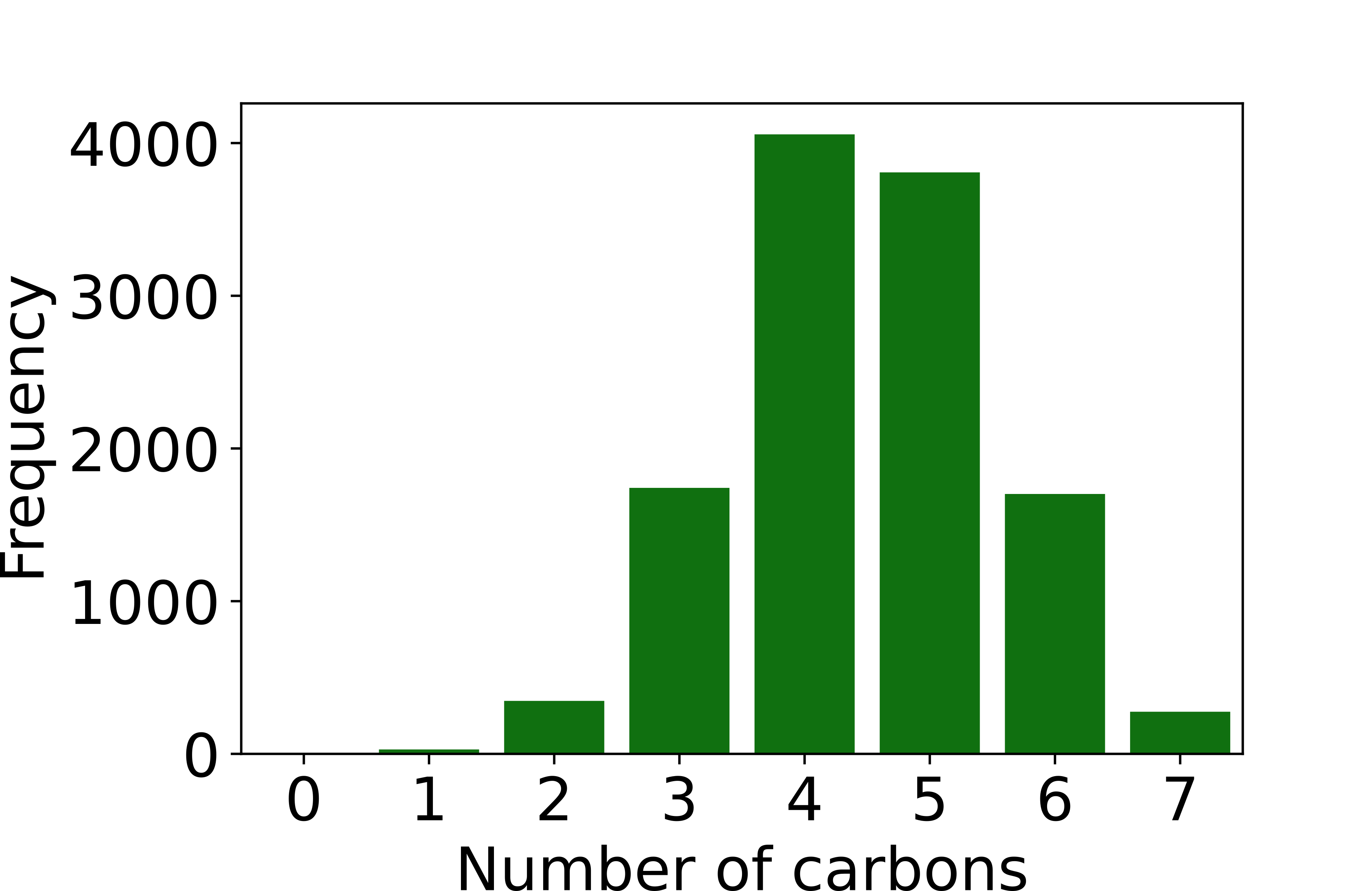}
\caption{Carbon population}
\label{subfig:nC}
\end{subfigure}
~~
\begin{subfigure}[t]{0.3\textwidth}
\centering
\includegraphics[width=\textwidth]{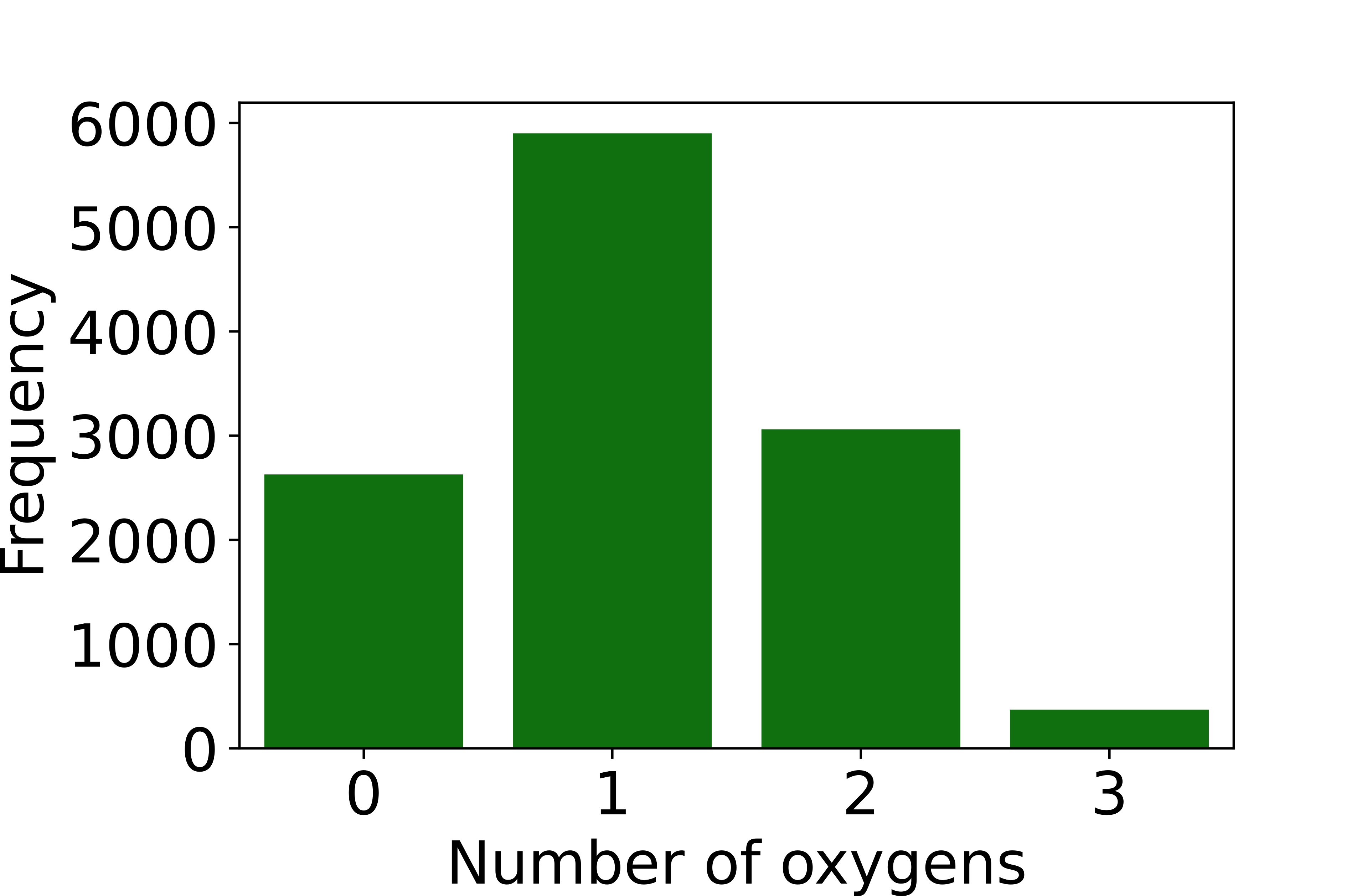}
\caption{Oxygen population}
\label{subfig:nO}
\end{subfigure}
~~
\begin{subfigure}[t]{0.3\textwidth}
\centering
\includegraphics[width=\textwidth]{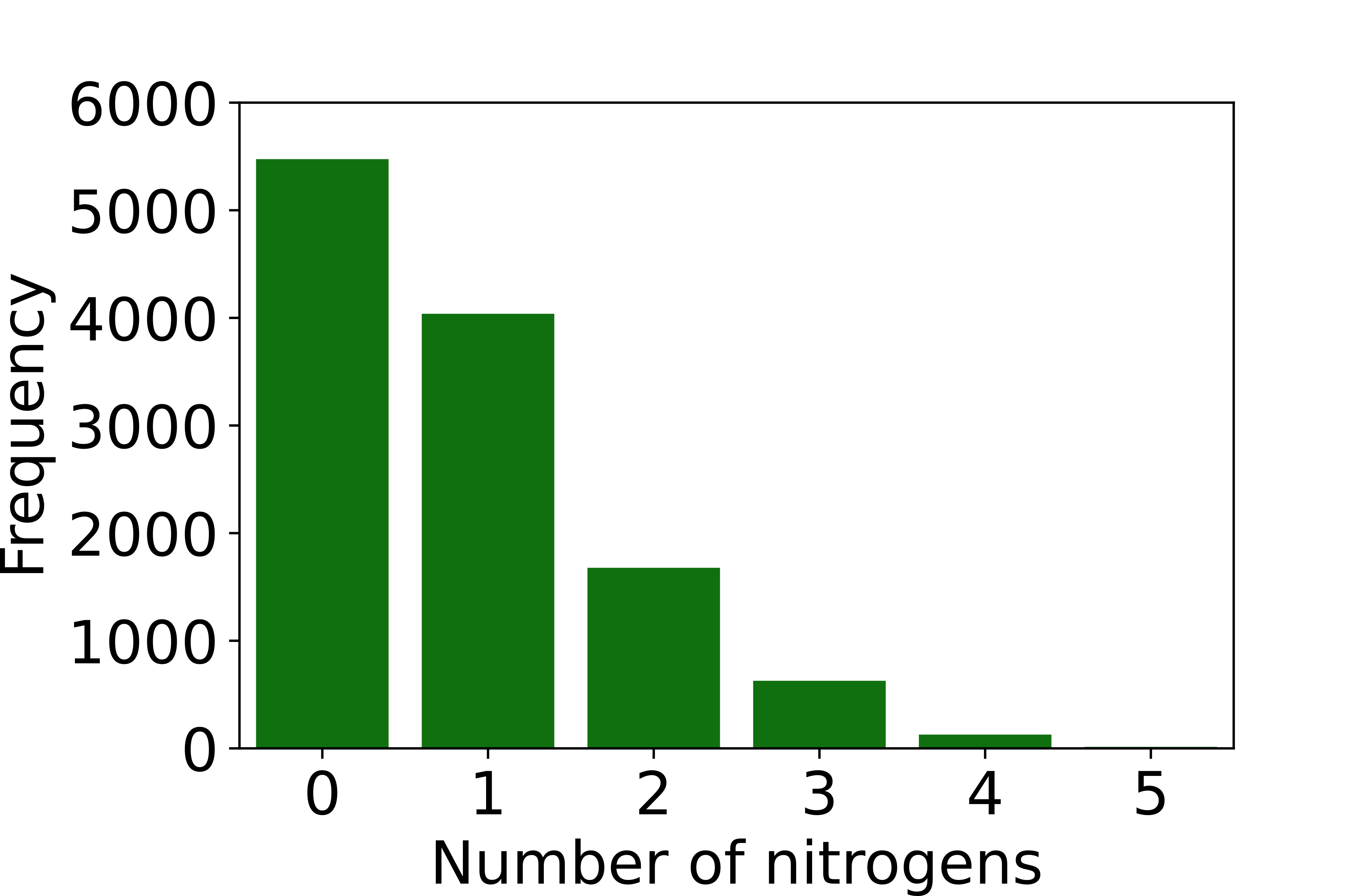}
\caption{Nitrogen population}
\label{subfig:nN}
\end{subfigure}

\caption{Overview of the composition of molecules in the dataset we will use to train the ML models. Histograms of heteroatoms, heavy atoms, and each atom type in molecules involved in the reactions of the dataset. Heteroatoms are considered to be any atom except carbon (C) or hydrogen (H).}
\label{fig:atom-types}
\rule[1ex]{\textwidth}{0.1pt}
\end{figure}

In Figure \ref{fig:atom-types}, we provide a comprehensive overview of the dataset. These figures show the distribution and diversity of chemical species within the dataset, shedding light on the types of reaction and chemical environments captured. Specifically, we include histograms that illustrate the distribution of different elements and display the number of occurrences for each type of atom.

\begin{figure}[h!]
\centering
    \begin{subfigure}[t]{0.32\textwidth}
    \centering
    \includegraphics[width=\textwidth]{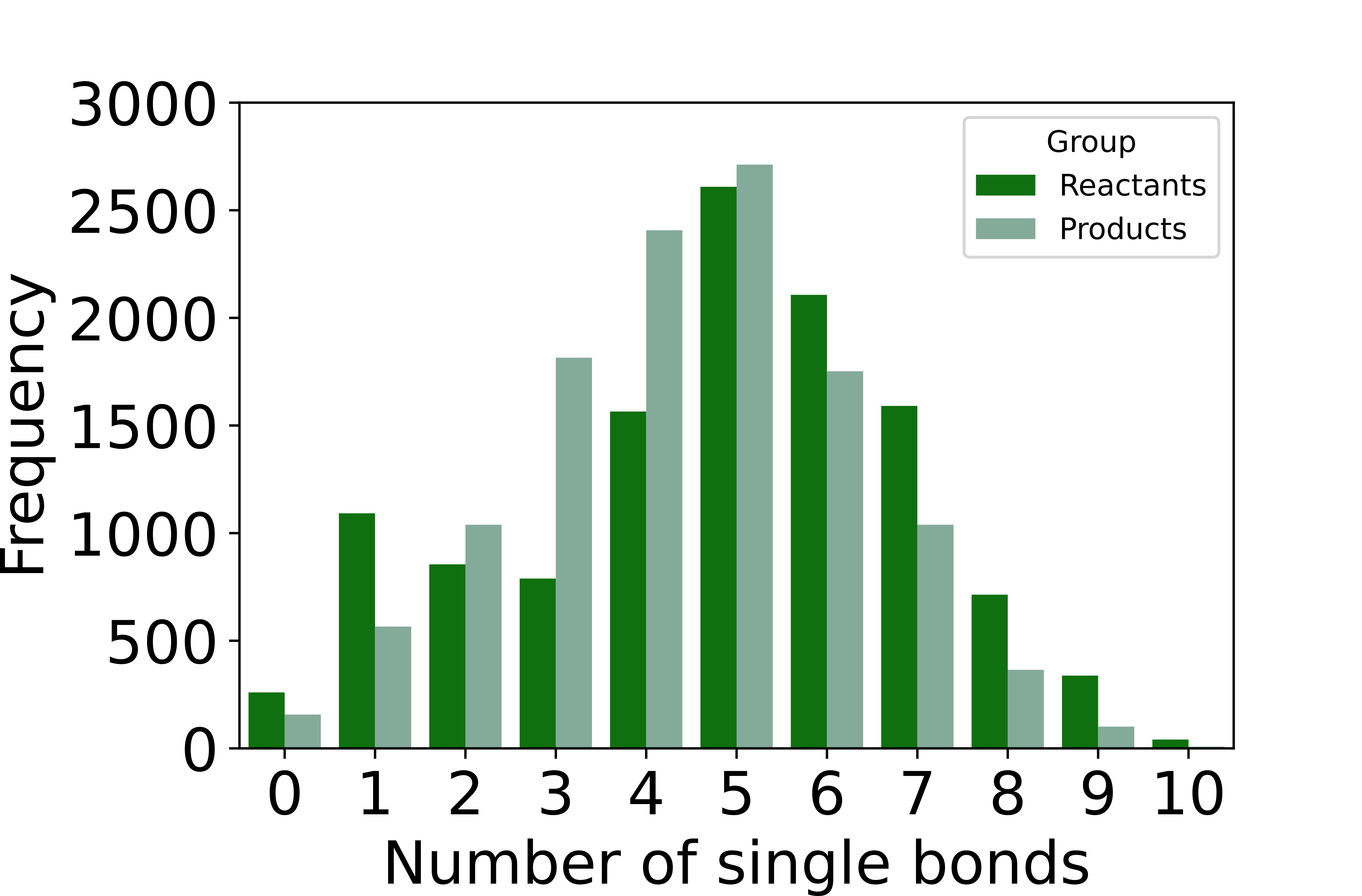}
    \caption{Single bonds}
    \label{subfig:sglbonds}
    \end{subfigure}%
    ~~
    \begin{subfigure}[t]{0.32\textwidth}
    \centering
    \includegraphics[width=\textwidth]{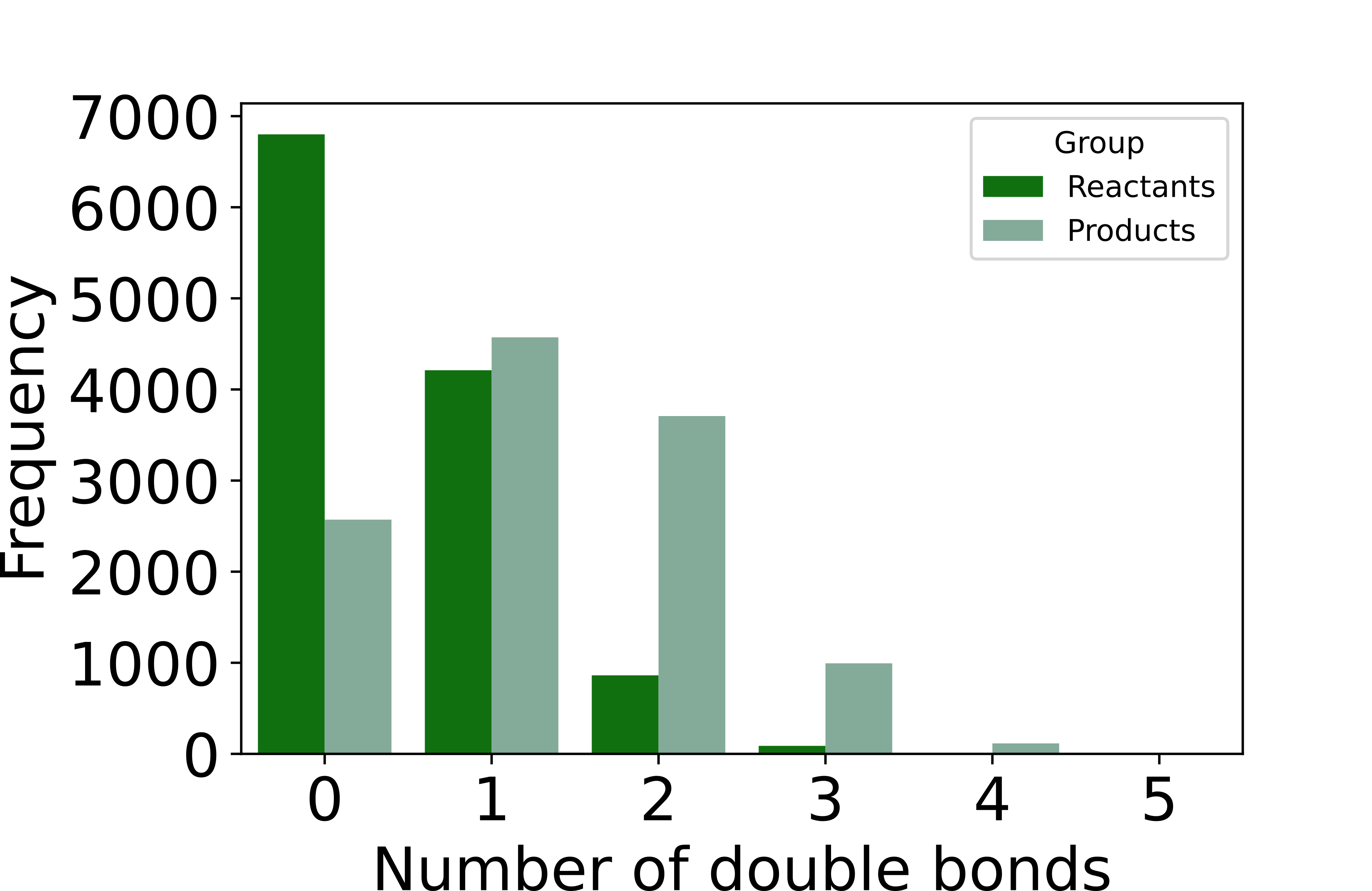}
    \caption{Double bonds}
    \label{subfig:dblbonds}
    \end{subfigure}
    ~~
    \begin{subfigure}[t]{0.32\textwidth}
    \centering
    \includegraphics[width=\textwidth]{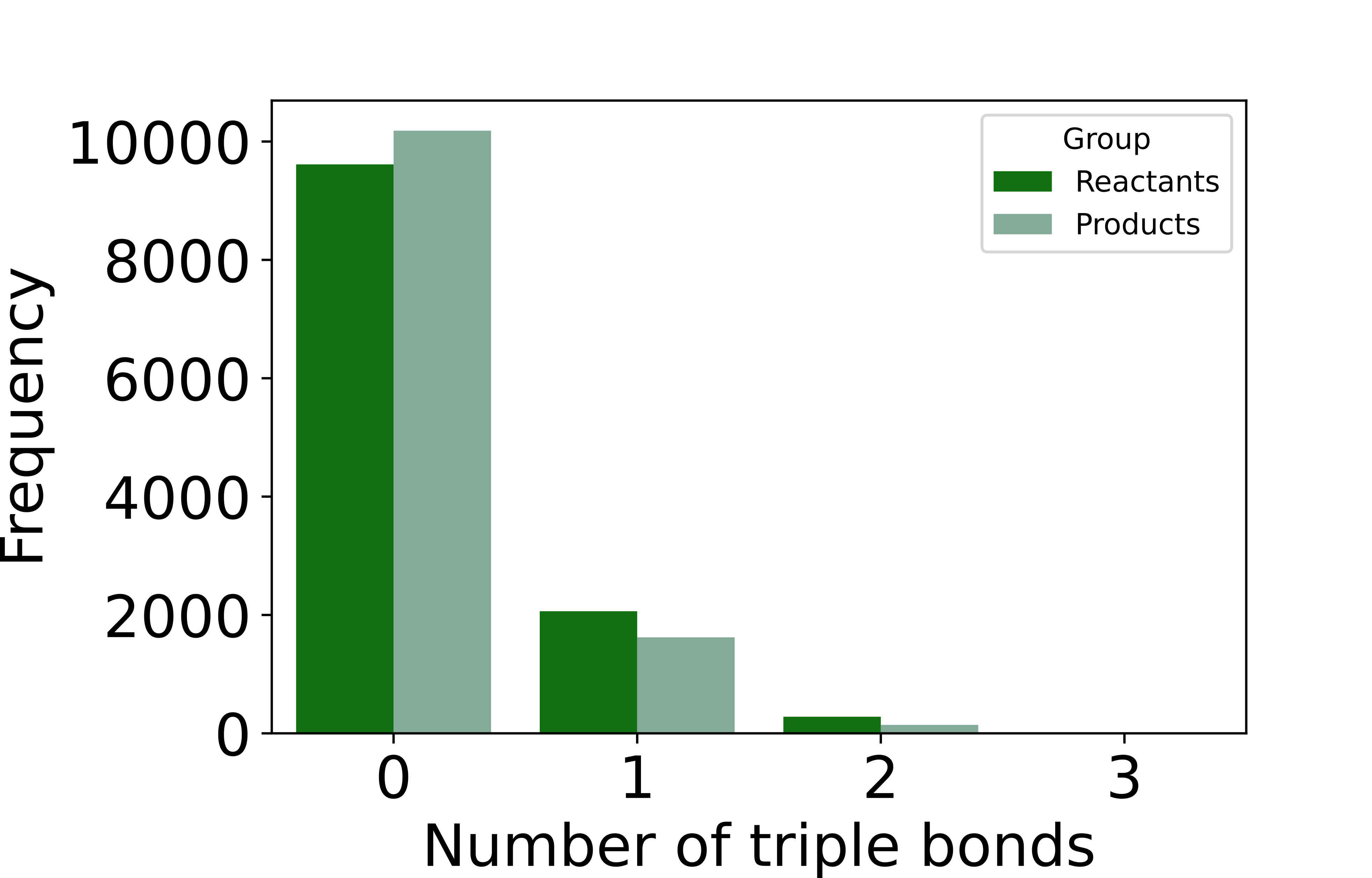}
    \caption{Triple bonds}
    \label{subfig:tplbonds}
    \end{subfigure}%
    ~~
    \begin{subfigure}[t]{0.32\textwidth}
    \centering
    \includegraphics[width=\textwidth]{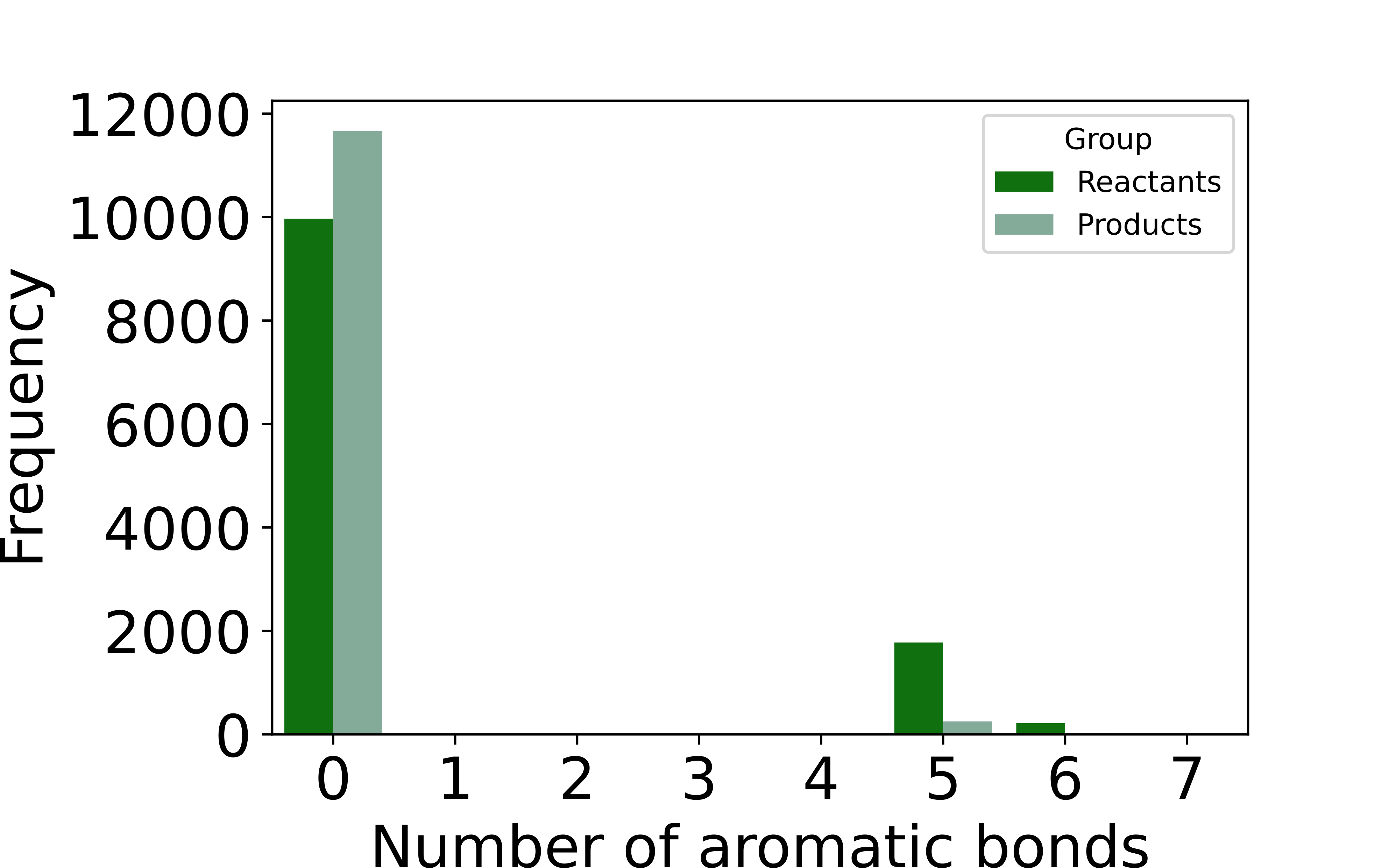}
    \caption{Aromatic bonds}
    \label{subfig:arobonds}
    \end{subfigure}
    
    \caption{Histograms containing the frequency of molecules with each number of single, double, triple, or aromatic bonds.}
    \label{fig:bond-counts}
\rule[1ex]{\textwidth}{0.1pt}
\end{figure}

In addition, we present histograms that categorize the number of single, double, triple, and aromatic bonds in the reactants and products in Figure \ref{fig:bond-counts}. These histograms provide insight into the bond types prevalent in the reactions, highlighting the nature of the chemical transformations captured in the dataset. By presenting these figures, we aim to provide a comprehensive overview of the dataset and its suitability in training ML/AI models and improving the accuracy of the Hessian-Trained ML-FF.

\clearpage

For training, we have calculated the Hessian at each point, which is defined as:

\begin{equation}
\label{eq:Hess}
\renewcommand*{\arraystretch}{1.5} 
\scalebox{0.68}{%
$
\Hess_E =
\begin{bmatrix}
    \diagE{x_1} & \ndiagE{x_1}{y_1} & \ndiagE{x_1}{z_1} & \ndiagE{x_1}{x_2} & \cdots & \ndiagE{x_1}{z_n} \\
    \ndiagE{y_1}{x_1} & \diagE{y_1} & \ndiagE{y_1}{z_1} & \ndiagE{y_1}{x_2} & \cdots & \ndiagE{y_1}{z_n} \\
    \ndiagE{z_1}{x_1} & \ndiagE{z_1}{y_1} & \diagE{z_1} & \ndiagE{z_1}{x_2} & \cdots & \ndiagE{z_1}{z_n} \\
    \ndiagE{x_2}{x_1} & \ndiagE{x_2}{y_1} & \ndiagE{x_2}{z_1} & \diagE{x_2} & \cdots & \ndiagE{x_2}{z_n} \\
    \vdots & \vdots & \vdots & \vdots & \ddots & \vdots \\
    \ndiagE{z_n}{x_1} & \ndiagE{z_n}{y_1} & \ndiagE{z_n}{z_1} & \ndiagE{z_n}{x_2} & \cdots & \diagE{z_n} \\
\end{bmatrix}
$
}
\end{equation}

where $H_E$ is called the Hessian matrix, $E$ represents the total energy of the system, and $x_i$, $y_i$, and $z_i$ represent each of the Cartesian coordinates of the atomic position of the atom $i$.

\clearpage

\section{MLIP Architecture}

The ANI-1 model uses modified Behler and Parrinello symmetry functions to capture the intricate chemical environments that surround individual atoms \cite{Behler2007,Behler2011}. These vectors then become input for a specialized form of High Dimensional Neural Network Potentials (HD-NNP) \cite{Behler2015}. Different NNPs are deployed for each atom type, each equipped with its own set of weights and biases. Architecturally, these HD-NNPs are structured as feed-forward neural networks, featuring multiple hidden layers and a variety of neurons. The outputs of each of these NNPs correspond to a partition per atom of the molecular potential energy. These values are summed up to obtain the potential energy.

\begin{figure*}[h!]
\captionsetup[subfigure]{justification=centering,font={stretch=0.8}}
\centering
\rule[1ex]{\textwidth}{0.1pt}
\begin{subfigure}{0.25\textwidth}
\centering
\includegraphics[scale=1.5]{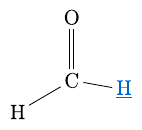}
\caption{Formaldehyde with $C_s$ symmetry. The two nonequivalent hydrogens are displayed in different colors.}
\label{subfig:formaldehyde}
\end{subfigure}
~~
\begin{subfigure}{0.7\textwidth}
\includegraphics[width=\textwidth]{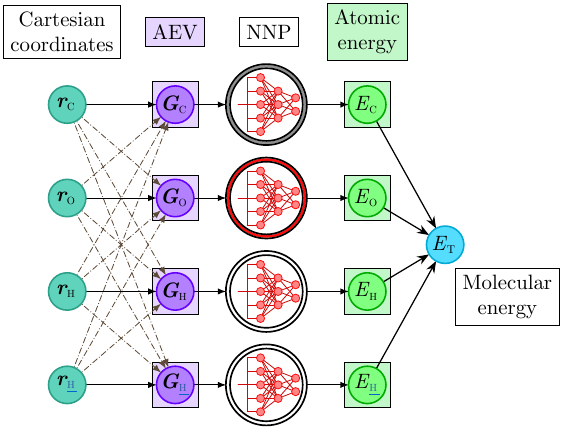}
\caption{A diagram of an HD-NNP for the formaldehyde molecule. The $G$-blocks calculate an \emph{atomic environment vector}, AEV, which is fed to the NNP. The molecular energy is the sum of the individual atomic energies.
This figure was adapted from figure 2 of Behler's paper \cite{Behler2011}.
}
\label{subfig:HD-NNP}
\end{subfigure}

\caption{Use of an array of NNPs to calculate the energy of a formaldehyde molecule. The formaldehyde molecule has three atom types, and therefore, three distinct NNPs are necessary. Each NNP is indicated by its ring color, and has its own set of $G$ functions and its own architecture.}
\label{fig:formaldehyde-HD-NNP}
\end{figure*}

\clearpage

\section{Single Transition State Training}

The xyz-formatted molecular geometry data for the reactant, transition state, and product of the reaction used in the single-TS training are presented below.

\setstretch{1.0}

\begin{verbatim}
13
Reactant
C                  1.76908200    0.61766900    0.45171600
O                  0.36988700    0.50871900    0.42121600
C                 -0.08120500   -0.61521500   -0.30205800
C                 -1.59249100   -0.50273600   -0.37993000
O                 -1.97798800    0.68798800   -1.02483500
H                  2.01367900    1.51467300    1.02468900
H                  2.23015800   -0.25486600    0.93926600
H                  2.18921100    0.71183100   -0.56146800
H                  0.34588800   -0.61102800   -1.31773600
H                  0.22082200   -1.55054000    0.19726500
H                 -1.99865900   -1.33489900   -0.96248500
H                 -2.00892300   -0.56304100    0.63778800
H                 -1.47946100    1.39144400   -0.58730300

13
Transition State
C                  1.10145000    0.62209100    0.49341400
O                  0.87410500   -0.65128500    1.07463600
C                  0.15253600   -1.30831400    0.09222800
C                 -0.98348800   -0.38881300   -0.25144600
O                 -0.68621400    1.24171600   -1.45195100
H                  1.31717700    1.34194100    1.28796500
H                  1.93978200    0.57254500   -0.22443100
H                  0.03395600    1.21570600   -0.68633100
H                  0.76052500   -1.52975300   -0.80418800
H                 -0.23399100   -2.27098400    0.47284000
H                 -1.69037300   -0.72898700   -1.00454300
H                 -1.43064800    0.09057200    0.61359400
H                 -1.34670400    1.90114800   -1.18191400

13
Product
C                  0.72266900    0.22324600    0.12956500
O                  1.36449700   -0.83581000    0.85470300
C                  0.29209400   -1.74302100    0.55981100
C                 -0.49894200   -0.65196000   -0.18137300
O                 -0.90831800    2.26604900   -1.48616600
H                  0.54062100    1.09954700    0.76515900
H                  1.30410900    0.53415100   -0.74610800
H                 -0.13413500    2.82747300   -1.36715600
H                  0.62824700   -2.58323500   -0.06077500
H                 -0.17038200   -2.13841400    1.47241700
H                 -0.69097700   -0.82569300   -1.24036400
H                 -1.42578300   -0.34401100    0.30719000
H                 -1.39232000    2.35211100   -0.65743700

\end{verbatim}

\setstretch{1.5}

\begin{figure}[h!]
    \centering
    \rule[1ex]{\textwidth}{0.1pt}
    \includegraphics[width=\textwidth]{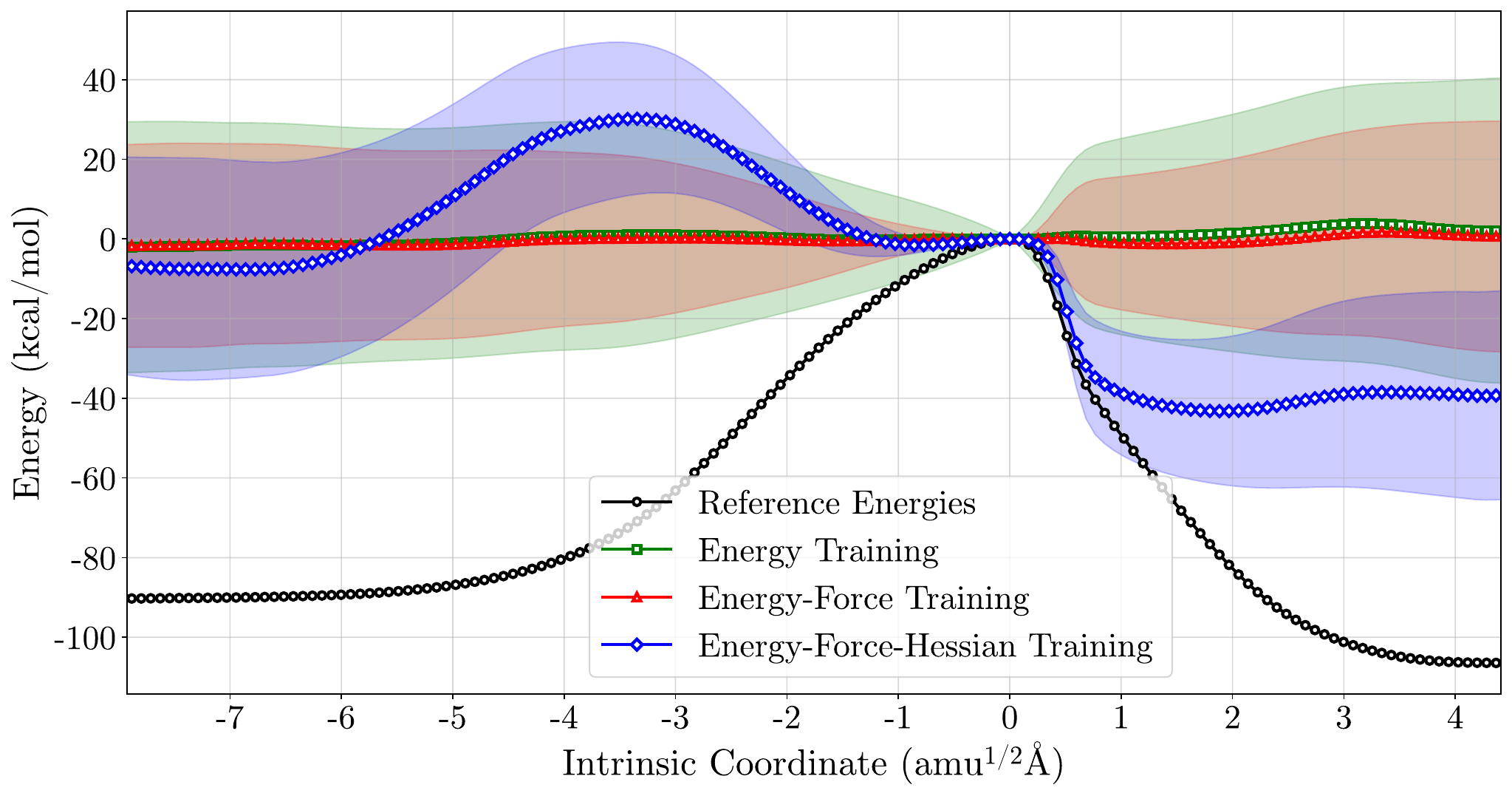}
    \caption{Energy predictions of molecular structures in IRC path of a reaction by models trained on only the TS point of the same reaction. The black circles correspond to the energies obtained by DFT calculations, which are considered to be the ground truth for our models. The colored markers represent the average energy predictions of the models trained to fit to energies; energies and forces; and energies, forces, and Hessian matrix of the data. The filled areas represent the standard deviation for each prediction by the ensemble of models.}
    \label{fig:TS-training}
\end{figure}

\clearpage

\section{Ensemble Predictions}

In this section, we present the ensemble predictions of our models for the energy values along the Intrinsic Reaction Coordinate (IRC) of a single reaction. Ensembles of models are used to enhance the robustness and reliability of the predictions by averaging the noise and reducing the variance inherent in single-model predictions.

For each type of model—energy fitting, energy-force fitting, and energy-force-Hessian fitting, we generate predictions using an ensemble of 100 models. Each model within the ensemble is represented by a line in the plots, with different fitting types distinguished by specific colors. This visualization allows for a clear comparison of the prediction performance and variability among the different fitting approaches.

Figure \ref{fig:ensemble-predictions} below illustrates the ensemble predictions of the energy values along the IRC for the chosen reaction. By examining these plots, we can assess how consistently each type of model predicts energy values and identify any trends or discrepancies that may arise.

\bigskip

\begin{figure}[h]
\centering
\rule[1ex]{\textwidth}{0.1pt}
\includegraphics[width=\textwidth]{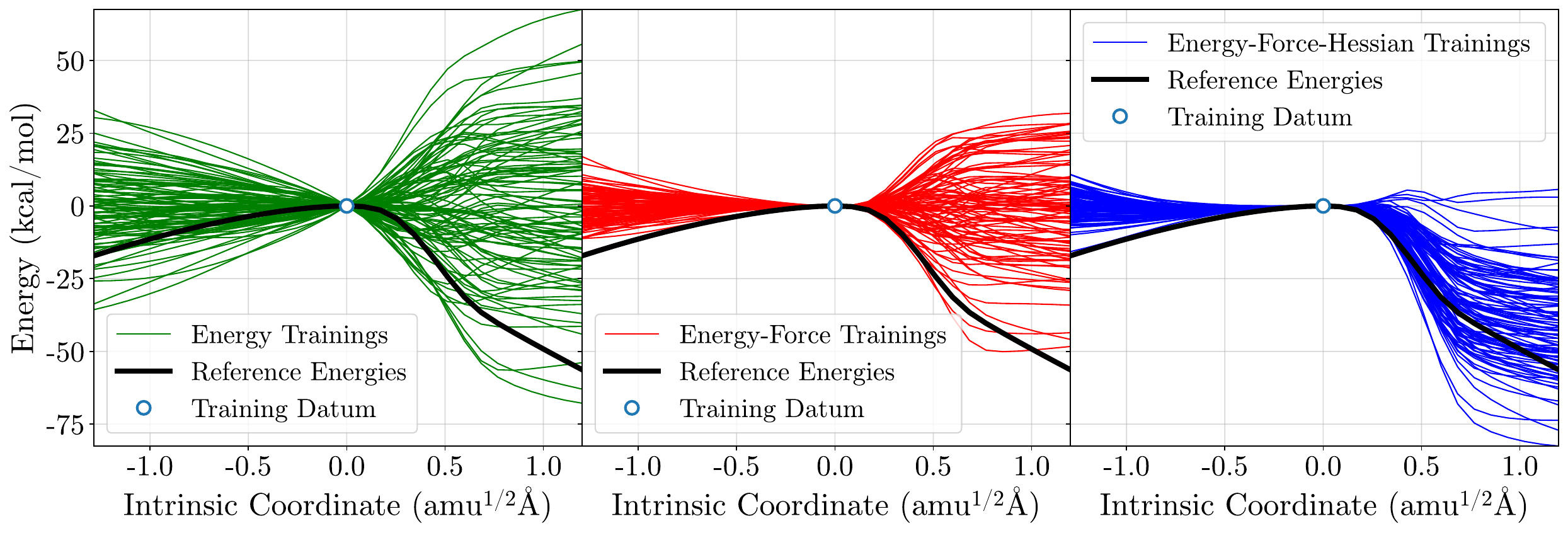}
\caption{Ensemble predictions of energy values along the Intrinsic Reaction Coordinate (IRC) for a single reaction. The plot displays the predictions of three types of models: Energy Fitting (green lines), Energy-Force Fitting (red lines), and Energy-Force-Hessian Fitting (blue lines) compared to the reference energies (black line) and the Transition State (light blue circle). Each line represents the prediction of an individual model within an ensemble of 100 models for each fitting type.}
\label{fig:ensemble-predictions}
\end{figure}

\clearpage

\section{Correlation plots}

Correlation plots serve as a valuable tool to assess the accuracy of our predictive models. They visually compare the reference values obtained from Density Functional Theory (DFT) calculations (x-axis) with the predicted values generated by our models (y-axis). In these graphs, the diagonal red dotted line represents the ideal case where the predicted values perfectly match the reference values (i.e. $y=x$). The closer the points are to this red dotted line, the more accurate the model predictions are. The density gradient in these plots indicates the concentration of data points, providing insight into areas where predictions are more consistent or where deviations are more frequent.

We have developed three types of models to enhance the prediction accuracy for different properties: Energy Fitting Models, Energy-Force Fitting Models, and Energy-Force-Hessian Fitting Models. All of these models were trained on the same dataset, which is composed of 35,087 reactants, transition states, and products that are part of 11,961 distinct reactions (R-TS-P dataset).

Each model is evaluated on three distinct datasets:

\begin{enumerate}
    \item \textbf{R-TS-P Dataset}: This dataset includes reactants, transition states, and products for 11,961 reactions (35,087 individual structures).
    \item \textbf{IRC Dataset}: This dataset consists of Intrinsic Reaction Coordinates (IRCs) for 2,000 reactions (225,963 individual structures).
    \item \textbf{NMS Dataset}: This dataset contains perturbed structures obtained from Normal Mode Sampling (NMS) of the IRC structures (58,219 individual structures).
\end{enumerate}

For each dataset and fitting type, the models produce three types of correlation plots, reflecting the predictions for energy, force, and Hessian values. These plots provide a comprehensive overview of the performance of the model in different datasets and prediction types.

\newpage
\subsection{Energy Fitting Models}


\begin{figure}[H]
    \centering
    \begin{subfigure}[b]{0.32\textwidth}
        \includegraphics[width=\textwidth]{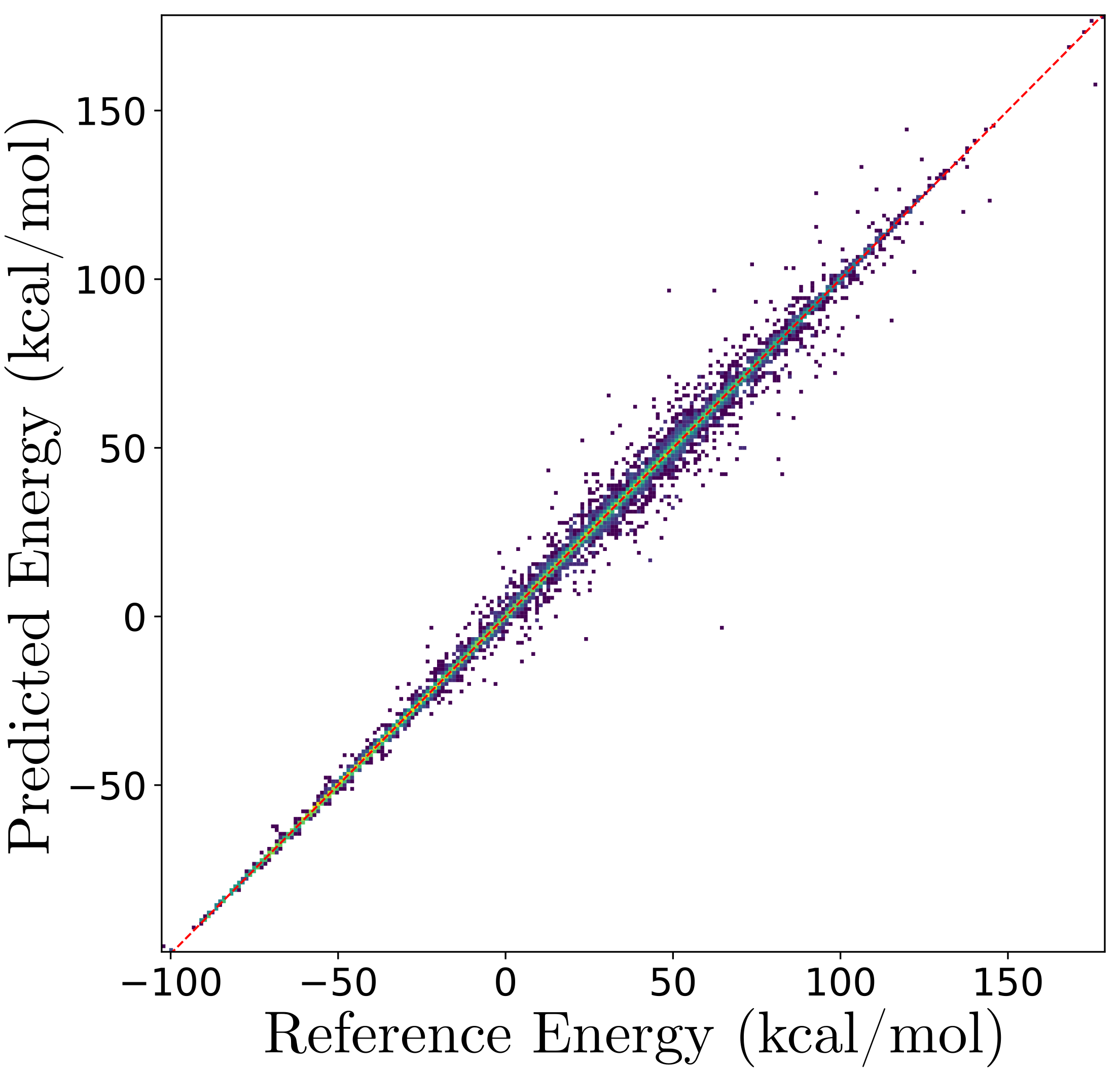}
        \caption{R-TS-P Dataset - Energy}
    \end{subfigure}
    \begin{subfigure}[b]{0.32\textwidth}
        \includegraphics[width=\textwidth]{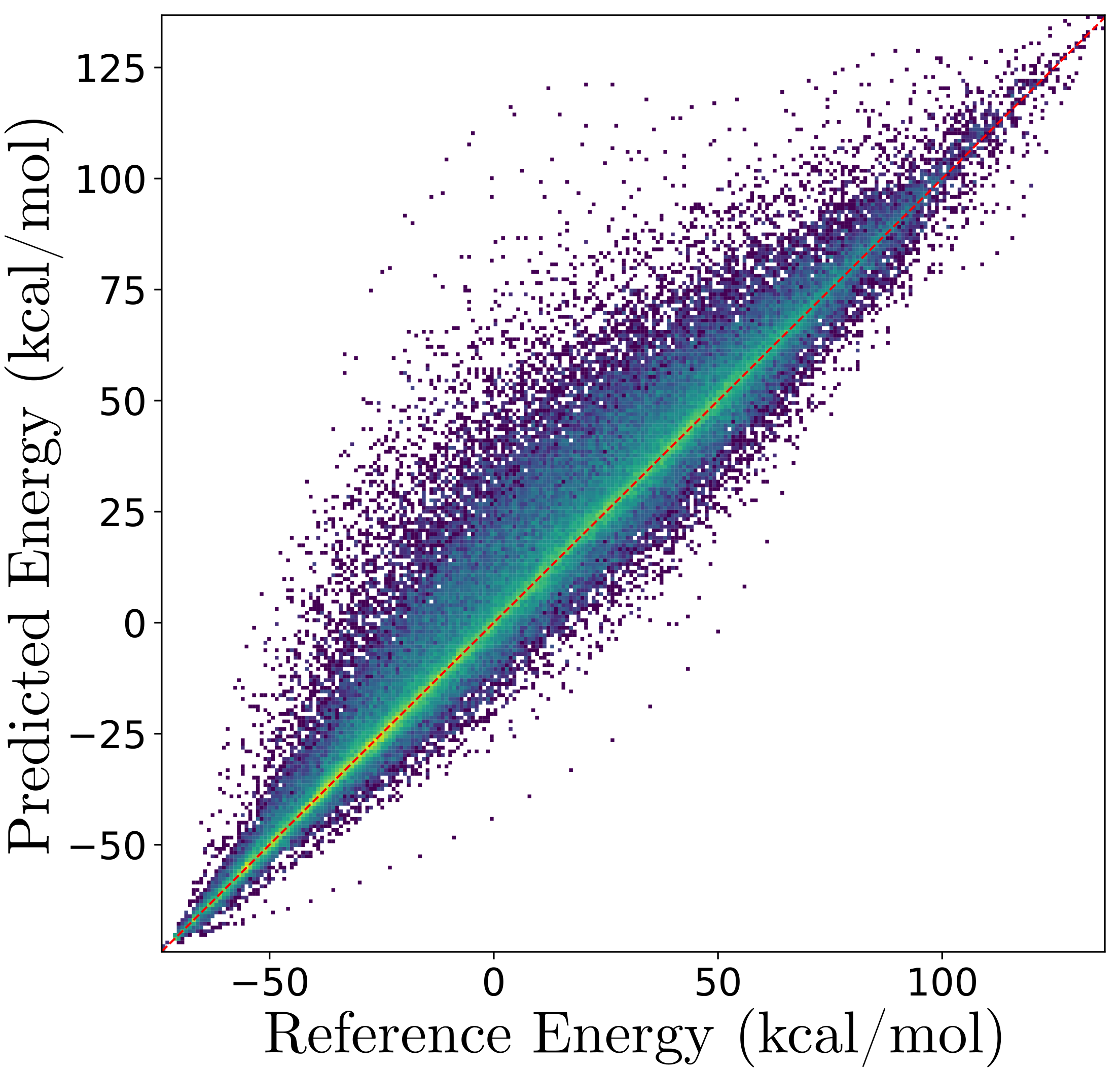}
        \caption{IRC Dataset - Energy}
    \end{subfigure}
    \begin{subfigure}[b]{0.32\textwidth}
        \includegraphics[width=\textwidth]{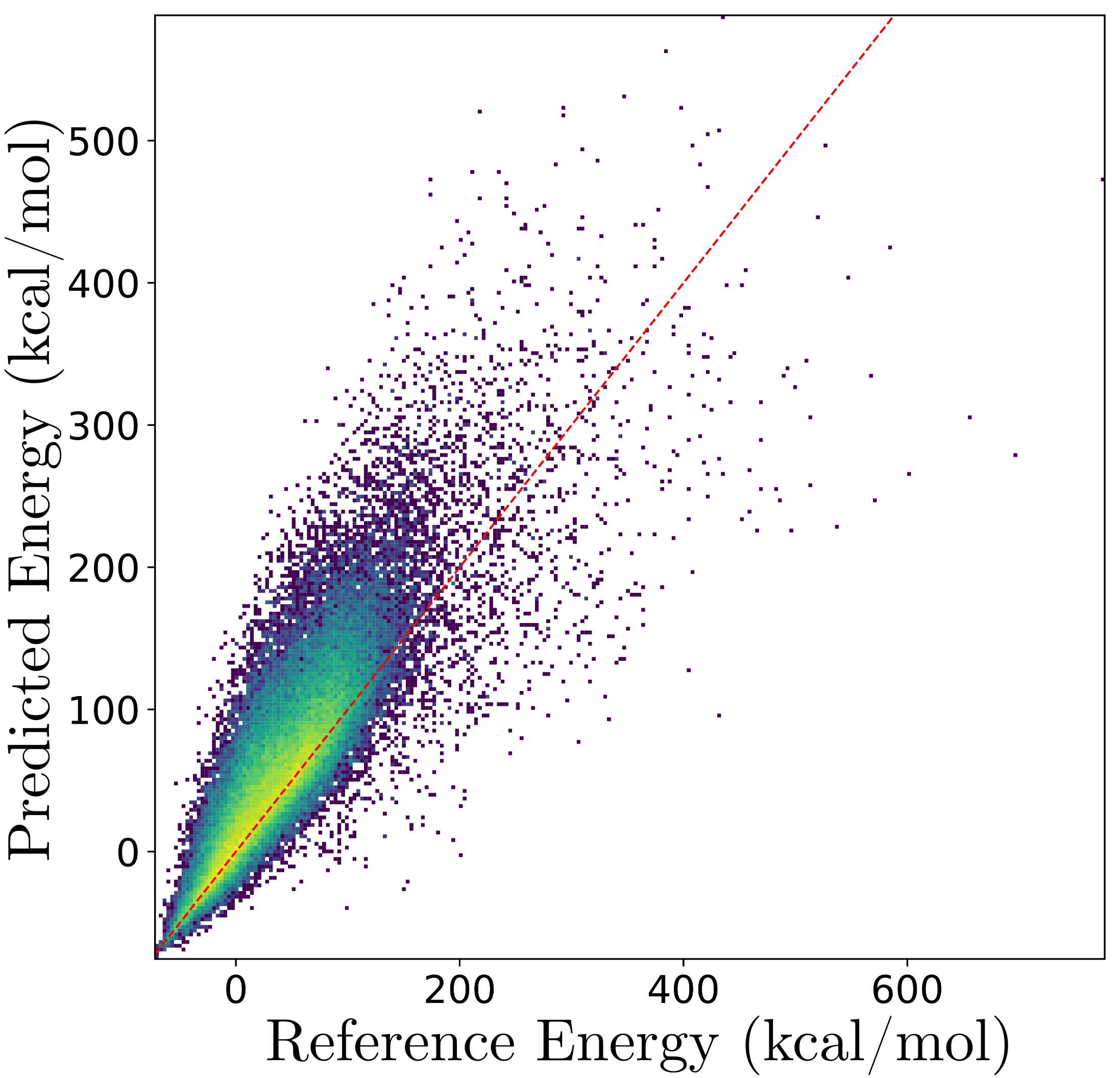}
        \caption{NMS Dataset - Energy}
    \end{subfigure}
    \caption{Energy Predictions of Energy Fitting Models}
    \label{fig:energy_fitting_E}
\end{figure}


\begin{figure}[H]
    \centering
    \begin{subfigure}[b]{0.32\textwidth}
        \includegraphics[width=\textwidth]{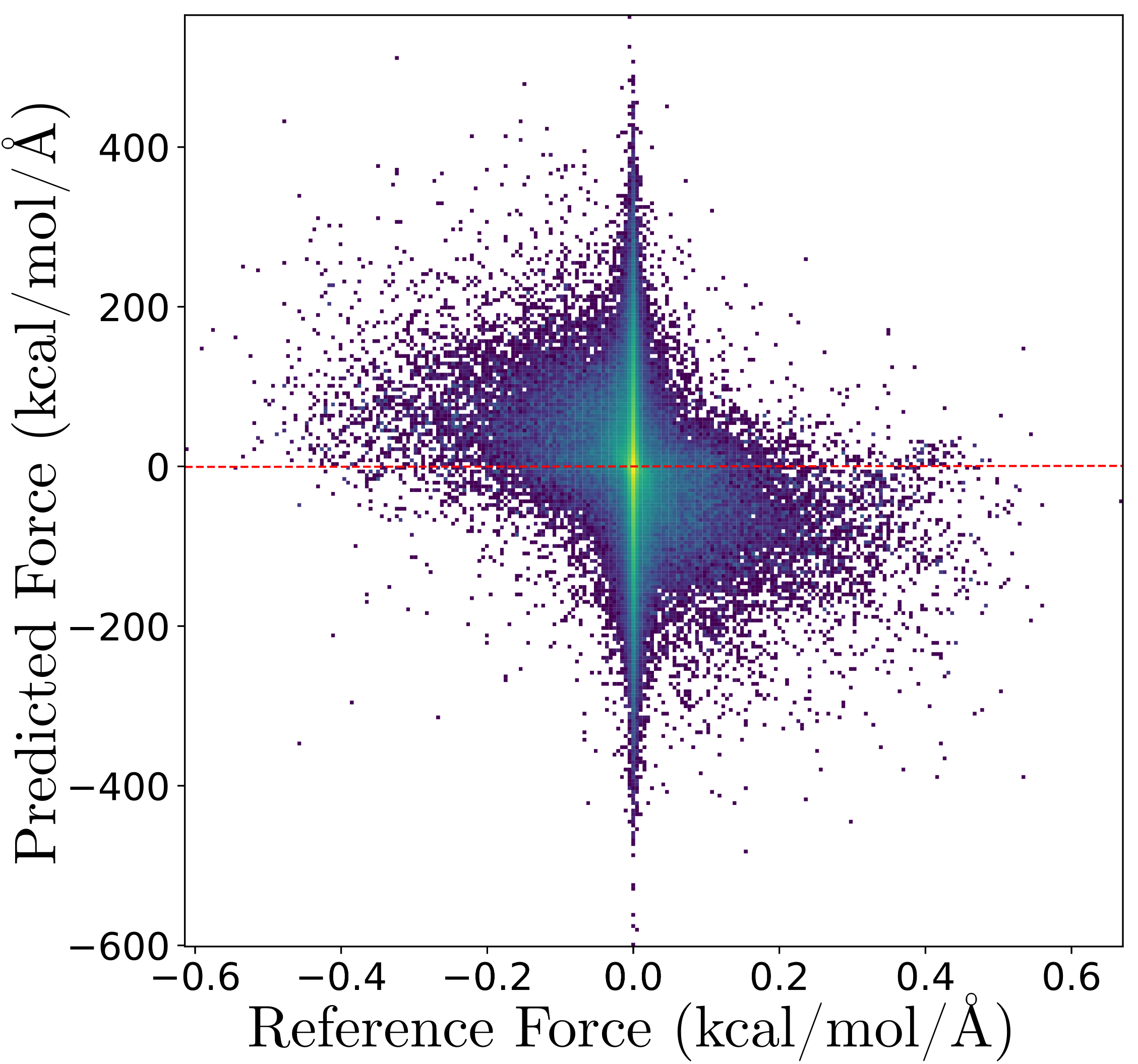}
        \caption{R-TS-P Dataset - Force}
    \end{subfigure}
    \begin{subfigure}[b]{0.32\textwidth}
        \includegraphics[width=\textwidth]{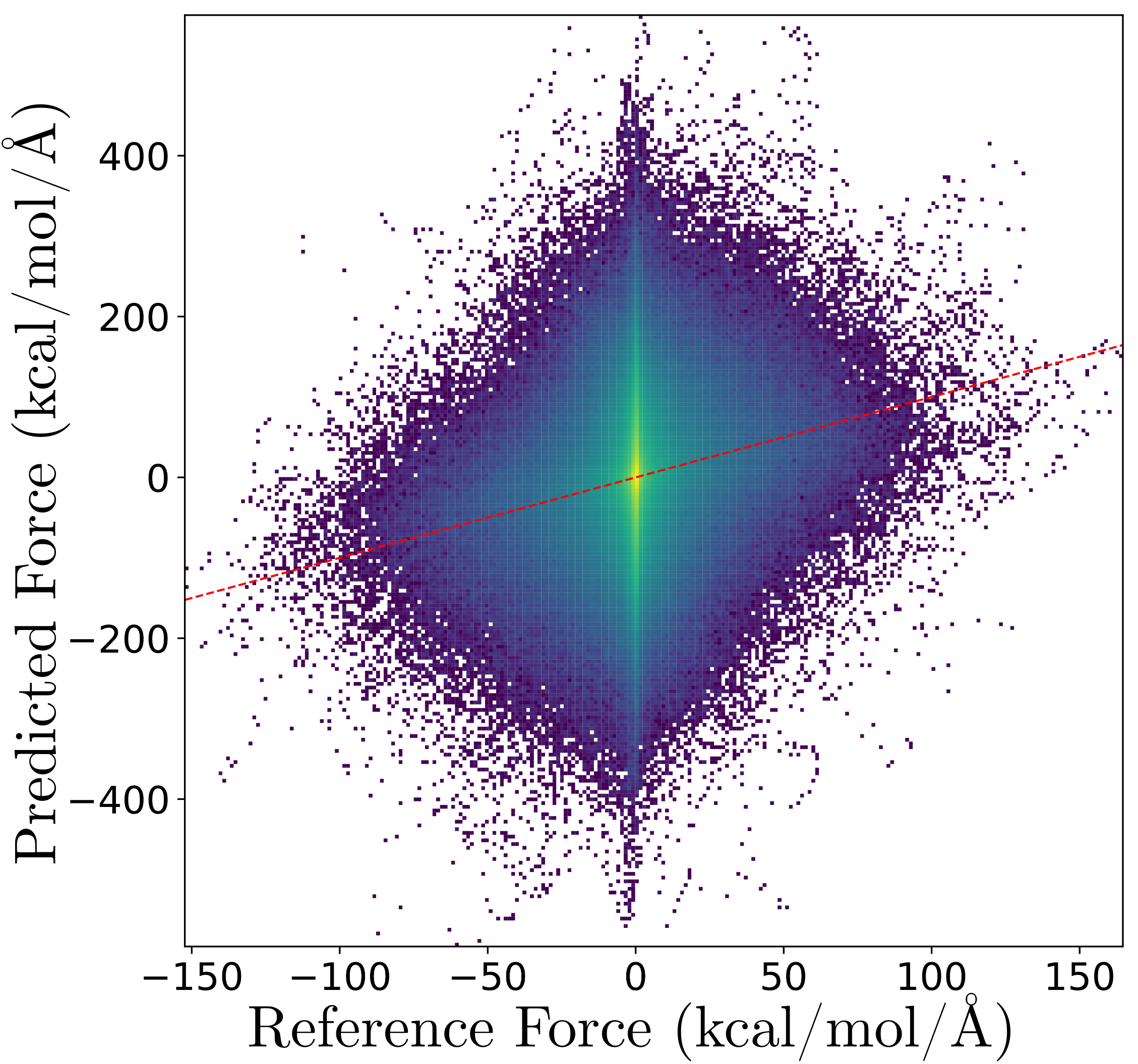}
        \caption{IRC Dataset - Force}
    \end{subfigure}
    \begin{subfigure}[b]{0.32\textwidth}
        \includegraphics[width=\textwidth]{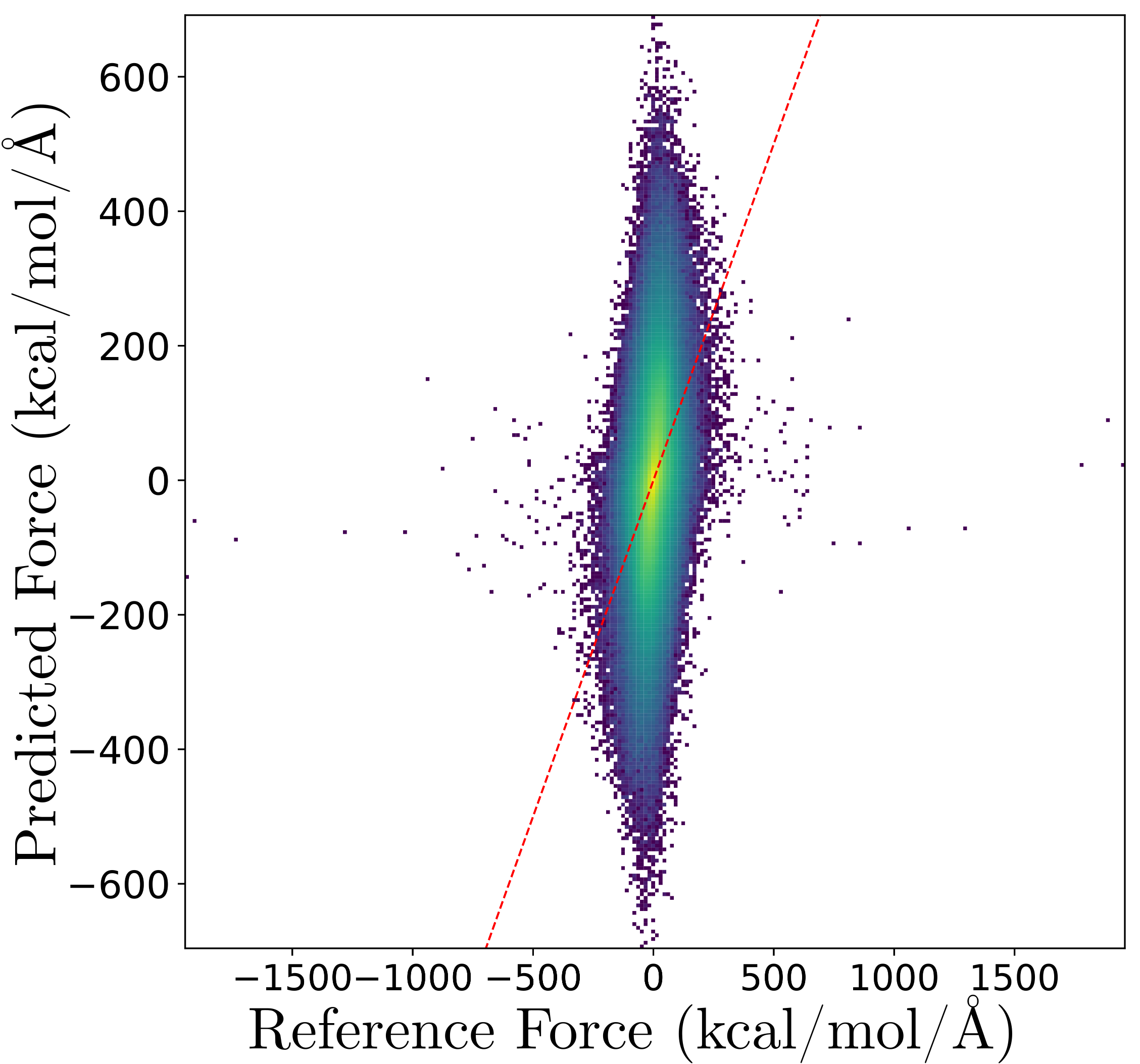}
        \caption{NMS Dataset - Force}
    \end{subfigure}
    \caption{Force Predictions of Energy Fitting Models}
    \label{fig:energy_fitting_F}
\end{figure}


\begin{figure}[H]
    \centering
    \begin{subfigure}[b]{0.32\textwidth}
        \includegraphics[width=\textwidth]{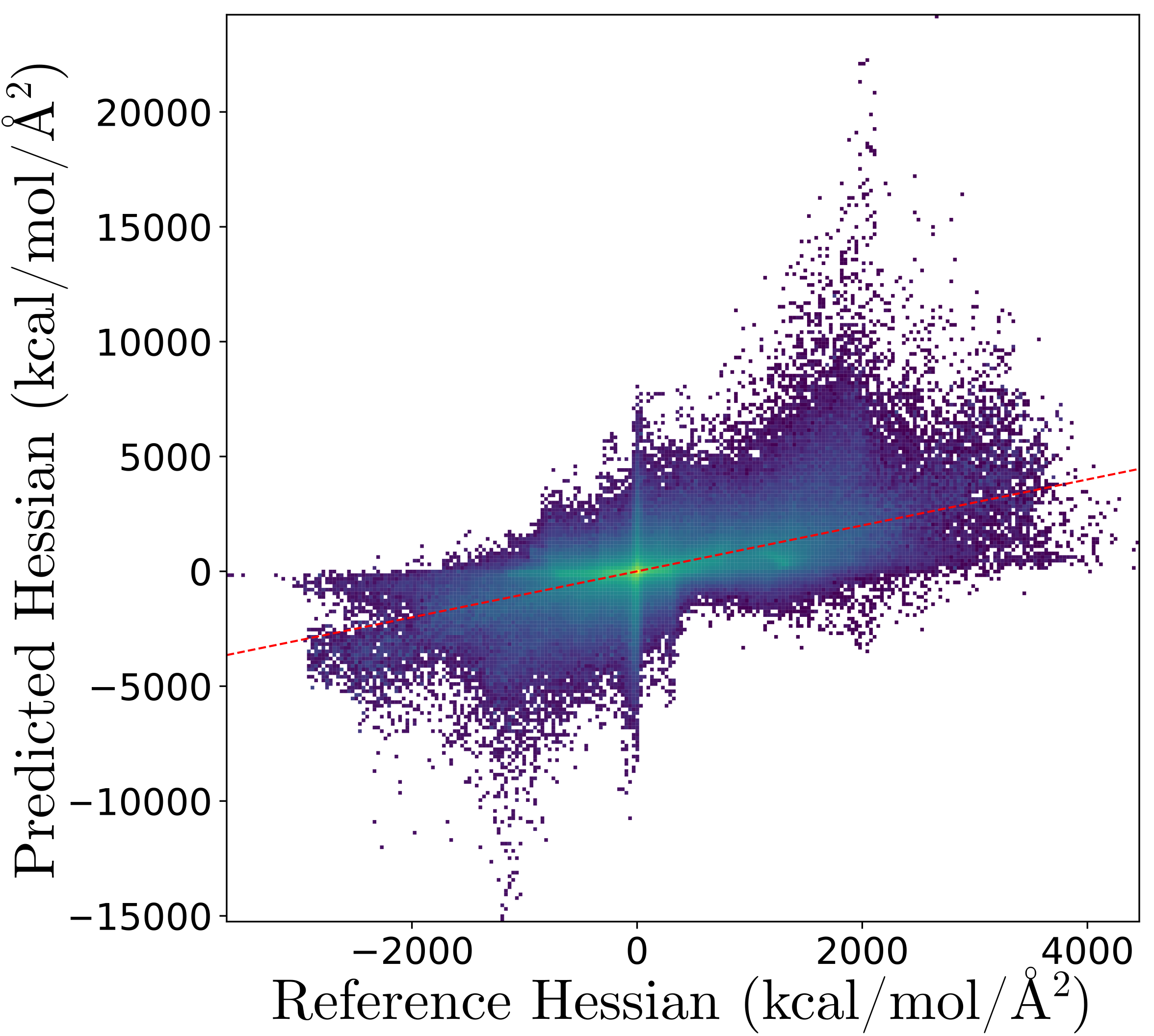}
        \caption{R-TS-P Dataset - Hessian}
    \end{subfigure}
    \begin{subfigure}[b]{0.32\textwidth}
        \includegraphics[width=\textwidth]{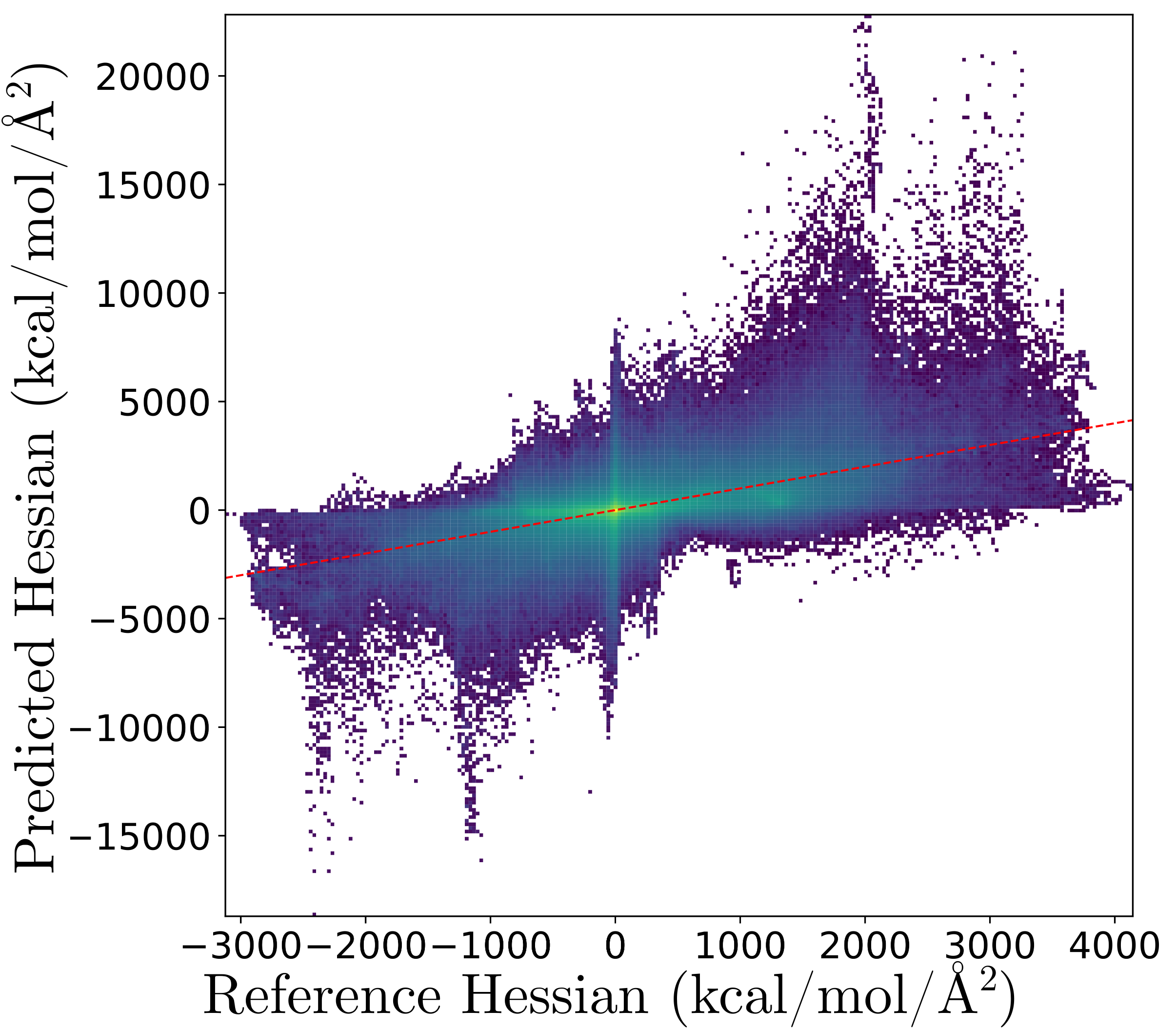}
        \caption{IRC Dataset - Hessian}
    \end{subfigure}
    \begin{subfigure}[b]{0.32\textwidth}
        \includegraphics[width=\textwidth]{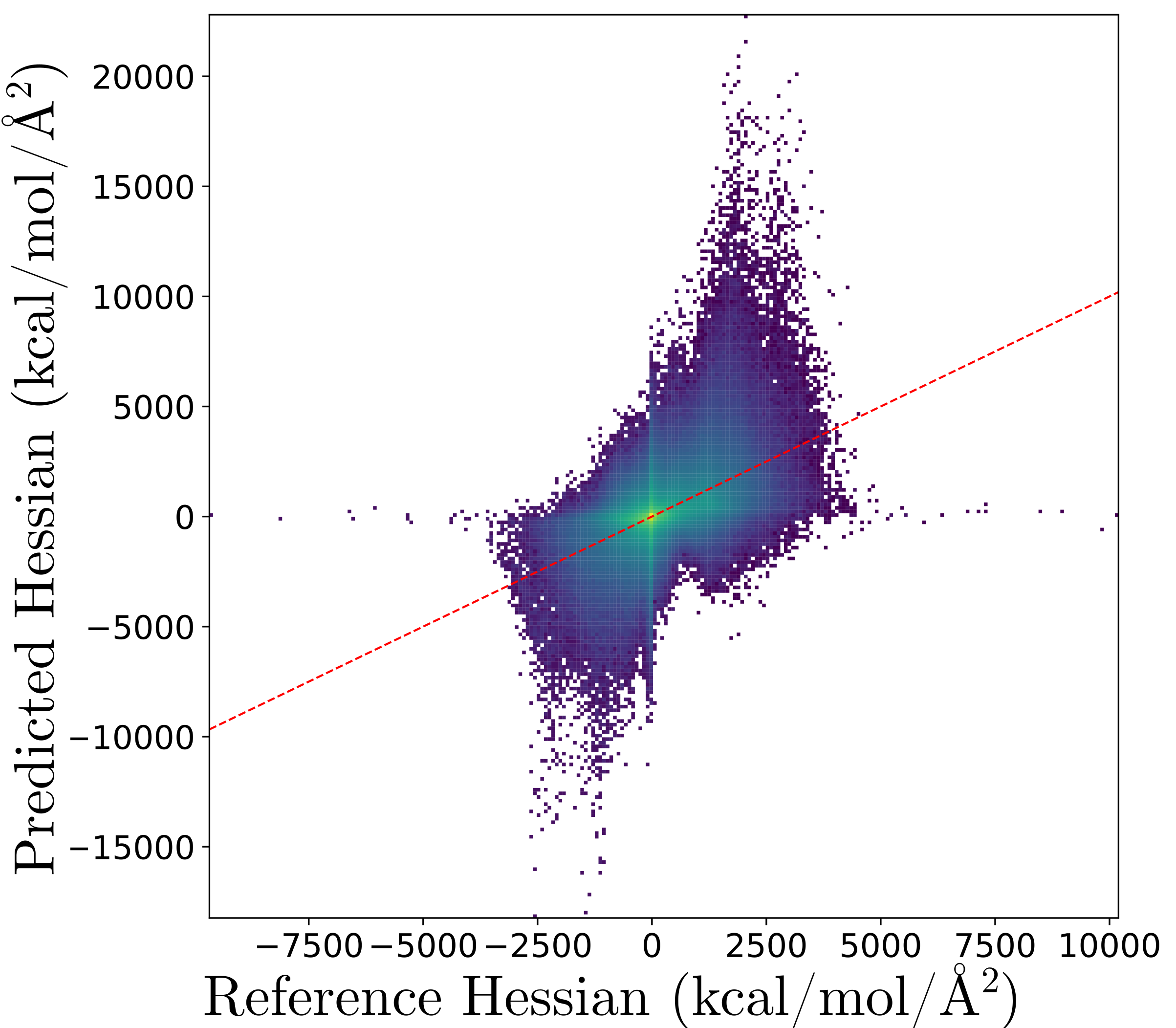}
        \caption{NMS Dataset - Hessian}
    \end{subfigure}
    \caption{Hessian Predictions of Energy Fitting Models}
    \label{fig:energy_fitting_H}
\end{figure}

\subsection{Energy-Force Fitting Models}


\begin{figure}[H]
    \centering
    \begin{subfigure}[b]{0.32\textwidth}
        \includegraphics[width=\textwidth]{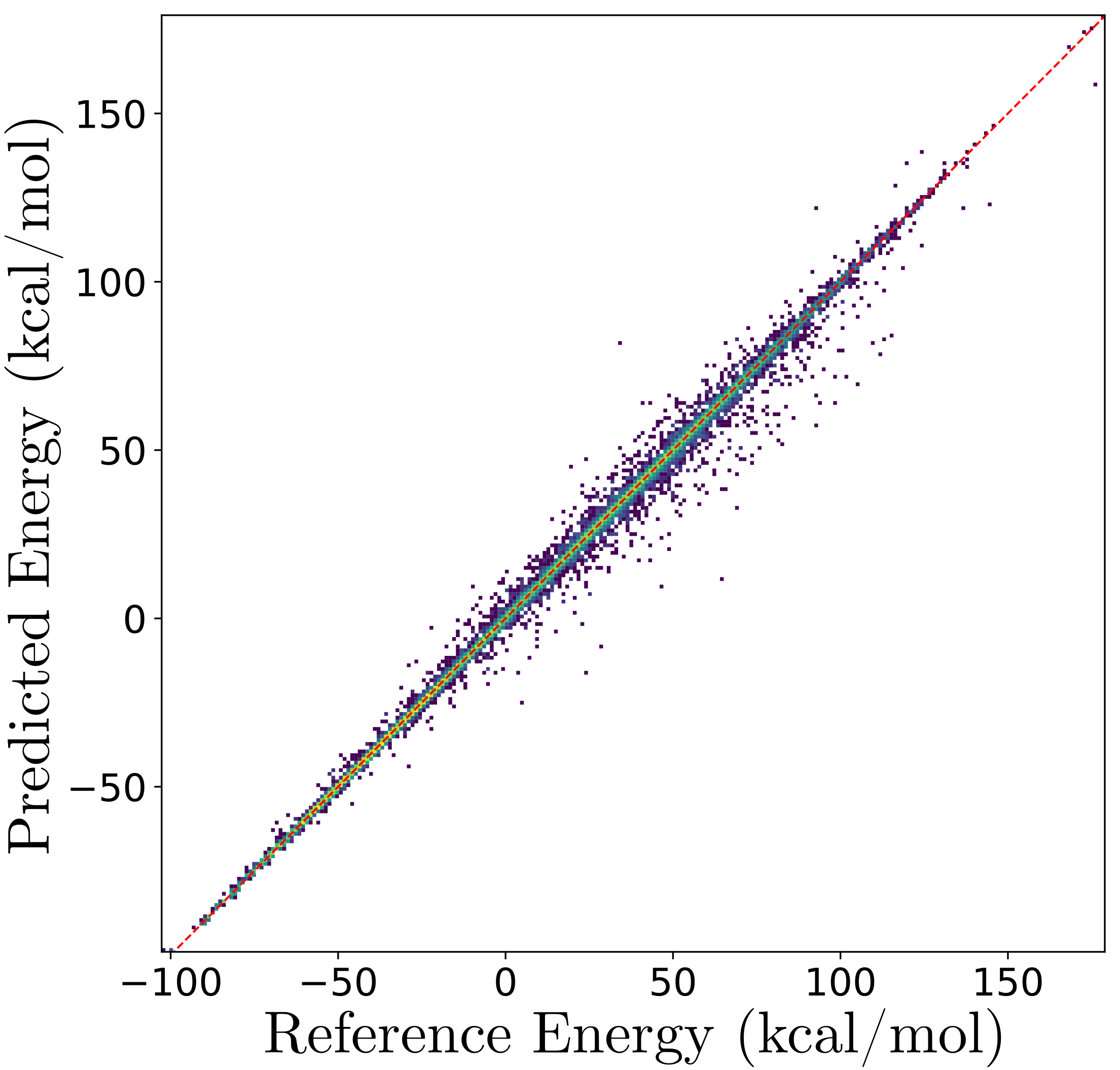}
        \caption{R-TS-P Dataset - Energy}
    \end{subfigure}
    \begin{subfigure}[b]{0.32\textwidth}
        \includegraphics[width=\textwidth]{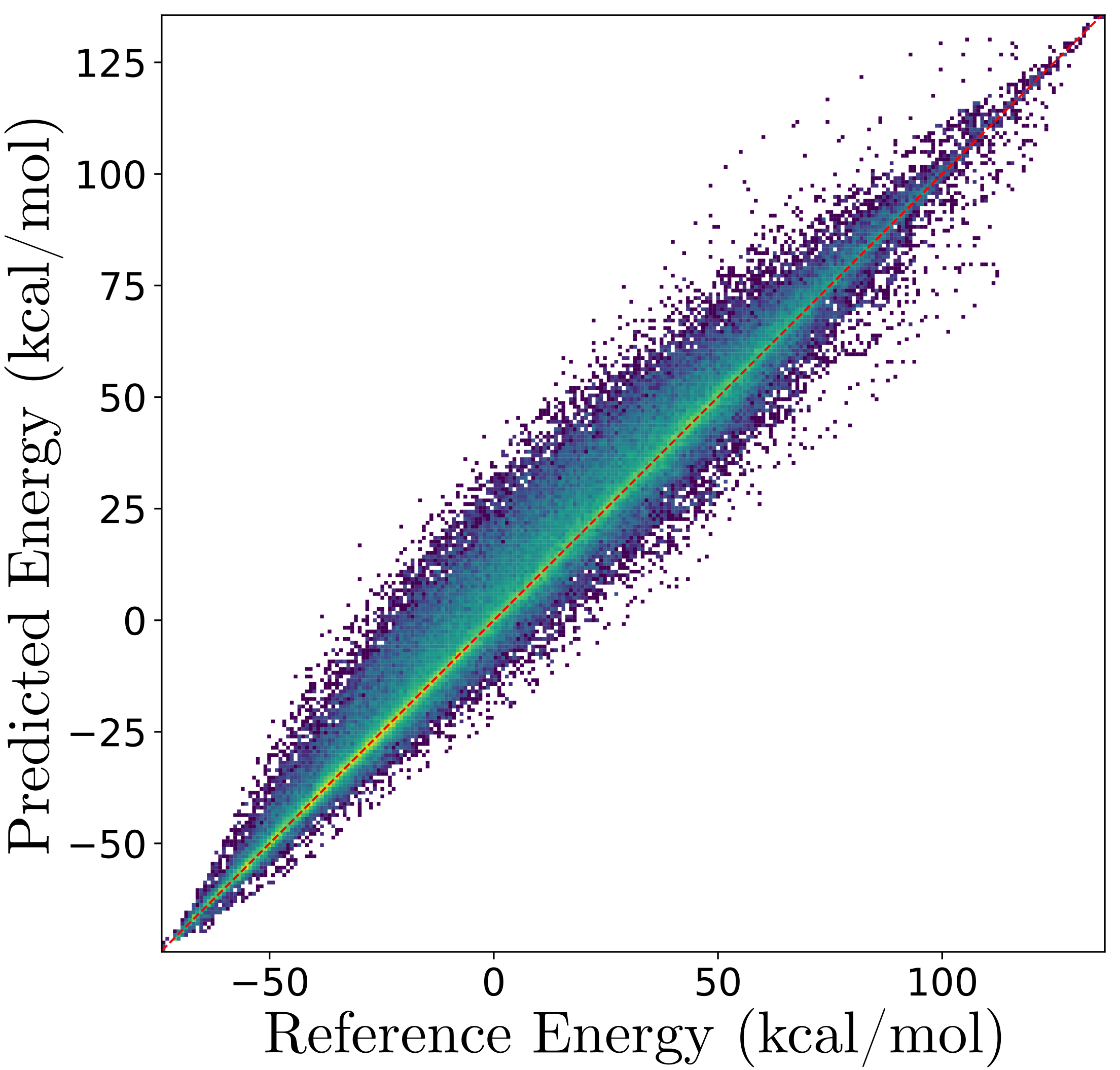}
        \caption{IRC Dataset - Energy}
    \end{subfigure}
    \begin{subfigure}[b]{0.32\textwidth}
        \includegraphics[width=\textwidth]{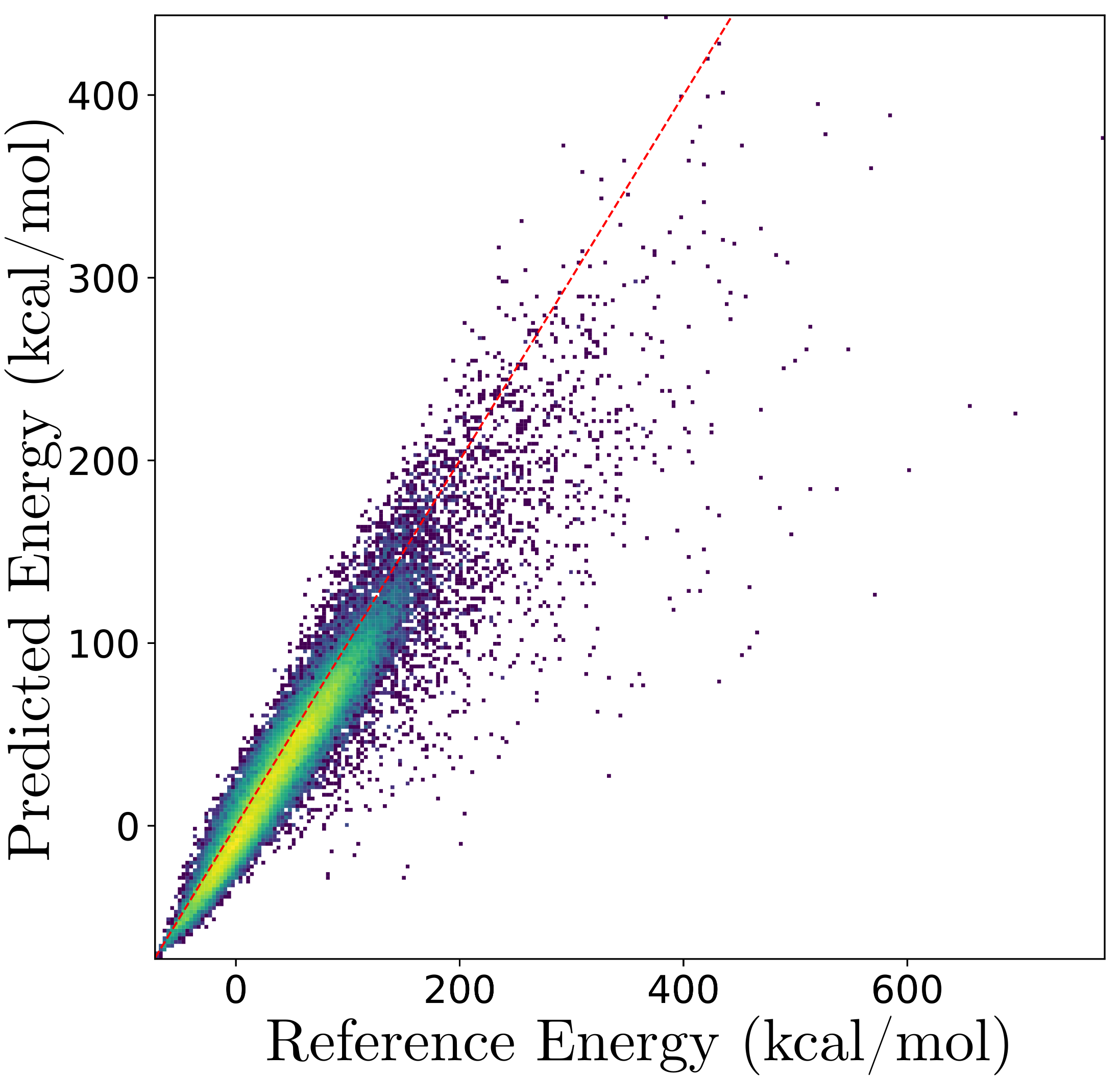}
        \caption{NMS Dataset - Energy}
    \end{subfigure}
    \caption{Energy Predictions of Energy-Force Fitting Models}
    \label{fig:energy_force_fitting_E}
\end{figure}


\begin{figure}[H]
    \centering
    \begin{subfigure}[b]{0.32\textwidth}
        \includegraphics[width=\textwidth]{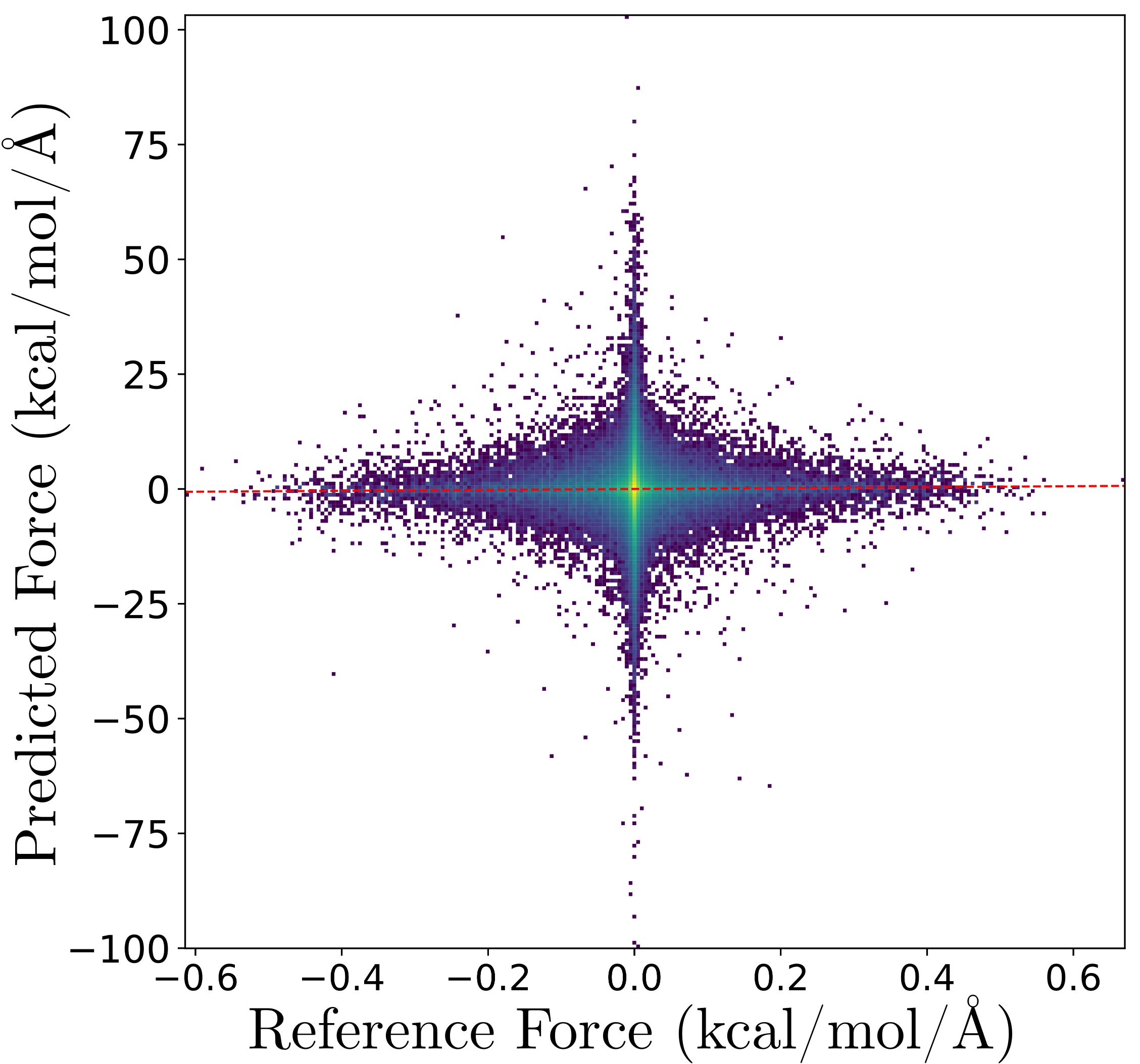}
        \caption{R-TS-P Dataset - Force}
    \end{subfigure}
    \begin{subfigure}[b]{0.32\textwidth}
        \includegraphics[width=\textwidth]{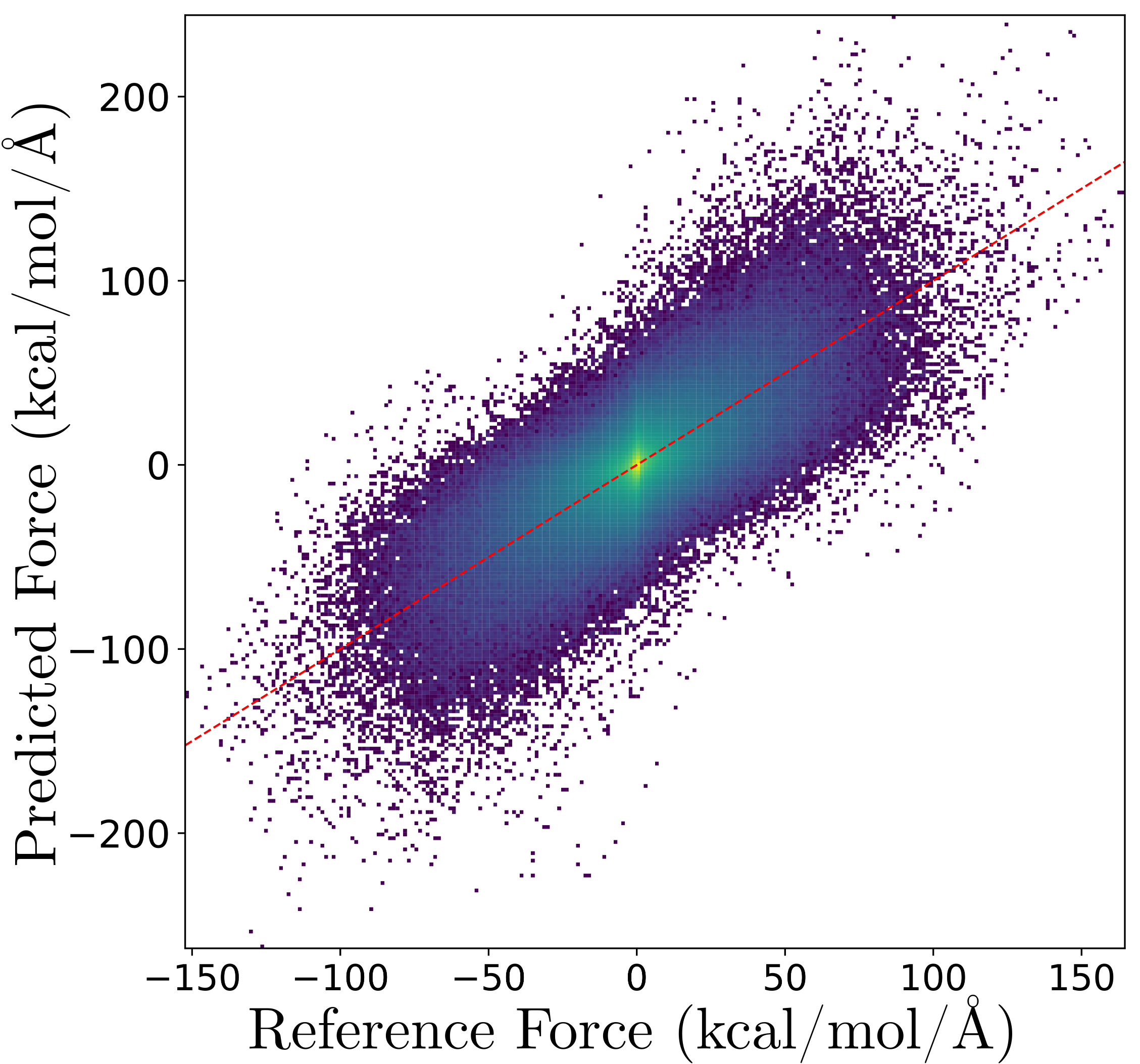}
        \caption{IRC Dataset - Force}
    \end{subfigure}
    \begin{subfigure}[b]{0.32\textwidth}
        \includegraphics[width=\textwidth]{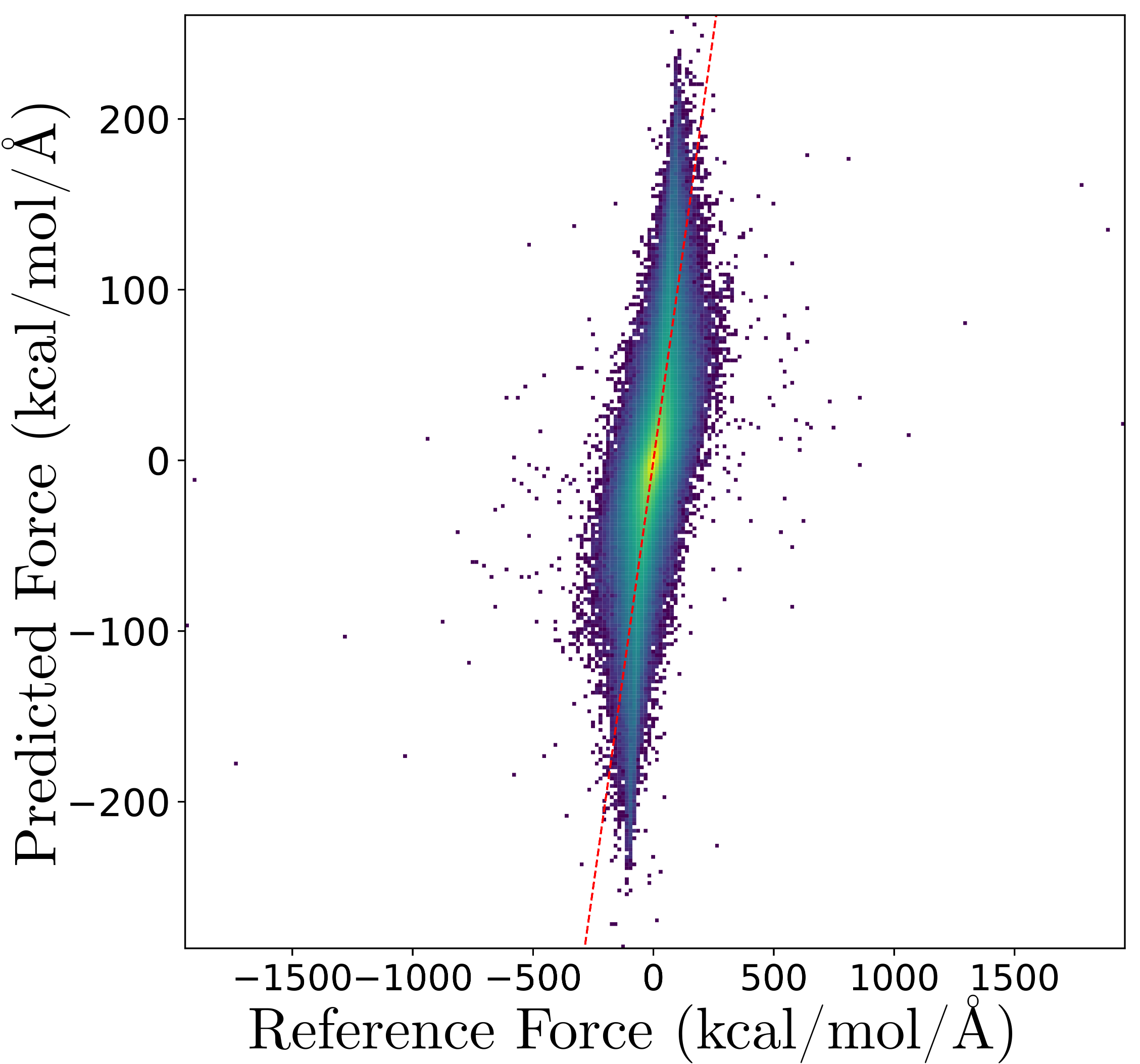}
        \caption{NMS Dataset - Force}
    \end{subfigure}
    \caption{Force Predictions of Energy-Force Fitting Models}
    \label{fig:energy_force_fitting_F}
\end{figure}


\begin{figure}[H]
    \centering
    \begin{subfigure}[b]{0.32\textwidth}
        \includegraphics[width=\textwidth]{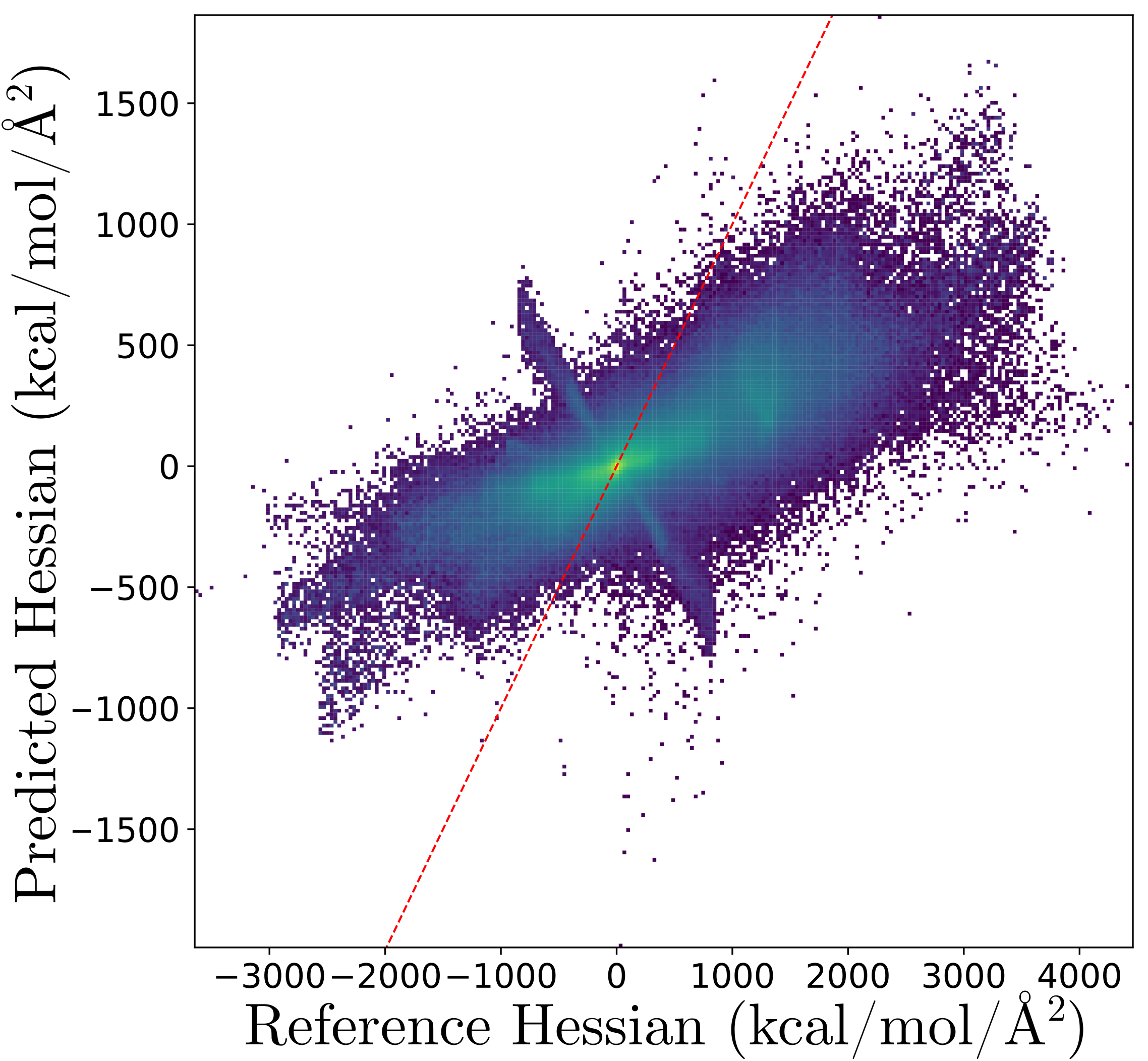}
        \caption{R-TS-P Dataset - Hessian}
    \end{subfigure}
    \begin{subfigure}[b]{0.32\textwidth}
        \includegraphics[width=\textwidth]{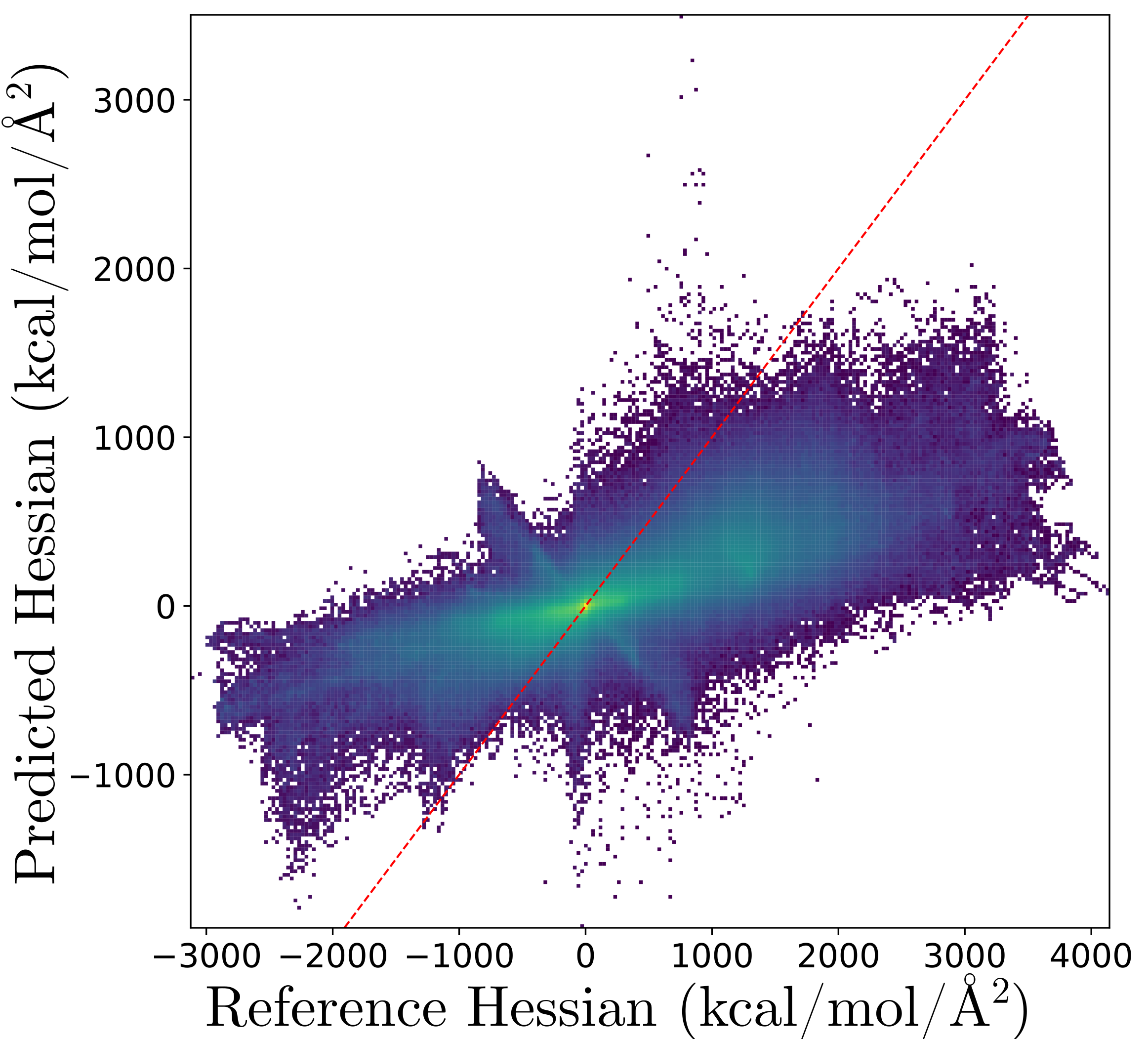}
        \caption{IRC Dataset - Hessian}
    \end{subfigure}
    \begin{subfigure}[b]{0.32\textwidth}
        \includegraphics[width=\textwidth]{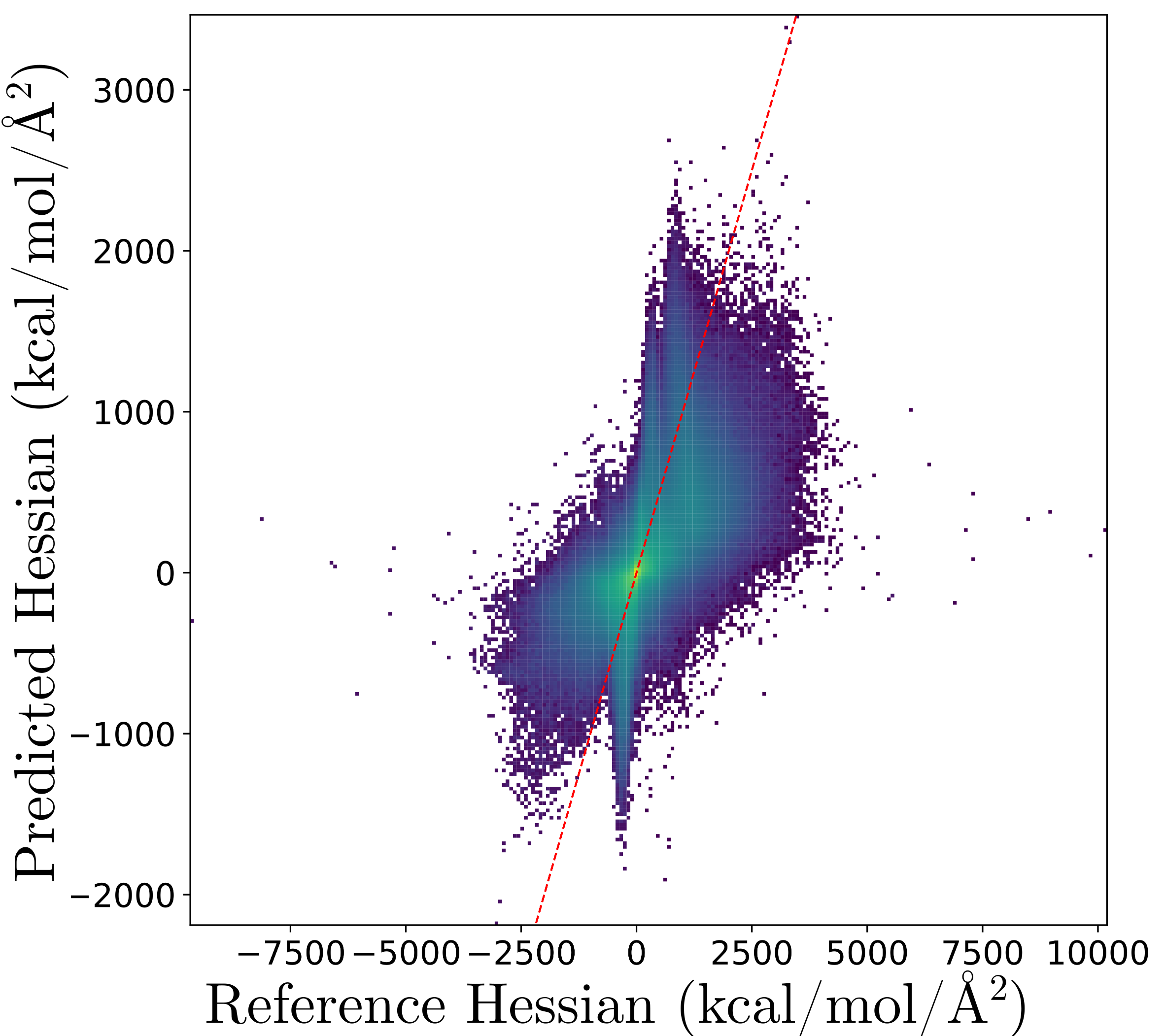}
        \caption{NMS Dataset - Hessian}
    \end{subfigure}
    \caption{Hessian Predictions of Energy-Force Fitting Models}
    \label{fig:energy_force_fitting_H}
\end{figure}

\subsection{Energy-Force-Hessian Fitting Models}


\begin{figure}[H]
    \centering
    \begin{subfigure}[b]{0.32\textwidth}
        \includegraphics[width=\textwidth]{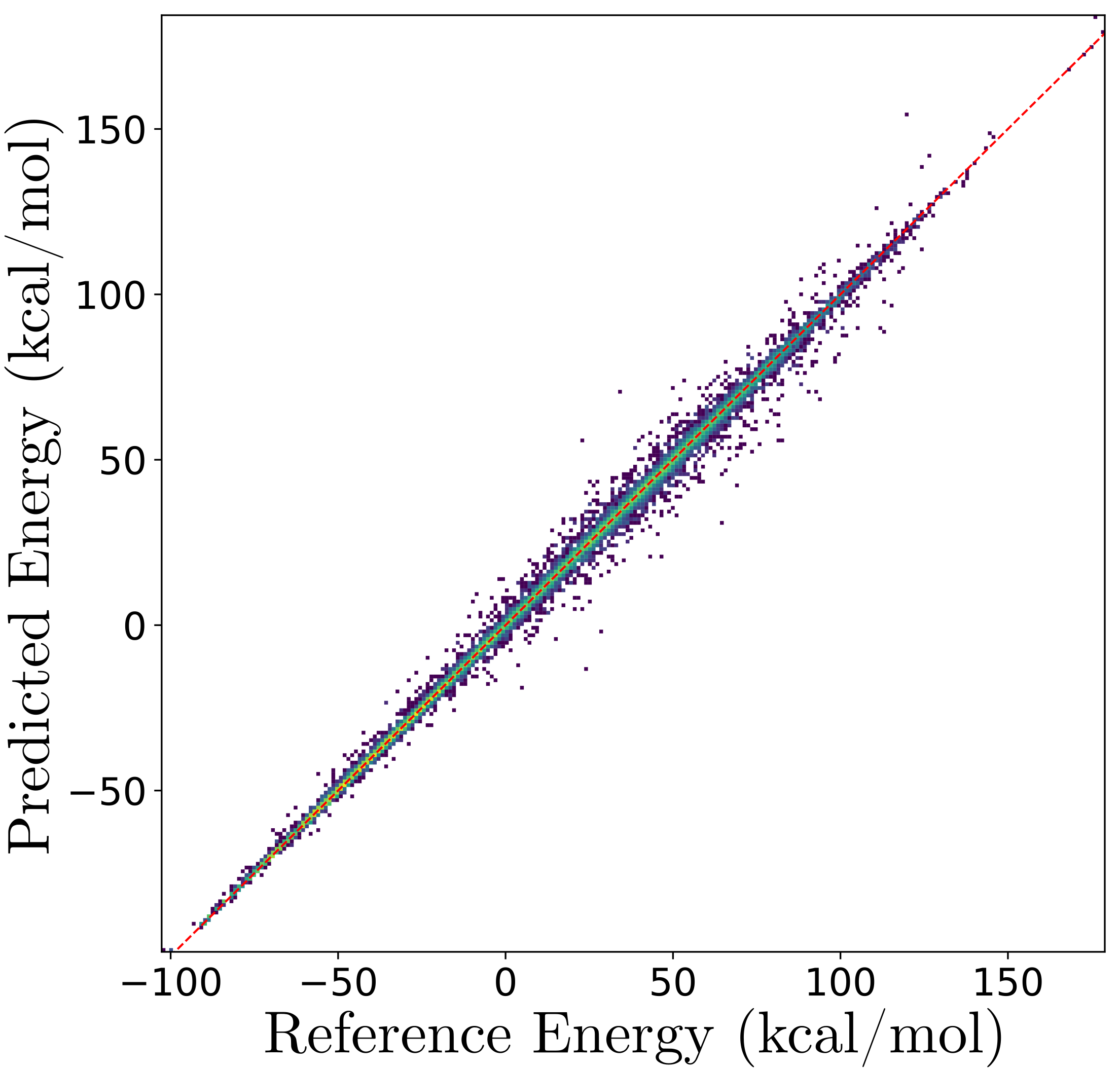}
        \caption{R-TS-P Dataset - Energy}
    \end{subfigure}
    \begin{subfigure}[b]{0.32\textwidth}
        \includegraphics[width=\textwidth]{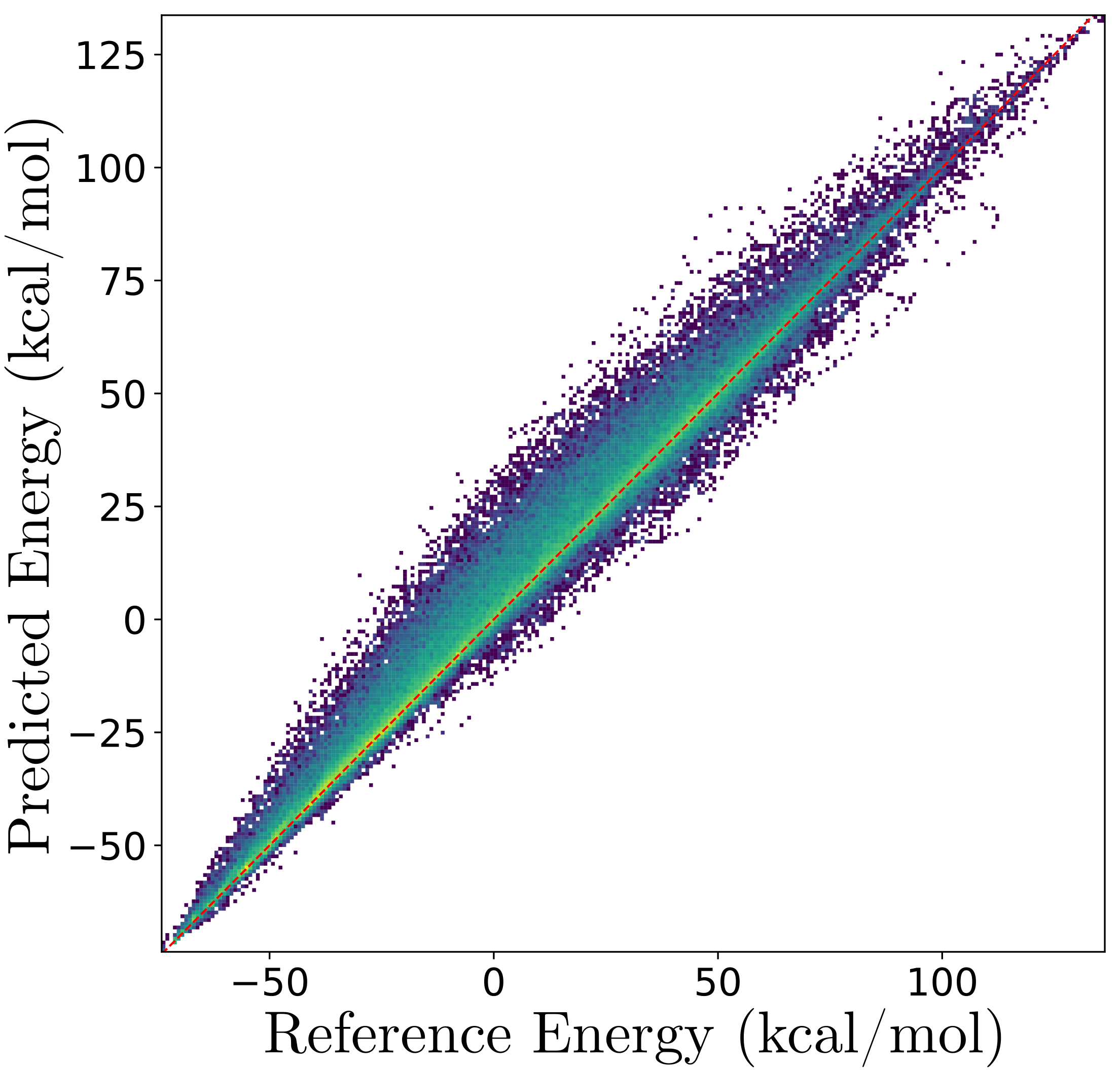}
        \caption{IRC Dataset - Energy}
    \end{subfigure}
    \begin{subfigure}[b]{0.32\textwidth}
        \includegraphics[width=\textwidth]{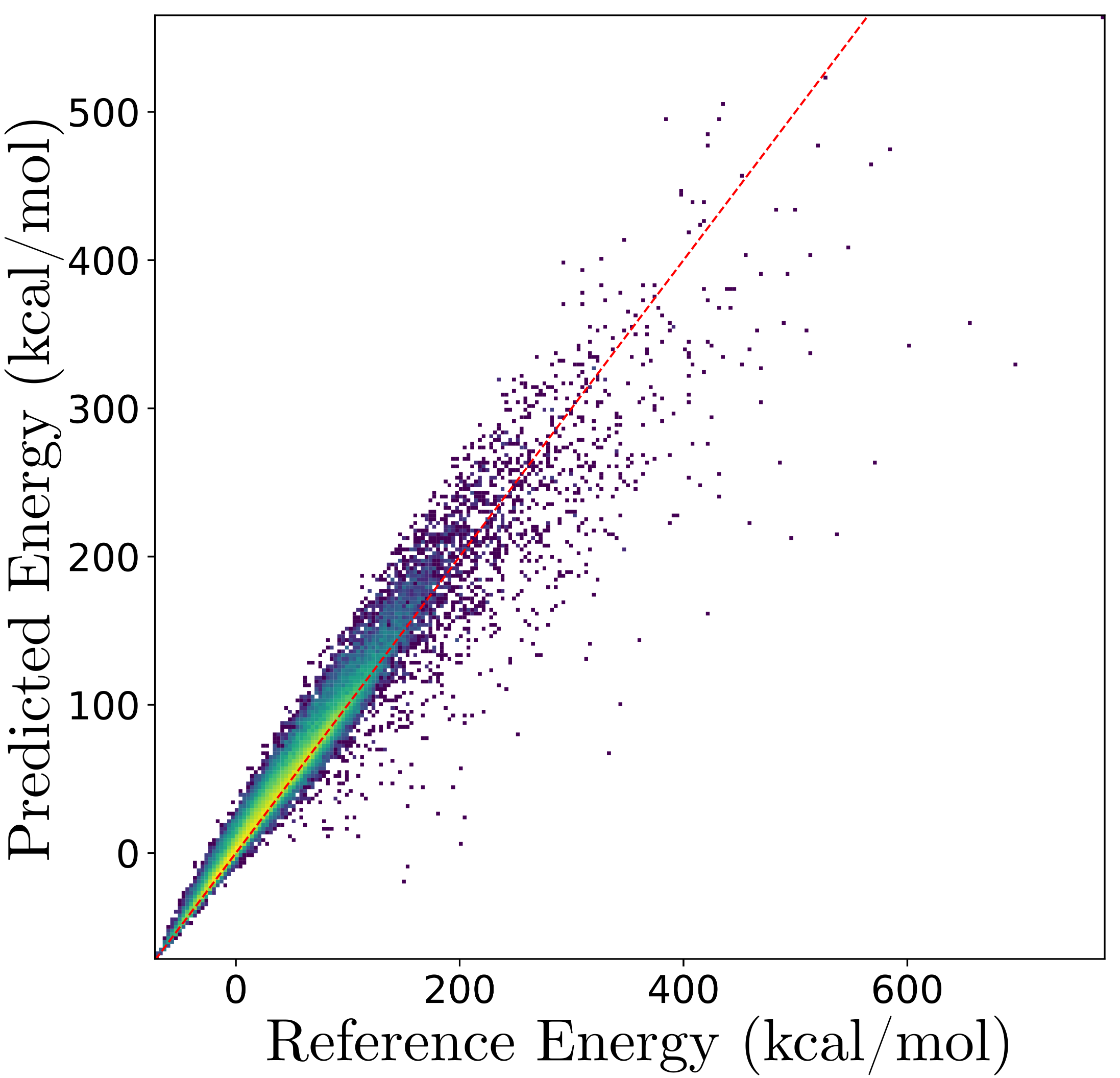}
        \caption{NMS Dataset - Energy}
    \end{subfigure}
    \caption{Energy Predictions of Energy-Force-Hessian Fitting Models}
    \label{fig:energy_force_Hessian_fitting_E}
\end{figure}


\begin{figure}[H]
    \centering
    \begin{subfigure}[b]{0.32\textwidth}
        \includegraphics[width=\textwidth]{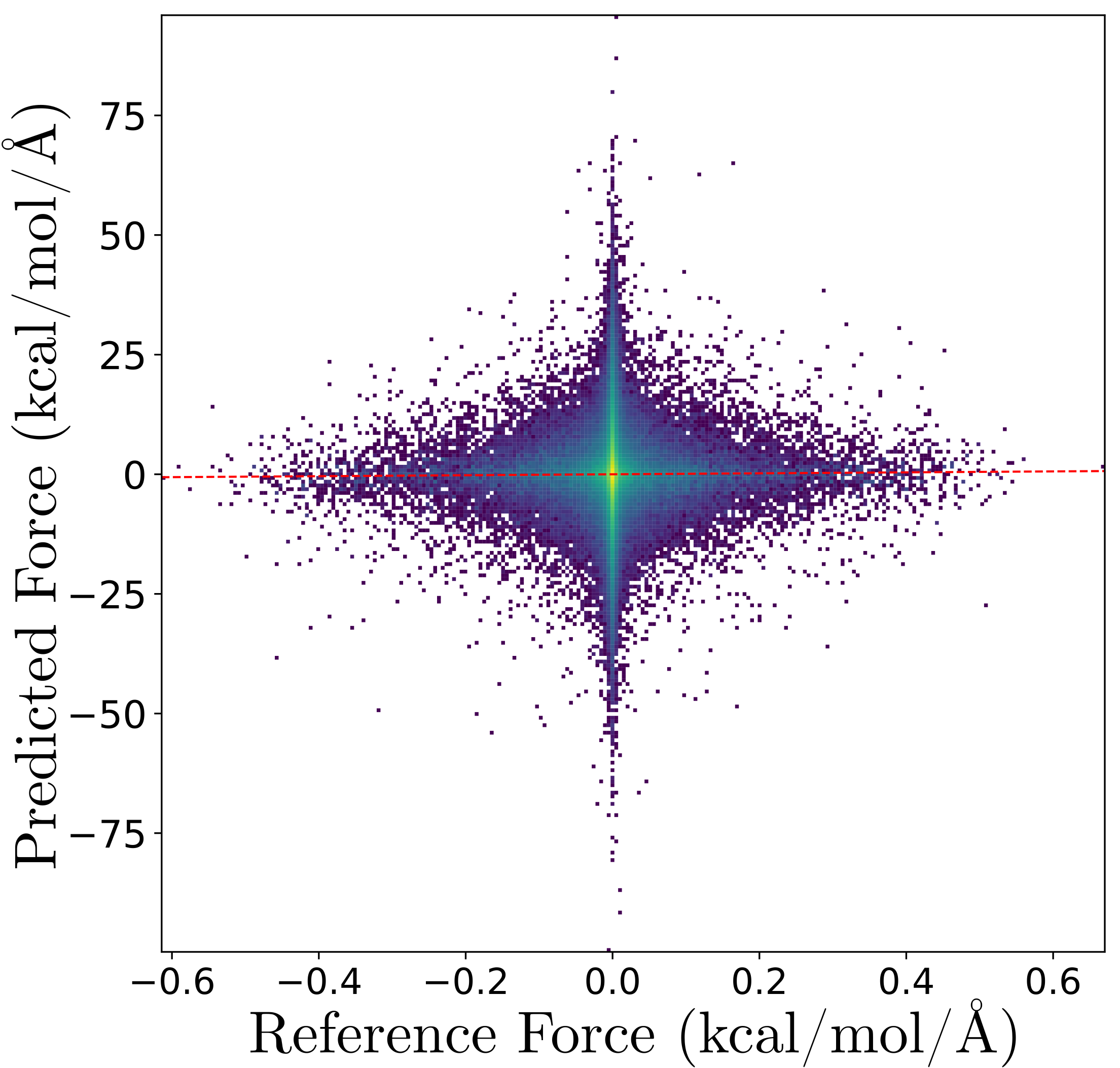}
        \caption{R-TS-P Dataset - Force}
    \end{subfigure}
    \begin{subfigure}[b]{0.32\textwidth}
        \includegraphics[width=\textwidth]{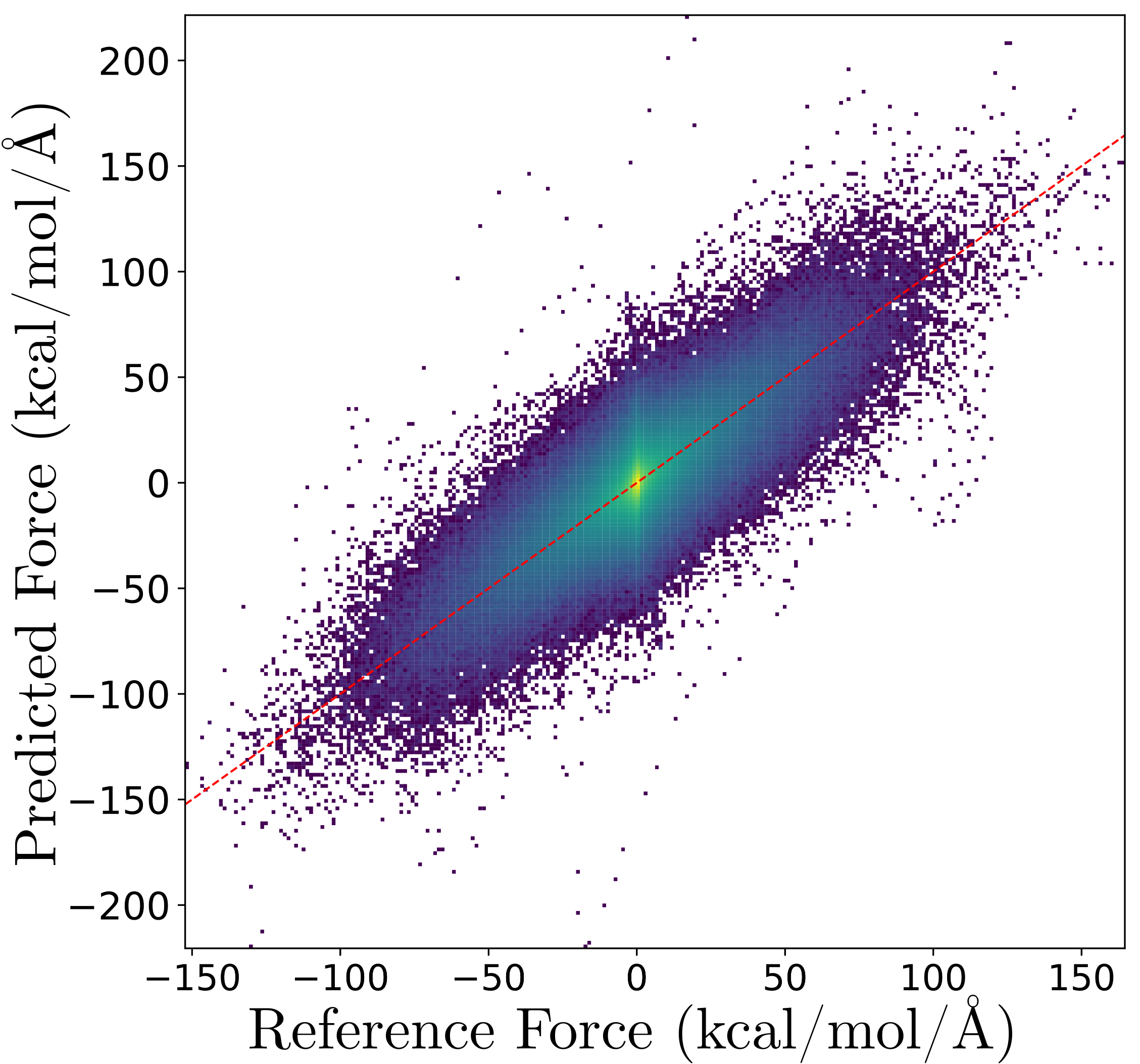}
        \caption{IRC Dataset - Force}
    \end{subfigure}
    \begin{subfigure}[b]{0.32\textwidth}
        \includegraphics[width=\textwidth]{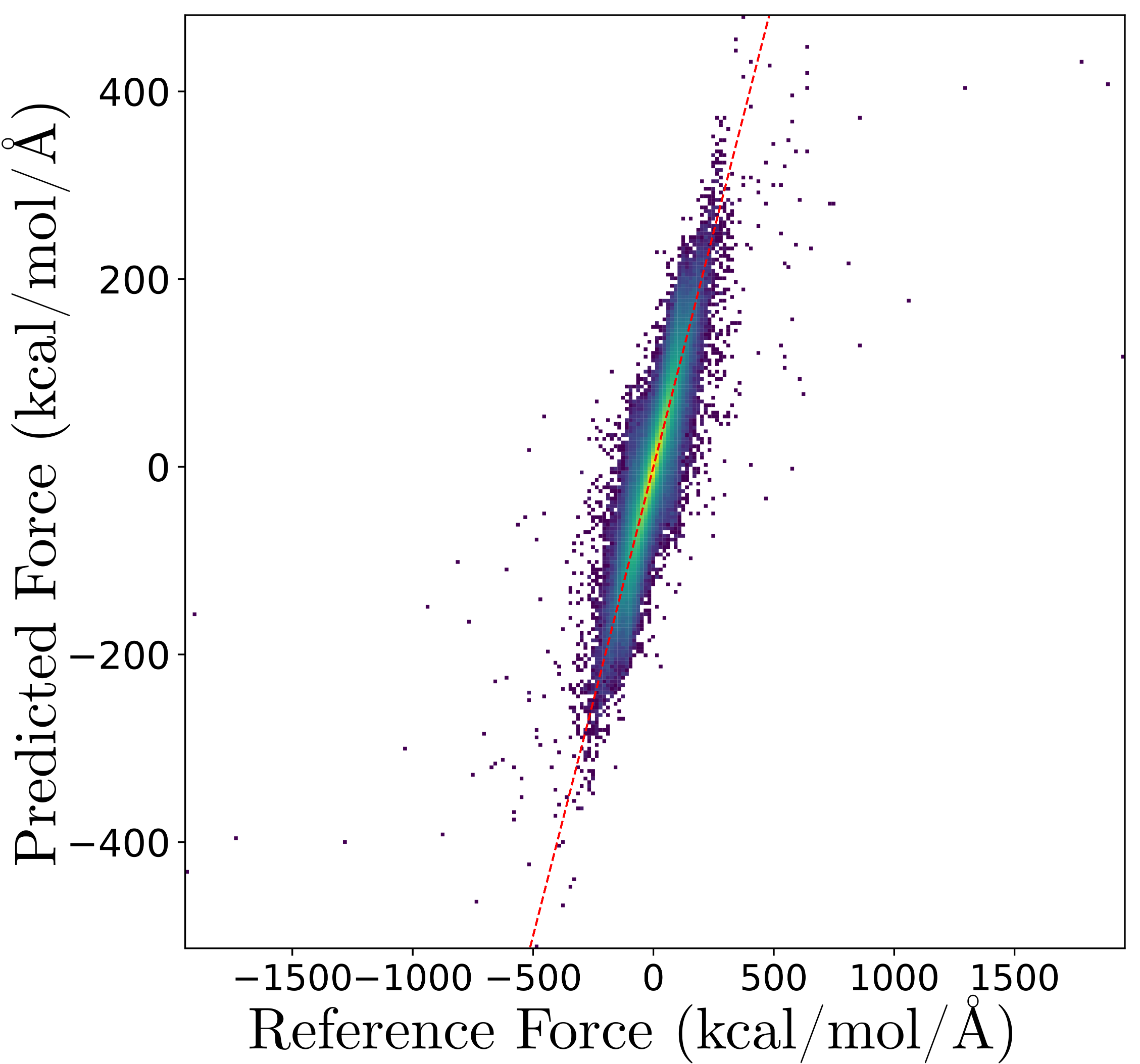}
        \caption{NMS Dataset - Force}
    \end{subfigure}
    \caption{Force Predictions of Energy-Force-Hessian Fitting Models}
    \label{fig:energy_force_Hessian_fitting_F}
\end{figure}


\begin{figure}[H]
    \centering
    \begin{subfigure}[b]{0.32\textwidth}
        \includegraphics[width=\textwidth]{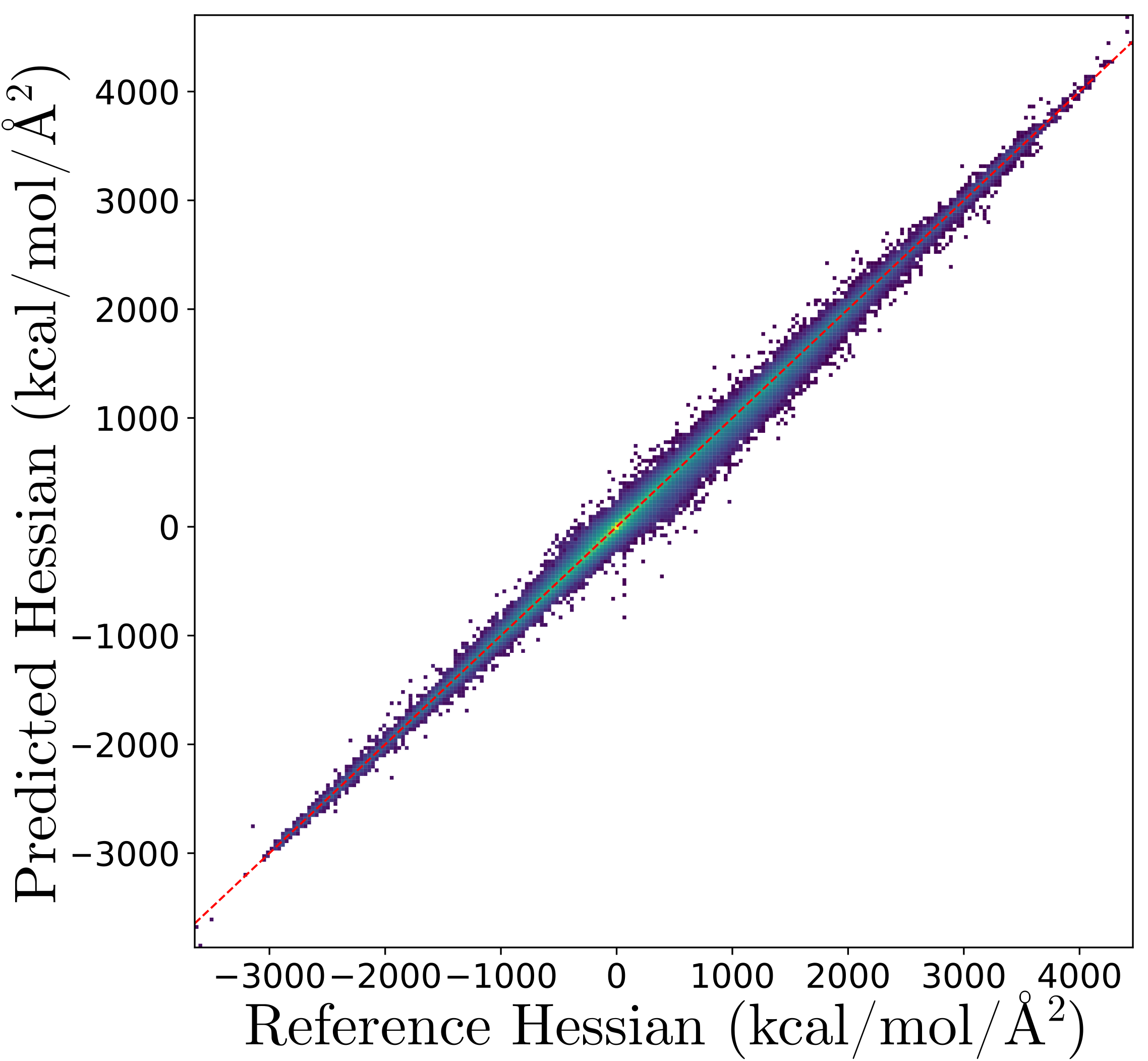}
        \caption{R-TS-P Dataset - Hessian}
    \end{subfigure}
    \begin{subfigure}[b]{0.32\textwidth}
        \includegraphics[width=\textwidth]{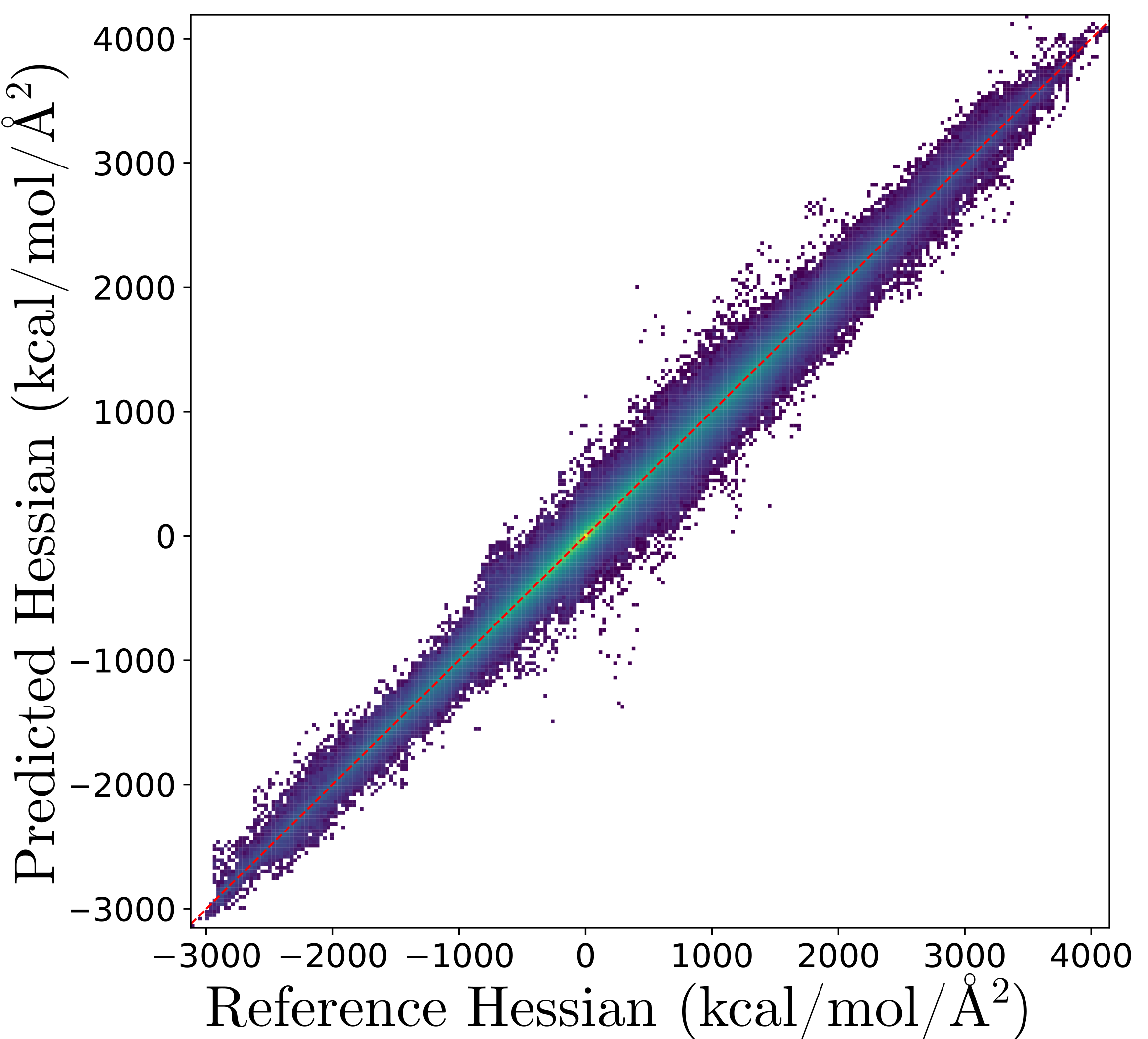}
        \caption{IRC Dataset - Hessian}
    \end{subfigure}
    \begin{subfigure}[b]{0.32\textwidth}
        \includegraphics[width=\textwidth]{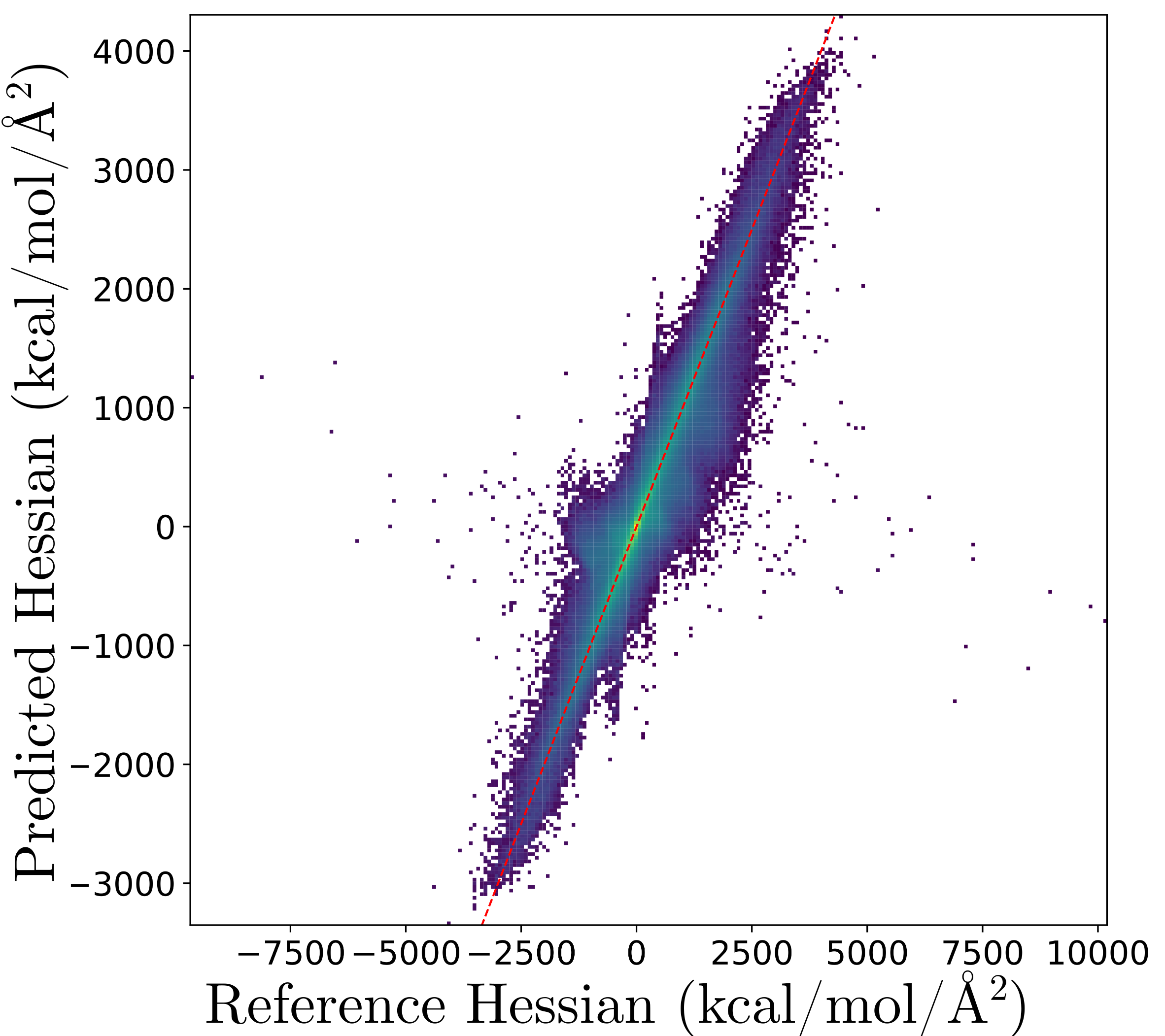}
        \caption{NMS Dataset - Hessian}
    \end{subfigure}
    \caption{Hessian Predictions of Energy-Force-Hessian Fitting Models}
    \label{fig:energy_force_Hessian_fitting_H}
\end{figure}

\section{Stability in Molecular Dynamics Simulations}

\begin{figure*}[h]
    \centering
    \includegraphics[scale=0.46]{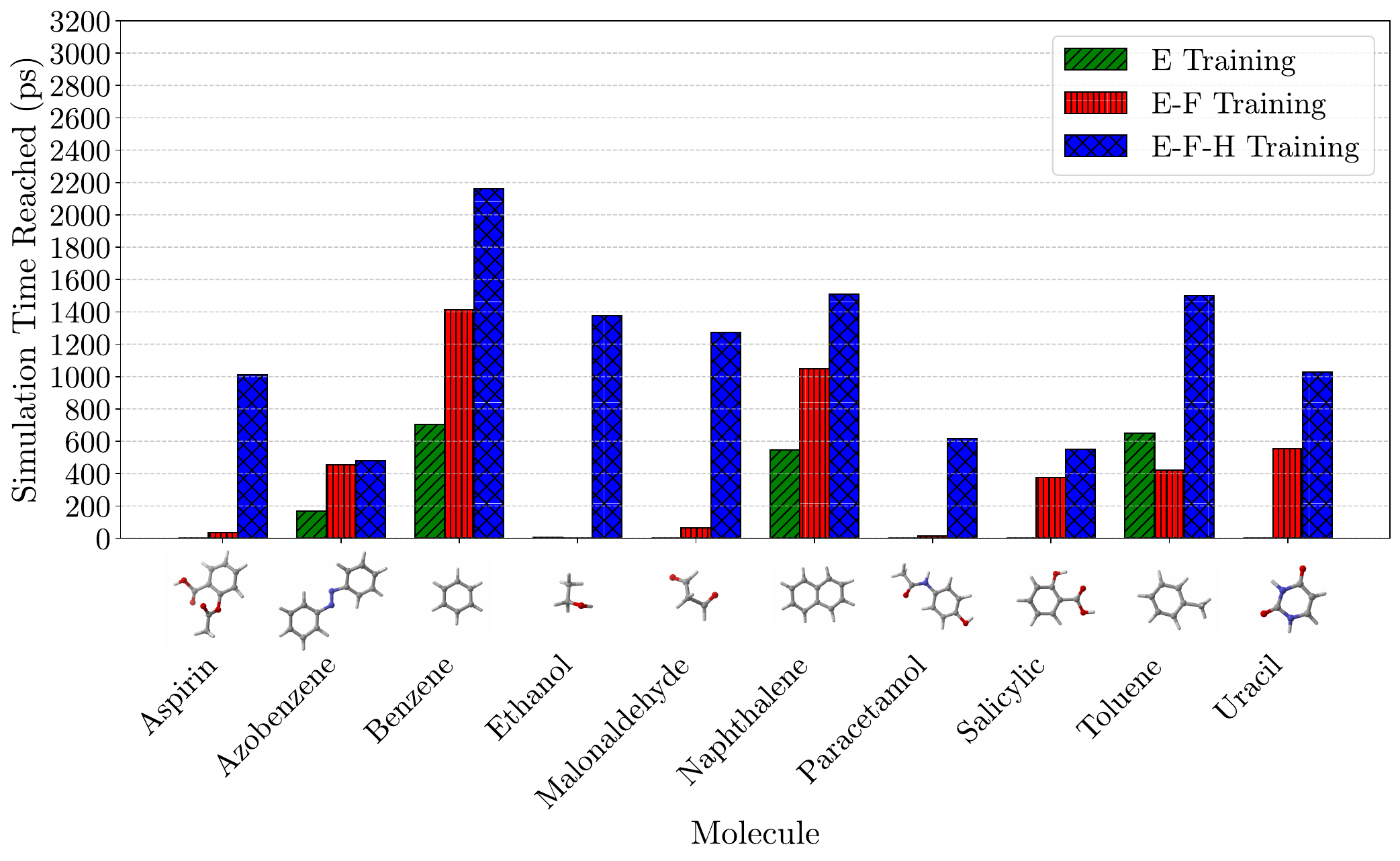}
    \caption{Simulation times reached before failure for each molecule in the MD17 dataset using ensembles of MLIP models trained with energy-only (E), energy-force (E-F), and energy-force-Hessian (E-F-H) loss functions.}
    \label{fig:md_times}
    \rule[1ex]{\textwidth}{0.1pt}
\end{figure*}

The results, visualized in Figure \ref{fig:md_times}, show the simulation times reached before failure for each molecule and highlight the significant differences between the three types of models. The E-F-H models exhibit substantially longer simulation times across all molecules, reflecting the enhanced stability achieved by incorporating Hessian information into the training process. The maximum temperatures and simulation times for each model type and molecule are summarized in Table S\ref{tab:md-results}, providing a detailed quantitative comparison of the model performance.

\begin{table*}[h!]
    \centering
    \begin{tabular}{l l r r}
        \toprule
        Molecule & Model & Temperature at Failure & Simulation Time at Failure\\
        & & {(\si{\kelvin})} & {(\si{\femto\second})}\\
        \midrule
        & E Training & $0$ & $0.0$\\
        Aspirin & E-F Training & $40$ & $35157.0$\\
        & E-F-H Training & $1015$ & $1011677.0$\\
        \midrule
        & E Training & $170$ & $167332.5$\\
        Azobenzene & E-F Training & $455$ & $453821.0$\\
        & E-F-H Training & $480$ & $478923.0$\\
        \midrule
        & E Training & $710$ & $705498.5$\\
        Benzene & E-F Training & $1415$ & $1412553.5$\\
        & E-F-H Training & $2160$ & $2159901.0$\\
        \midrule
        & E Training & $5$ & $4736.5$\\
        Ethanol & E-F Training & $0$ & $0.0$\\
        & E-F-H Training & $1380$ & $1377289.0$\\
        \midrule
        & E Training & $0$ & $0.0$\\
        Malonaldehyde & E-F Training & $70$ & $65582.0$\\
        & E-F-H Training & $1275$ & $1272913.0$\\
        \midrule
        & E Training & $550$ & $545645.0$\\
        Naphtalene & E-F Training & $1050$ & $1047211.5$\\
        & E-F-H Training & $1510$ & $1508725.0$\\
        \midrule
        & E Training & $0$ & $0.0$\\
        Paracetamol & E-F Training & $15$ & $13492.5$\\
        & E-F-H Training & $620$ & $617084.5$\\
        \midrule
        & E Training & $0$ & $0.0$\\
        Salicylic & E-F Training & $380$ & $377569.5$\\
        & E-F-H Training & $555$ & $551635.5$\\
        \midrule
        & E Training & $655$ & $650086.0$\\
        Toluene & E-F Training & $425$ & $423360.5$\\
        & E-F-H Training & $1505$ & $1501559.0$\\
        \midrule
        & E Training & $0$ & $0.0$\\
        Uracil & E-F Training & $560$ & $555103.0$\\
        & E-F-H Training & $1030$ & $1029069.0$\\
        \bottomrule
    \end{tabular}
    \caption{Maximum temperatures and total simulation times reached before failure for ensembles of MLIP models trained with energy-only (E), energy-force (E-F), and energy-force-Hessian (E-F-H) loss functions on each molecule in the MD17 dataset. Simulations that failed on optimization of the initial molecular geometry have a simulation time of 0.0 fs.}
    \label{tab:md-results}
\end{table*}

\clearpage

\section{NEB Analysis: Multi-Process Reaction}

\begin{figure}[h]
\centering
\includegraphics[scale=0.42]{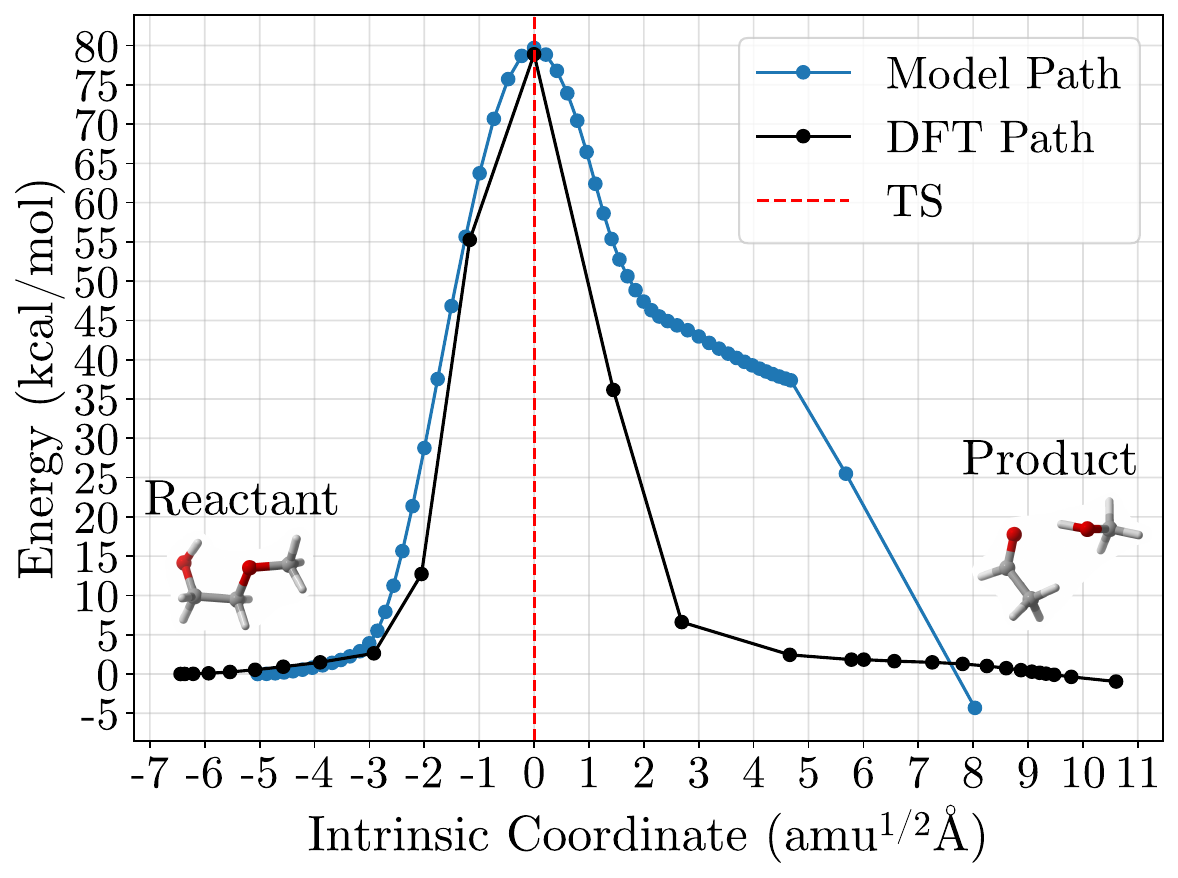}
\caption{Energy profiles obtained from NEB calculations using models trained with an E-F-H loss functions (blue dots) alongside DFT reference values (black dots). The x axis represents the geometric distances between intermediates as Intrinsic Coordinates, where a reference value of zero was assigned to the TS geometry (dashed vertical red line). The y axis represents the energies in \si{\kilo\cal\per\mole}. The E-only trained model and the E-F trained model were not able to converge in the NEB analysis of this reaction. This NEB represents a multi-process reaction where the EFH model does not completely reproduce. However, the energy barrier prediction compared to the DFT barrier is accurate. The atom coloring follows the CPK convention (red for oxygen, grey for carbon, and white for hydrogen).}
\label{fig:NEB_plot_SI}
\hrulefill
\end{figure}

To further assess the predictive accuracy and extrapolation capabilities of the Hessian-trained MLIP, we analyzed a multi-process reaction using Nudged Elastic Band (NEB) calculations. This reaction consists of multiple simultaneous processes, making it an ideal test case for evaluating how well the model captures reaction barriers and the overall shape of the potential energy surface (PES) beyond equilibrium structures.

While the activation energy predictions from the E-F-H model align well with DFT (78.88 \si{\kilo\cal\per\mole} from the model compared to 79.68 \si{\kilo\cal\per\mole} from DFT), significant deviations in the PES shape appear on the product side of the reaction coordinate. The MLIP-predicted energy profile does not smoothly reproduce the post-transition-state energy relaxation observed in DFT. Instead, the product basin is distorted, suggesting that the model does not fully capture the long-range relaxation effects or secondary reaction events that occur after the transition state.

\clearpage

\section{Data Efficiency for IRC Dataset}

\begin{figure*}[h!]
\captionsetup[subfigure]{justification=centering,font={stretch=0.8}}
\centering
\begin{subfigure}[t]{0.46\textwidth}
\centering
\includegraphics[scale=0.4]{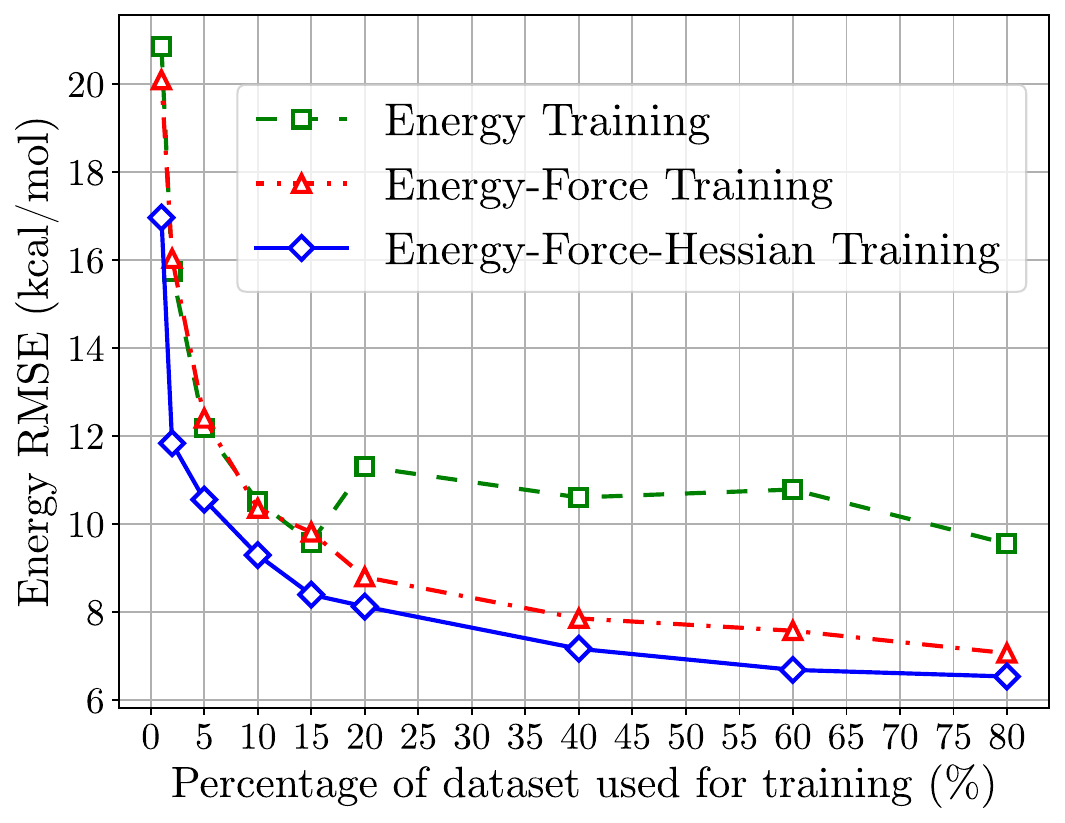}
\caption{Energy RMSEs}
\label{subfig:RMSE-E-IRC}
\end{subfigure}
~~
\begin{subfigure}[t]{0.46\textwidth}
\includegraphics[scale=0.4]{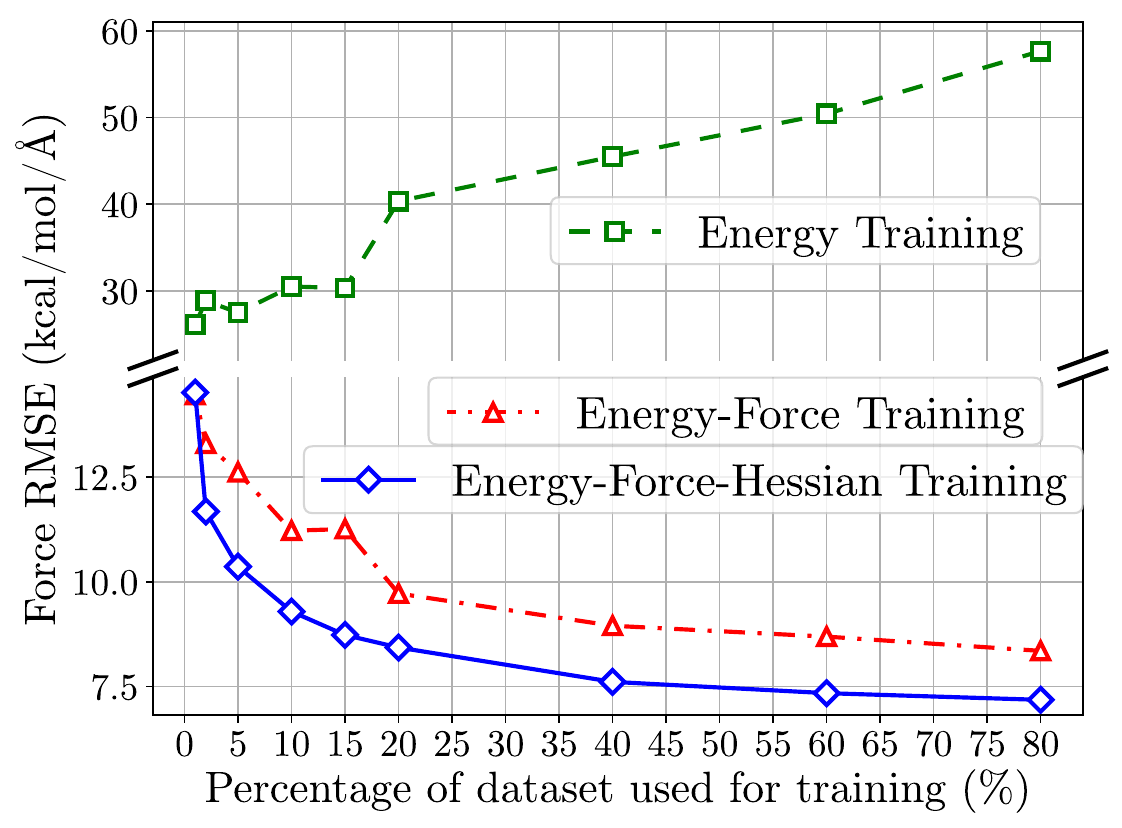}
\caption{Force RMSEs}
\label{subfig:RMSE-F-IRC}
\end{subfigure}
\\
\medskip
\begin{subfigure}[t]{0.46\textwidth}
\centering
\includegraphics[scale=0.4]{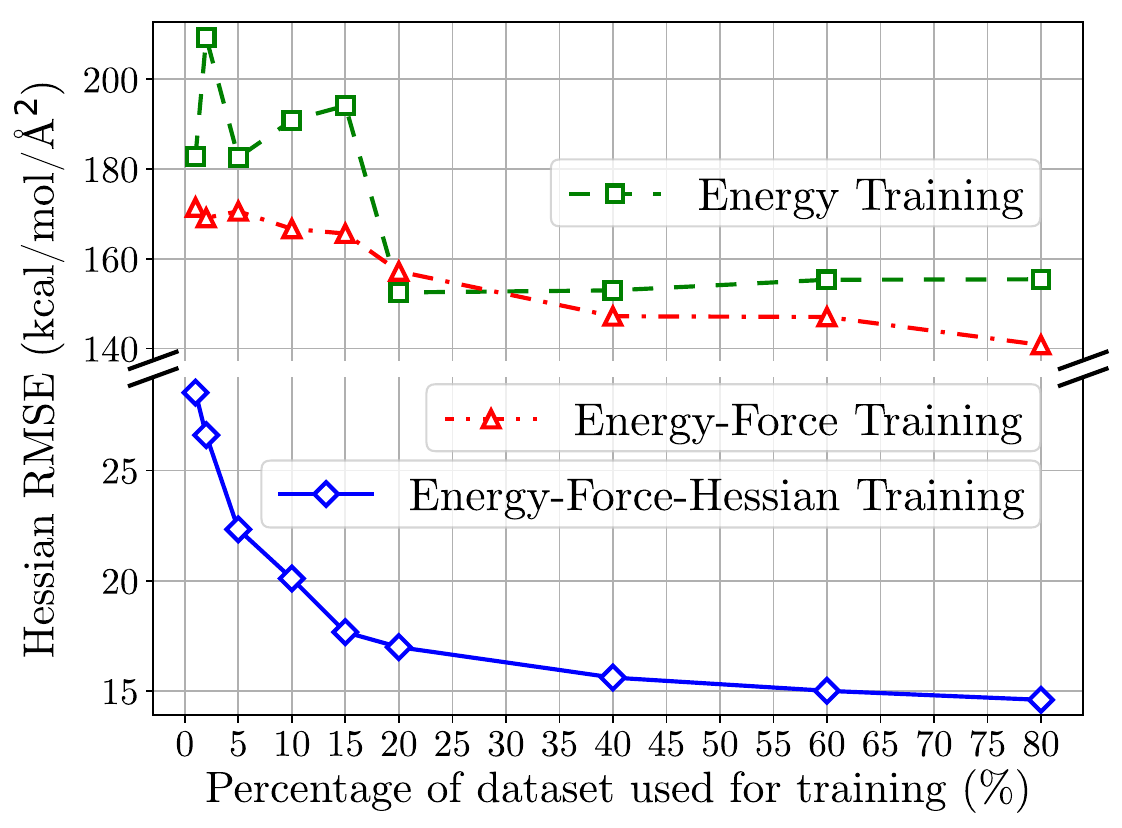}
\caption{Hessian RMSEs}
\label{subfig:RMSE-H-IRC}
\end{subfigure}

\caption{Energy, force, and Hessian root mean squared errors and average training time per epoch versus training data volume. The models shown in these figures were trained on reactants, transition states, and product of 11,961 elementary chemical reactions and tested on IRC structures of the first 2,000 reactions in the dataset.}
\label{fig:data-efficiency_IRC}
\rule[1ex]{\textwidth}{0.1pt}
\end{figure*}

Examining the dashed blue lines in Figure \ref{subfig:RMSE-E}, it becomes evident that the inclusion of Hessian information significantly improves the data efficiency of the MLIP models. Remarkably, the model trained on energies, forces and Hessian matrices (energy-force-Hessian fitting model) attains a comparable RMSE in energy predictions of intermediate structures in IRC paths using just 20\% of the total dataset volume, a performance parity achieved by the energy-force fitting model only when utilizing twice that amount of data. Furthermore, when the data volume for the Hessian energy force fitting model is increased to 40\% of the total volume of the data set, it almost achieves the RMSE performance in the energy predictions of the energy force fitting model trained with 80\% of the dataset. These observations underscore a profound implication: incorporating Hessian data into the training process potentially doubles the data efficiency of MLIP models.

\end{document}